\documentclass[11pt]{article}
\usepackage{bbm}
\usepackage{mathrsfs}
\usepackage{amsfonts}
\usepackage{amssymb}
\usepackage{epsfig}
\usepackage{amsfonts,amssymb, mathrsfs, amsmath,   amssymb,  theorem,  float}

\newtheorem{theorem}{Theorem}[section]

\newtheorem{cor}[theorem]{Corollary}
\newtheorem{lemma}[theorem]{Lemma}
\newtheorem{definition}[theorem]{Definition}
\newtheorem{remark}[theorem]{Remark}
\newtheorem{example}[theorem]{Example}

\floatstyle{ruled}
\newfloat{algorithm}{t}{loa}
\floatname{algorithm}{Algorithm}
\newcommand{\Inp}[1]
  {\noindent\begin{tabular}{@{}p{1.8cm}@{}p{13.2cm}@{}}
   {\bf Input: }&#1 \end{tabular}}
\newcommand{\Outp}[1]
  {\noindent\begin{tabular}{@{}p{1.8cm}@{}p{13.2cm}@{}}
   {\bf Output: }&#1 \end{tabular}}

\def\and{\cap}

\def\bref#1{(\ref{#1})}
\def\proof{{\noindent\em Proof:} }

\DeclareFontFamily{U}{fsy}{} \DeclareFontShape{U}{fsy}{m}{n}{<->s*[.
9]psyr}{} \DeclareSymbolFont{der@m}{U}{fsy}{m}{n}
\DeclareMathSymbol{\diff}{\mathord}{der@m}{182}

\newcommand{\SPC}{\hspace*{15pt}}
\newcommand{\qedd}{\hspace*{\fill}$\Box$\medskip}

\floatstyle{ruled}
\newfloat{algorithm}{t}{loa}
\floatname{algorithm}{Algorithm}

\def\X{{\mathbb{X}}}
\def\Y{{\mathbb{Y}}}
\def\U{{\mathbb{U}}}
\def\V{{\mathbb{V}}}
\def\L{{\mathbb{L}}}
\def\J{{\mathcal{J}}}

\def\A{{\mathcal A}}

\def\C{{\mathcal C}}

\def\CQ{\mathcal{Q}}
\def\PK{\P^{[\overrightarrow{K}]}}

\def\YK{\Y^{[\overrightarrow{K}]}}

\def\buk{\bu^{[\overrightarrow{K}]}}
\def\kk{\mathcal{K}}

\def\ff{\mathcal{F}}
\def\ee{\mathcal{E}}

\def\qq{\mathbb{Q}}

\def\PS{{\mathbb P}}

\def\SS{{\mathbb{S}}}
\def\CC{{\mathbb{S}}}

\def\G{{\mathbb G}}
\def\CI{{\mathcal {I}}}

\def\SP{{\mathbb{S}}}

\def\P{{\mathbb{P}}}
\def\Q{{\mathbb{Q}}}
\def\BF{{\mathbb{F}}}

\def\bu{{\mathbf{u}}}
\def\bv{{\mathbf{v}}}
\def\bc{{\mathbf{c}}}
\def\bz{{\mathbf{z}}}
\def\dzero{\mathbb{V}}

\def\sat{\hbox{\rm{sat}}}

\def\max{\hbox{\rm{max}}}

\def\Res{\hbox{\rm Res}}
\def\Span{\hbox{\rm Span}}
\def\vol{\hbox{\rm{vol}}}
\def\deg{\hbox{\rm{deg}}}

\def\init{\hbox{\rm{I}}}
\def\ord{\hbox{\rm{ord}}}
\def\Eord{\hbox{\rm{Eord}}}

\def\H{\hbox{\rm{H}}}
\def\lead{\hbox{\rm{ld}}}
\def\sep{\hbox{\rm{S}}}

\def\dim{\hbox{\rm{dim}}}

\def\mod{\hbox{\rm{mod}}}

\def\ord{\hbox{\rm{ord}}}
\def\rk{\hbox{\rm{rk}}}
\def\Jac{\hbox{\rm{Jac}}}

\def\chow{\hbox{\rm{Chow}}}

\def\coeff{\hbox{\rm{coeff}}}
\def\conv{\hbox{\rm{conv}}}
\def\I{\mathcal{I}}

\def\NP{\hbox{\rm{NP}}}

\def\SR{{\mathbf{R}}}

\def\Q{{\mathbb Q}}

\def\CC{{\mathbb C}}
\def\NN{{\mathbb N}}
\def\F{{\mathcal{F}}}

\def\TT{{\mathbb{T}}}
\def\JJ{{\mathbb{J}}}

\def\trdeg{\hbox{\rm{tr.deg}}}

\def\dtrdeg{\hbox{\rm{d.tr.deg}}}

\def\and{\cap}

\newcounter{bean}
\def\bl{\begin{list}{Step \arabic{bean}}{\usecounter{bean}}\labelwidth=34pt}
\def\el{\end{list}}

\def\deg{{\rm deg}}

\def\init{{\rm I}}

\def\normalization1{{\rm normalization1}}
\def\normalization{{\rm normalization}}

\def\irrfactor1{{\rm irrfactor1}}
\def\irrfactor{{\rm irrfactor}}

\def\card{\rm card}

\def\sat{{\rm sat}}

\def\rank{{\rm rk}}

\def\codim{{\rm Codim}}
\def\vf{\vskip5pt}
\def\vt{\vskip10pt}

\begin{document}

\title{Sparse Differential Resultant for\\ Laurent Differential Polynomials\thanks{Partially
       supported by a National Key Basic Research Project of China (2011CB302400) and  by grants from NSFC
       (60821002,11101411). Part of the results in this paper was reported in ISSAC
       2011~\cite{li} and received the Distinguished Paper Award.}}
\author{Wei Li, Chun-Ming Yuan, Xiao-Shan Gao\thanks{To whom correspondence should be addressed.}\\
KLMM, Academy of Mathematics and Systems Science\\ Chinese Academy of Sciences, Beijing 100190, China\\
Email: \{liwei,cmyuan,xgao\}@mmrc.iss.ac.cn }
\date{}

\maketitle

\begin{abstract}
\noindent In this paper, we first introduce the concept of Laurent
differentially essential systems and give a criterion for Laurent
differentially essential systems in terms of their supports. Then
the sparse differential resultant for a Laurent differentially
essential system is defined and its basic properties are proved. In
particular, order and degree bounds for the sparse differential
resultant are given. Based on these bounds, an algorithm to compute
the sparse differential resultant is proposed, which is single
exponential in terms of the number of indeterminates, the Jacobi
number of the system, and the size of the system.
%
%The Jacobi number and the size reflect the sparseness of the system


\end{abstract}

\tableofcontents

\section{Introduction}

The multivariate resultant, which gives conditions for an
over-determined  system of polynomial equations to have common
solutions, is a basic concept in algebraic geometry
\cite{eisenbud,gelfand,hodge,joun1,joun2,ph1,sturmfels}. In recent
years, the multivariate resultant is emerged as one of the most
powerful computational tools in elimination theory due to its
ability to eliminate several variables simultaneously without
introducing much extraneous solutions. Many algorithms with best
complexity bounds for problems such as polynomial equation solving
and first order quantifier elimination, are based on the
multivariate resultant
\cite{brownawell,canny1,emiris1,emiris2005,renegar1}.

In the theory of multivariate resultants, polynomials are assumed to
contain all the monomials with degrees up to a given bound. In
practical problems, most polynomials are sparse in that they only
contain certain fixed monomials. For such sparse polynomials, the
multivariate resultant often becomes identically zero and cannot
provide any useful information.

As a major advance in algebraic geometry and elimination theory, the
concept of sparse resultant was introduced by Gelfand, Kapranov,
Sturmfels, and Zelevinsky \cite{gelfand,sturmfels}.
The degree of the sparse resultant is the
Bernstein-Kushnirenko-Khovanskii (BKK) bound \cite{bkk} instead of
the Be\'{z}out bound \cite{gelfand, Pedersen,sturmfels2}, which
makes the computation of the sparse resultant more efficient.
The concept of sparse resultant is originated from the work of
Gelfand, Kapranov, and Zelevinsky on generalized hypergeometric
functions, where the central concept of $\mathcal {A}$-discriminant
is studied \cite{Gelfand10}.
Kapranov, Sturmfels, and Zelevinsky introduced the concept of
$\mathcal {A}$-resultant \cite{Kapranov1}.
Sturmfels further introduced the general mixed sparse resultant and
gave a single exponential algorithm to compute the sparse resultant
\cite{sturmfels,sturmfels2}.
%
%Basic properties for sparse resultant were proved
%\cite{sturmfels,sturmfels2}.
%
Canny and Emiris showed that the sparse resultant is a factor of the
determinant of a Macaulay style matrix and gave an efficient
algorithm to compute the sparse resultant based on this matrix
representation \cite{emiris0,emiris1}.
%\cite{canny2,emiris1}, which was shown to be asymptotically best.
D'Andrea further proved that the sparse resultant is the quotient of
two Macaulay style determinants similar to the multivariate
resultant \cite{dandrea1}.
%
%Other properties of sparse resultants can be found in
%\cite{dandrea1,gelfand,Pedersen,sturmfels,sturmfels2}.

Using the analogue between ordinary differential operators and
univariate polynomials, the differential resultant for two linear
ordinary differential operators was implicitly given by Ore
\cite{ore} and then studied by Berkovich and Tsirulik
\cite{berkovich} using Sylvester style matrices.
The subresultant theory was first studied by Chardin \cite{chardin1}
for two differential operators and then by Li \cite{zmli} and Hong
\cite{hhong} for the more general Ore polynomials.

For nonlinear differential polynomials, the differential resultant
is more difficult to define and study.
The differential resultant for two nonlinear differential
polynomials in one variable was defined by Ritt in
\cite[p.47]{ritt0}.
In \cite[p.46]{handbook}, Zwillinger proposed to define the
differential resultant of two differential polynomials as the
determinant of a matrix following the idea of algebraic multivariate
resultants, but did not give details.
General differential resultants were defined by Carr\`a-Ferro using
Macaulay's definition of algebraic resultants \cite{dres1}. But, the
treatment in \cite{dres1} is not complete. For instance, the
differential resultant for two generic differential polynomials with
positive orders and degrees greater than one is always identically
zero if using the definition in \cite{dres1}.
In \cite{yang-dixon}, Yang, Zeng, and Zhang used the idea of
algebraic Dixon resultant to compute the differential resultant.
Although efficient, this approach is not complete, because it is not
proved that the differential resultant can always be computed in
this way.
Differential resultants for linear ordinary differential polynomials
were studied by Rueda-Sendra \cite{lres1,sonia-arkiv}.
In \cite{gao}, a rigorous definition for the differential resultant
of $n+1$ differential polynomials in $n$ variables was first
presented and its properties were proved.
A generic differential polynomial with order $o$ and degree $d$
contains an exponential number of differential monomials in terms of
$o$ and $d$. Thus it is meaningful to study the sparse differential
resultant which is the main focus of this paper.

Our first observation is that the sparse differential resultant is
related with the non-polynomial solutions of algebraic differential
equations, that is, solutions with non-vanishing derivatives to any
order. As a consequence, the sparse differential resultant should be
more naturally defined for Laurent differential polynomials. This is
similar to the algebraic sparse resultant \cite{gelfand,sturmfels2},
where non-zero solutions of Laurent polynomials are considered.

Consider $n+1$ Laurent differential polynomials in $n$ differential
indeterminates $\Y=\{y_1,\ldots,y_n\}$:
\begin{equation} \label{eq-i01}
\P_i=\sum\limits_{k=0}^{l_i}u_{ik} M_{ik}\,(i=0,\ldots,n),
\end{equation}
where $u_{ik}\in\ee$ are differentially independent over $\qq$ and
$M_{ik}$ are Laurent differential monomials in $\Y$.
As explained later in this paper, we can assume that $M_{ik}$ are
monomials with non-negative exponent vectors $\alpha_{ik}$.
Let $s_i=\ord(\P_i,\Y)$ and denote
$M_{ik}/M_{i0}=\prod_{j=1}^n\prod_{l=0}^{s_i}(y_j^{(l)})^{t_{ikjl}}$
$\triangleq (\Y^{[s_i]})^{\alpha_{ik}-\alpha_{i0}}$, where
$y_j^{(l)}$ is the $l$-th derivative of $y_j$ and  $\Y^{[s_i]}$ is
the set $\{y_j^{(l)}:\,1\leq j\leq n, 0\leq l\leq s_i\}$. Let
$\bu_i=(u_{i0},u_{i1},\ldots,u_{il_i})\,(i=0,\ldots,n)$ be the
coefficient vector of $\P_i$.

The concept of Laurent differentially essential system is
introduced, which is a necessary and sufficient condition for the
existence of sparse differential resultant. $\P_0,\ldots,\P_n$ are
called Laurent differentially essential if
$[\P_0,\ldots,\P_n]\cap\Q\{\bu_0\ldots,\bu_n\}$ is a prime
differential ideal of codimension one, where $[\P_0,\ldots,\P_n]$ is
a differential ideal generated in
$\qq\{\Y,\Y^{-1};\bu_0,\ldots,\bu_n\}$.
This concept is similar to (but weaker than) the concept of
essential supports introduced by Sturmfels in \cite{sturmfels2}, but
its properties are more complicated. Precisely, we have
\begin{theorem}\label{th-i0}
For $\P_i$ given in \bref{eq-i01}, let $q_j =
\max_{i=0}^n{\ord(\P_i,y_j)}$ and
 $d_{ij} = \sum\limits_{k=0}^{l_i} u_{ik} \sum\limits_{l=0}^{q_j}
t_{ikjl}x_j^{l}\,$ $(i=0,\ldots,n; j=1,\ldots,n)$ where $x_j$ are
algebraic indeterminates. Denote
\[M_\P=\left(\begin{array}{cccccccc}
d_{01} & \,d_{02} & \,\ldots & \,d_{0n} \\
d_{11} & \,d_{12} & \,\ldots & \,d_{1n} \\
&  & \ddots & \\
d_{n1} & \,d_{n2} & \,\ldots & \,d_{nn}
\end{array}\right) \]
to be the {\em symbolic support matrix} of \bref{eq-i01}. Then the
following assertions hold. \begin{itemize}
\item[$1)$] The differential transcendence degree of
$\Q\langle\bu_0\ldots,\bu_n\rangle\langle\frac{\P_0}{M_{00}},\ldots,\frac{\P_n}{M_{n0}}\rangle$
over $\Q\langle\bu_0\ldots,\bu_n\rangle$ equals $\rank(M_\P)$.

\item[$2)$]
$[\P_0,\ldots,\P_n]\cap\Q\{\bu_0,\ldots,\bu_n\}$ is a prime
differential
 ideal of codimension $n+1-\rank(M_{\P})$.
So $\P_0,\ldots,\P_n$ form a Laurent differentially essential system
if and only if $\rank(M_\P)=n$.

\item[$3)$] $\P_0,\ldots,\P_n$ form a Laurent differentially essential
system if and only if there exist $k_i\,(1\le k_i\le l_i)$ such that
$\rank(M_{k_0,\ldots,k_n})=n$ where $M_{k_0,\ldots,k_n}$ is the
symbolic support matrix for the Laurent differential monomials
$M_{0k_0}/M_{00},\ldots, M_{nk_n}/M_{n0}$.

\end{itemize}
\end{theorem}
With the above theorem, computing the differential transcendence
degree of certain differential polynomials is reduced to computing
the rank of certain symbolic matrix. Similar to the case of linear
equations, this result provides a useful tool to study generic
differential polynomials.
As an application of the above result, the differential dimension
conjecture \cite[p.178]{ritt} for a class of generic differential
polynomials is proved.
%As a consequence, checking of Laurent differentially essential
%system is reduced to computing the rank of a symbolic matrix.
%
For the $n+1$ Laurent differential monomials
$M_{0k_0}/M_{00},\ldots, M_{nk_n}/M_{n0}\,(1\le k_i\le l_i)$
mentioned in 3) of Theorem \ref{th-i0}, a more efficient algorithm
to compute their differential transcendence degree over $\Q$ is
given by reducing their symbolic support matrix to a standard form
called T-shape.

\vskip 5pt Before introducing the properties of the sparse
differential resultant, the concept of Jacobi number is given below.
Let $\G=\{g_1,\ldots,g_n\}$ be $n$ differential polynomials in
$\Y=\{y_1,\ldots,y_{n}\}$. Let $s_{ij} = \ord(g_i,y_j)$ be the order
of $g_i$ in $y_j$ if $y_j$ occurs effectively in $f_i$ and $s_{ij} =
-\infty$ otherwise.
Then the {\em Jacobi bound}, or the {\em Jacobi number}, of $\G$,
denoted as $\Jac(\G)$, is the maximum number of the summations of
all the diagonals of $S$. Or equivalently,
$$\Jac(\G) = \max\sum_{i=1}^n s_{i\sigma(i)},$$
where $\sigma$ is a permutation of $\{1,\ldots,n\}$.
{\em Jacobi's Problem} conjectures that the order of the zero
dimensional component of $\G$ is bounded by the Jacobi number of
$\G$ \cite{ritt-jacobi}.

The properties of the sparse differential resultant are summarized
in the following theorem.
\begin{theorem}\label{th-i1}
The sparse differential resultant
$\SR(\bu_0,\ldots,\bu_n)\in\Q\{\bu_0,\ldots,\bu_n\}$ of
$\P_0,\ldots,\P_n$ has the following properties.
\begin{itemize}
\item[$1)$]
$\SR(\bu_0,\ldots,\bu_n)$ is differentially homogenous in each
$\bu_i\,(i=0,\ldots,n)$.

\item[$2)$]
 $h_i = \ord(\SR,\bu_i)\leq J_i=\Jac(\P_{\hat{i}})$ where $\P_{\hat{i}}=\{\P_0^N,\ldots,\P_n^N\}\backslash\{\P^N_i\}$.
 %, where $\Jac(S)$ is the Jacobi number of a differential polynomial set $S$.

\item[$3)$]
Let $\mathcal{Z}_0(\P_0,\ldots,\P_n)$ be the set of all
specializations of the coefficients $u_{ik}$ of $\P_i$ under which
$\P_i=0\,(i=0,\ldots,n)$ have a common non-polynomial solution and
$\overline{\mathcal{Z}_0(\P_0,\ldots,\P_n)}$ the Kolchin
differential closure of $\mathcal{Z}_0(\P_0,\ldots,\P_n)$. Then
$\overline{\mathcal{Z}_0(\P_0,\ldots,\P_n)}=\V\big(\sat(\SR )\big)$.

\item[$4)$]
Assume that $\P_i\,(i=0,\ldots,n)$ have the same set $\A$ of
monomials. The differential toric variety $X_\A$ associated with
$\A$ is defined and is shown to be an irreducible projective
differential variety of dimension $n$. Furthermore, the differential
Chow form \cite{gao,li} of $X_\A$ is $\SR$.

\item[$5)$] {\rm (Poison Type Product Formula)}
Let $\bu_0$ appear in $\SR$ and $t_0=\deg(\SR,u_{00}^{(h_0)})$. Then
there exist $\xi_{\tau k}$ in certain differential field $\F_\tau$
($\tau=1,\ldots,t_0$) such that
 $$ \SR=A\prod_{\tau=1}^{t_0} (u_{00}+\sum\limits_{k=1}^{l_0} u_{0k}\xi_{\tau k})^{(h_0)},$$
 where $A$ is a polynomial in
$\qq\langle\bu_1,\ldots,\bu_n\rangle[\bu_0^{[h_0]}\backslash
u_{00}^{(h_0)}]$.
Furthermore, if 1) any $n$  of the $\P_i\,(i=0,\ldots,n)$ form a
differentially independent set over
$\Q\langle\bu_0,\ldots,\bu_n\rangle$ and 2) for each $j=1,\ldots,n$,
$\textbf{e}_j\in\Span_{\mathbb{Z}}\{\alpha_{ik}-\alpha_{i0}:k=1,\ldots,l_i;i=0,\ldots,n\}$,
then there exist $\eta_{\tau k}\in \F_\tau$
\,$(\tau=1,\ldots,t_0;\,k=1,\ldots,n)$ such that
\begin{align} \label{eq-fac2}
\SR  & =
A\prod_{\tau=1}^{t_0}\bigg[\frac{\P_0(\eta_\tau)}{M_{00}(\eta_\tau)}\bigg]^{(h_0)},
\quad \nonumber
\end{align}
where  $\eta_\tau=(\eta_{\tau1},\ldots,\eta_{\tau n})$ and
$\textbf{e}_i$ is the exponent vector of $y_i$. Moreover,
$\eta_\tau\,(\tau=1,\ldots,t_0)$ are generic points of the prime
differential ideal
$[\P_1,\ldots,\P_n]:\mathbbm{m}\subset\ff\langle\bu_0,\ldots,$
$\bu_n\rangle\{\Y\}$, where $\mathbbm{m}$ is the set of all
differential monomials in $\Y$.

\item[$6)$]
$\deg(\SR)\leq \prod_{i=0}^n
    (m_i+1)^{h_i+1}\leq (m+1)^{\sum_{i=0}^n(J_i+1)}\leq (m+1)^{J+n+1}$, where  $m_i=\deg(\P_i,\Y)$,
    $m=\max_i\{m_i\}$,
    and $J = \sum_{i=0}^n J_i$.

\item[$7)$] Let $s_i=\ord(\P_i,\Y)$. Then
$\SR$ has a representation
 $$
 \prod_{i=0}^n M_{i0}^{(h_i+1)\deg(\SR)}\cdot
 \SR=\sum_{i=0}^n\sum_{j=0}^{h_i}G_{ij}\big(\P_{i}\big)^{(j)}
 $$
where
 $G_{ij}\in \Q[\bu_0^{[h_0]},\ldots,\bu_n^{[h_n]},\Y^{[h]}]$
with $h=\max\{h_i+s_i\}$ such that $\deg(G_{ij}(\P_{i})^{(j)})\leq
[m+1+\sum_{i=0}^n(h_i+1)\deg(M_{i0})]\deg(\SR)$.
\end{itemize}
\end{theorem}

Although similar to the properties of algebraic sparse resultants,
each property given above is an essential extension of its algebraic
counterpart. For instance, it needs lots of efforts to obtain the
Poison type product formula. Property 2) is unique for the
differential case and reflects the sparseness of the system in
certain sense.

More properties for the sparse differential resultant are proved in
this paper. For instance, the explicit condition for the equation
system  \bref{eq-i01} to have a unique solution for $\Y$ is given.
The sparse resultant for differential polynomials with non-vanishing
degree terms are also defined, which gives conditions for the
existence of solutions instead of non-polynomial solutions.

Let $\P_i\,(i=0,\ldots,n)$ in \bref{eq-i01} be generic differential
polynomials containing all monomials with order $\le s_i$ and degree
$\le m_i$ and $\SR(\bu_0,\ldots,\bu_n)$  the differential resultant
of $\P_0,\ldots,\P_n$. Then a BKK style degree bound is given:
\begin{theorem}\label{th-i3}
For each $i\in\{0,1,\ldots,n\}$,
 $$\deg(\SR,\bu_i)\leq\sum_{k=0}^{s-s_i}\mathcal
{M}\big((\CQ_{jl})_{j\neq i,0\leq l\leq
s-s_j},\CQ_{i0},\ldots,\CQ_{i,k-1},\CQ_{i,k+1},\ldots,\CQ_{i,s-s_i}\big)$$
where $\CQ_{jl}$ is the Newton polytope of $(\P_j)^{(l)}$ as a
polynomial in $y^{[s]}_1,\ldots,y^{[s]}_n$ and $\mathcal {M}(S)$ is
the mixed volume for the polytopes in $S$.
\end{theorem}

In principle, the sparse differential resultant can be computed with
the characteristic set method for differential polynomials via
symbolic computation \cite{ritt,bc1,ardm1,sit,wu1}.
%
%, and in particular with the change of order algorithms given by
%Boulier-Lemaire-Maza \cite{boulier2010} and
%Golubitsky-Kondratieva-Ovchinnikov \cite{alexey}. The Laurent
%differentially essential system already forms a triangular set when
%considering $u_{i0}$ as the leading variables, and the sparse
%differential resultant is the first element of the characteristic
%set of the prime ideal generated by the differentially essential
%system under a different special ranking. Therefore, the change of
%order strategy proposed in \cite{boulier2010,alexey} can be used.
%
But in general, differential elimination procedures based on
characteristic sets do not have an elementary complexity bound
\cite{alexey2}.

Based on the order and degree bounds given in 2) and 6) of Theorem
\ref{th-i1}, a single exponential algorithm to compute the sparse
differential resultant $\SR$ is proposed. The idea of the algorithm
is to compute $\SR$ with its order and degree increasing
incrementally and to use linear algebra to find the coefficients of
$\SR$ with the given order and degree. The order and degree bounds
serve as the termination condition. Precisely, we have

\begin{theorem}\label{th-i2}
With notations introduced in Theorem \ref{th-i1}, the sparse
differential resultant of $\P_0,\ldots,\P_n$ can be computed with at
most
$O\big(\frac{(J+n+2)^{O(l(J+1))}m^{O(l(J+1)(J+n+1))}}{n^{n}}\big)$
%$O(m^{O(nlJ^2)}(nJ)^{O(lJ)})$
$\Q$-arithmetic operations, where $l=\sum_{i=0}^n (l_i+1)$,
$m=\max_{i=0}^n m_i$, and $J=\sum_{i=0}^n J_i$.
%(*** Could we use
%$J=\sum_{i=0}^n J_i$ instead of $J=\max_{i=0}^n J_i$ in the
%complexity analysis?)
\end{theorem}
From Theorem \ref{th-i2}, the complexity of this algorithm is single
exponential in terms of %$n$,
$l$, and $J$. The sparseness is
reflected in the quantity $l$ which is called the size of the system
and the Jacobi number $J$.
%
%In the generic case, $l = {(s+1)n+m \choose m}$.
%
Note that even the complexity of computing the algebraic sparse
resultant is  single exponential \cite{sturmfels,emiris1}.
%, although it is much better than the complexity given in Theorem \ref{th-i2}.
%
%In \cite{emiris2005}, a probabilistic algorithm is given to compute
%the numerical evaluation of the algebraic resultant $\SR$, whose
%complexity is linear in $\deg(\SR)$.
%
The  algorithm  seems to be the first one to eliminate several
variables from nonlinear differential polynomials with a single
exponential complexity.
%
%The main difficulty to obtain better complexity bounds is due to the
%fact there exists no matrix representations for the sparse
%differential resultant. This issue will be discussed further in the
%conclusion section of this paper.

The rest of the paper is organized as follows.
In Section 2, preliminary results are introduced.
In Section 3, the sparse differential resultant for Laurent
differentially essential systems is defined.
In Section 4, Theorem \ref{th-i0} is proved.
In Section 5, properties 1) - 5) of Theorem \ref{th-i1} are proved.
In Section 6, properties 6) and 7) of Theorem \ref{th-i1} and
Theorems \ref{th-i3} and \ref{th-i2} are proved.
%
%In Section 7, properties of sparse differential resultants for
%differential polynomials with non-vanishing degree zero terms are given.
%
In Section 7, the paper is concluded and several unsolved problems
for differential sparse resultant are proposed.

\section{Preliminaries}
  In this section,   some basic notations and preliminary results in
 differential algebra will be given. For more details about differential algebra,
 please refer to \cite{ritt,kol,bc1,sit,gao}.

\subsection{Differential polynomial algebra and Kolchin topology}
Let $\mathcal {F}$ be a fixed ordinary differential field of
characteristic zero, with a derivation operator  $\delta$. An
element $c\in\ff$ such that $\delta c=0$ is called a constant of
$\ff.$ In this paper, unless otherwise indicated, $\delta$ is kept
fixed during any discussion and we use primes and exponents $(i)$ to
indicate derivatives under $\delta$. Let $\Theta$ denote the free
commutative semigroup with unit (written multiplicatively) generated
by $\delta$.

A typical example of differential field is $\mathbb{Q}(x)$ which is
the field of rational functions in a variable $x$ with
$\delta=\frac{d}{dx}$.

Let $S$ be a subset of a  differential field $\mathcal{G}$ which contains
$\mathcal {F}$.   We will   denote respectively by  $\mathcal
{F}[S]$, $\mathcal {F}(S)$, $\mathcal {F}\{S\}$, and $\mathcal
{F}\langle S\rangle$    the smallest subring, the smallest subfield,
the smallest differential subring, and the smallest  differential subfield of
$\mathcal    {G}$ containing $\mathcal {F}$ and $S$.  If we denote
$\Theta(S)$ to be the    smallest subset of $\mathcal {G}$
containing $S$ and stable under $\delta$,    we have $\mathcal
{F}\{S\}=\mathcal    {F}[\Theta(S)]$ and $\mathcal {F}\langle
S\rangle=\mathcal    {F}(\Theta(S))$. A differential extension field
$\mathcal{G}$ of $\ff$ is said to be finitely  generated if $\mathcal{G}$ has a
finite subset $S$ such that $\mathcal{G}=\ff\langle S\rangle$.

A subset $\Sigma$ of a  differential extension field $\mathcal {G}$
of $\mathcal {F}$ is said to be {\em differentially dependent} over
$\mathcal {F}$ if the set $(\theta\alpha)_{\theta \in \Theta,
\alpha\in\Sigma}$ is algebraically dependent over $\mathcal {F}$,
and is said to be {\em differentially independent} over $\mathcal
{F}$, or to be a family of {\em differential indeterminates} over
$\mathcal {F}$ in the contrary case.
In the case $\Sigma$ consists of one element $\alpha$, we say that
$\alpha$ is differentially algebraic or differentially
transcendental over $\mathcal {F}$ respectively. The maximal subset
$\Omega$ of $\mathcal {G}$ which are differentially independent over
$\mathcal {F}$ is said to be a differential transcendence basis of
$\mathcal {G}$ over $\mathcal {F}$. We use $\dtrdeg \,\mathcal
{G}/\mathcal {F}$ (see \cite[p.105-109]{kol}) to denote the {\em
differential transcendence degree} of $\mathcal {G}$ over $\mathcal
{F}$, which is the cardinal number of $\Omega$. Considering
$\mathcal {F}$ and $\mathcal {G}$ as ordinary algebraic fields, we
denote the algebraic transcendence degree of $\mathcal {G}$ over
$\mathcal {F}$ by $\trdeg\,\mathcal {G}/\mathcal {F}$.

A homomorphism $\varphi$ from a differential ring $(\mathcal
{R},\delta)$ to a differential ring $(\mathcal {S},\delta_1)$ is a
{\em differential homomorphism} if $\varphi\circ
\delta=\delta_1\circ \varphi$. If $\mathcal {R}_0$ is a common
differential subring of $\mathcal {R}$ and $\mathcal {S}$ and the
homomorphism $\varphi$ leaves every element of $\mathcal {R}_0$
invariant, it is said to be over $\mathcal {R}_0$. If, in addition
$\mathcal {R}$ is an integral domain and $\mathcal {S}$ is a
differential field, $\varphi$ is called a {\em differential
specialization} of $\mathcal {R}$ into $\mathcal {S}$ over $R_0$.
The following property about differential specialization will be
needed in this paper, which can be proved similarly to Theorem~2.16
in \cite{gao}.
\begin{lemma}\label{lm-special}
Let  $P_{i}(\U, \Y)\in \mathcal {F}\langle \Y\rangle\{\U\}$ $(i=1,
\ldots, m)$ where $\U=(u_{1},\ldots,u_{r})$ and $\Y=(y_{1}, \ldots,
y_{n})$ are sets of differential indeterminates.   If the set
$(P_{i}(\U, \Y))^{(\sigma_{ij})}\,(i=1, \ldots, m;
j=1,\ldots,n_{i})$ are algebraically dependent over $\mathcal
{F}\langle \U \rangle$, then for  any differential specialization
$\U$ to $\U^0\subset\mathcal {F}$ over $\ff$, $(P_{i}(\U^0,
\Y))^{(\sigma_{ij})}(i=1, \ldots, m; j=1,\ldots,n_{i})$ are
algebraically dependent over $\mathcal {F}$.
In particular, if $P_{i}(\U, \Y)$ $(i=1, \ldots, m)$ are
differentially dependent over $\mathcal {F}\langle \U \rangle$, then
for any differential  specialization $\U$ to
$\overline{\U}\subset\mathcal {F}$ over $\F$,
$P_{i}(\overline{\U},\Y) \, (i=1, \ldots,  m)$ are differentially
dependent over $\mathcal {F}$.
\end{lemma}

A differential extension field  $\mathcal {E}$ of $\ff$ is called a
{\em universal differential extension field}, if for any finitely
generated differential extension field $\ff_1$ of $\ff$ in $\mathcal
{E}$ and any finitely generated differential extension field $\ff_2$
of $\ff_1$ not necessarily in $\mathcal {E}$, $\ff_2$ can be
embedded in $\mathcal {E}$ over $\ff_1$, i.e. there exists a
differential extension field $\ff_3$ in $\mathcal {E}$ that is
differentially isomorphic to $\ff_2$ over $\ff_1$. Such a
differential universal extension field of $\ff$ always exists
(\cite[Theorem 2, p. 134]{kol}). By definition, any finitely
generated differential extension field of $\ff$ can be embedded over
$\ff$ into $\mathcal {E}$, and $\mathcal {E}$ is a universal
differential extension field of every finitely generated
differential extension field of $\ff$. In particular, for any
natural number $n$, we can find in $\ee$ a subset  of cardinality
$n$ whose elements are differentially independent over $\ff.$
Throughout the present paper, $\ee$ stands for a fixed universal
differential extension field of $\ff$.

Now suppose $\Y=\{y_{1}, y_{2}, \ldots, y_{n}\}$ is a set of
differential indeterminates over $\ee$. For any $y\in\Y$, denote
$\delta^ky$ by $y^{(k)}.$ The elements of $\mathcal
{F}\{\Y\}=\mathcal {F}[y_j^{(k)}:j=1,\ldots,n;k\in \mathbb{N}]$ are
called {\em differential polynomials} over $\ff$ in $\Y$,
 and $\mathcal {F}\{\Y\}$ itself is called the {\em differential polynomial ring } over $\ff$ in $\Y$.
A differential polynomial ideal $\mathcal {I}$ in $\mathcal {F}\{\Y\}$
is an ordinary algebraic ideal which is closed under derivation,
 i.e. $\delta(\mathcal {I})\subset\mathcal {I}$.
 And a prime (resp. radical) differential ideal is a
differential ideal which is prime (resp. radical) as an ordinary
algebraic polynomial ideal.
For convenience, a prime differential ideal is assumed not to be the
unit ideal in this paper.

By a {\em differential affine space} we mean any one of the sets
$\ee^n\,(n\in \mathbb{N}).$ An element $\eta=(\eta_1,\ldots,\eta_n)$
of $\ee^n$ will be called a point. Let $\Sigma$ be a subset of
differential polynomials in $\mathcal {F}\{\Y\}$. A point
$\eta=(\eta_{1},\ldots,\eta_{n}) \in \mathcal {E}^n$ is called a
differential zero of $\Sigma$ if $f(\eta)=0$ for any $f \in\Sigma$. The
set of differential zeros of $\Sigma$ is denoted by $\dzero(\Sigma)$,
which is called a {\em differential variety} defined over $\ff$.  The  differential varieties in $\ee^n$ (resp. the
 differential varieties in $\ee^n$ that are defined over $\ff$) are the
closed sets in a topology called the {\em Kolchin topology} (resp.
the Kolchin $\ff$-topology).

For a differential variety $V$ which is defined over $\ff$,  we denote
$\mathbb{I}(V)$ to be the set of all differential polynomials in
$\ff\{\Y\}$ that vanish at every point of $V$. Clearly,
$\mathbb{I}(V)$ is a radical differential ideal in $\ff\{\Y\}$. And
there exists a bijective correspondence between Kolchin $\ff$-closed
sets and radical  differential ideals in $\ff\{\Y\}$. That is, for any
 differential  variety $V$ defined over $\ff$, $\dzero(\mathbb{I}(V))=V$
and for any radical differential ideal $\mathcal{I}$ in $\ff\{\Y\}$,
$\mathbb{I}(\dzero(\mathcal {I}))=\mathcal {I}$.

Similarly as in algebraic geometry, an $\ff$-irreducible
differential variety can be defined. And there is a bijective
correspondence between $\ff$-irreducible differential varieties and
prime differential ideals in $\ff\{\Y\}$.
A point $\eta\in\V(\CI)$ is called a {\em generic point} of a prime
ideal $\CI\subset\ff\{\Y\}$, or of the irreducible variety
$\V(\CI)$, if for any polynomial $P\in\ff\{\Y\}$ we have $P(\eta)=0
\Leftrightarrow P\in\CI$.
%
%A prime differential ideal or an $\ff$-irreducible variety can be
%described by its generic point.
%
It is well known that \cite[p.27]{ritt}
%\begin{lemma}\label{lm-gp}
a non-unit differential ideal is prime if and only if it has a
generic point.
%\end{lemma}
%

Let $\mathcal {I}$ be a prime differential ideal in $\ff\{\Y\}$ and
$\xi=(\xi_{1},\ldots,\xi_{n})$ a generic point of $\mathcal {I}$
\cite[p.19]{kol}.
The {\em dimension} of $\mathcal {I}$ or of $\dzero(\mathcal {I})$
is defined to be the differential transcendence degree of  the
differential extension field $\mathcal {F}\langle
\xi_{1},\ldots,\xi_{n}\rangle$ over $\mathcal {F}$, that is,
$\dim(\mathcal {I})=\dtrdeg\, \mathcal {F}\langle
\xi_{1},\ldots,\xi_{n}\rangle/\mathcal {F}$.

We will conclude this section by introducing some basic concepts in
projective differential algebraic geometry which will be used in
Section \ref{sec-toric}. For more details, please refer to
\cite{kol74,li1}. And unless otherwise stated, in the whole paper,
we only consider the affine differential case.

For each $l\in\mathbb{N}$, consider a projective space
$\textbf{P}(l)$ over $\ee$. By a {\em differential projective space}
we mean any one of the sets $\textbf{P}(l)\,(l\in \mathbb{N}).$
Denote $z_0,z_1,\ldots,z_l$ to be the homogenous coordinates.
%Following Kolchin \cite{kol4}, we now introduce the concept of
%differentially homogenous polynomials.
%\begin{definition} \label{d-homogenous}
%A differential polynomial $p \in \mathcal {F}\{y_{0},\ldots,y_{n}\}$
%is called differentially homogenous of degree $m$ if for a new
%differential indeterminate $\lambda$, we have $p(\lambda
%y_{0},\lambda y_{1}\ldots,\lambda
%y_{n})=\lambda^{m}p(y_{0},y_{1},\ldots,y_{n}) $.
%\end{definition}
%
%
%The differential analog of Euler's theorem related to homogenous
%polynomials is valid.
%\begin{theorem}\cite{kol4} \label{th-dhomo}\,
% $f \in \mathcal{F}\{y_{0},y_{1},\ldots,y_{n}\}$ is differentially
%homogenous of degree $m$ if and only if \newline \[\sum_{j=0}^{n}
%\sum_{k \in \mathbb{N}} {k+r \choose r} y_{j}^{(k)} \frac{\partial
%f(y_{0},\ldots,y_{n})}{\partial y_{j}^{(k+r)} } = \left\{
%\begin{array}{ccc} mf & & r = 0 \\ 0 & & r \neq 0 \\ \end{array} \right.\]
%\end{theorem}
Let $\I$ be a differential ideal of $\ff\{\bz\}$ where
$\bz=\{z_0,z_1,\ldots,z_l\}$. Denote
$\I:\bz=\{f\in\ff\{\bz\}|\,z_jf\in\I\, \text{for each
$j=0,\ldots,l$}\}$.

\begin{definition}\label{def-diffhomo}
Let $\I$ be a differential ideal of $\ff\{\bz\}$. $\I$ is called a
{\em differentially homogenous differential ideal} of $\ff\{\bz\}$
if $\I:\bz=\I$ and for every $P\in\I$ and a differential
indeterminate $\lambda$ over $\ff\{\bz\}$,
$P(\lambda\bz)\in\ff\{\lambda\}\I$ in the differential ring
$\ff\{\lambda,\bz\}$.

\end{definition}
%For prime differential ideals, it is easier to detect their
%differential homogeneity \cite{kol74,li1}.
%\begin{lemma}
%\label{le-prime-homogenous} Let $\CI$ be a prime differential ideal
%of $\ff\{\bz\}$. Then  $\CI$ is differentially homogenous if and
%only if $\CI:\bz=\CI$ and for every zero $(\xi_0,\ldots,\xi_l)$ of
%$\CI$ in $\ee^{l+1}$ and each $s\in\ee\backslash\{0\}$,
%$(s\xi_0,\ldots,s\xi_l)$ is a zero of $\CI$.
%\end{lemma}

Consider a differential polynomial $P\in\ee\{\bz\}$ and a point
$\alpha\in \textbf{P}(l)$. Say that $P$ vanishes at $\alpha$, and
that $\alpha$ is a zero of $P$, if $P$ vanishes at $\lambda\alpha$
for every $\lambda$ in $\ee$. For a subset $\mathscr{M}$ of
$\textbf{P}(l)$, let $\mathbb{I}(\mathscr{M})$ denote the set of
differential polynomials
 in $\ff\{\bz\}$ that vanishes on  $\mathscr{M}$.
Let $\V(S)$ denote the set of points of $\textbf{P}(l)$ that are
zeros of the subset $S$ of $\ee\{\bz\}$. And a
subset $V$ of $\textbf{P}(l)$ is called a %{\em
%projective Differential  variety} (resp.
{\em projective differential $\ff$-variety}
 if there exists $S\subset\ff\{\bz\}$ %(resp. $S\subset\qq\{\Y_1,\ldots,\Y_p\}$)
 such that $V=\V(S)$.
There exists a one-to-one correspondence between projective
differential varieties and perfect differentially homogenous
differential ideals.  And  a projective differential $\ff$-variety $
V$ is $\ff$-irreducible if and only if $\mathbb{I}(V)$ is prime.

Let $\I$ be a prime differentially homogenous ideal and
$\xi=(\xi_0,\xi_1,\ldots,\xi_l)$ be a generic point of $\I$ with
$\xi_0\neq0$. Then the differential dimension of $\V(\I)$ is defined
to be the differential transcendence degree of
$\ff\langle(\xi_0^{-1}\xi_k)_{1\leq k\leq l}\rangle$ over $\ff$.

\subsection{Characteristic sets of a  differential polynomial system}

     Let $f$ be a differential polynomial in $\ff\{\Y\}$.
     We define the order of $f$ w.r.t. $y_i$ to be the greatest number $k$ such that $y_{i}^{(k)}$
appears effectively in $f$, which is denoted by $\ord(f,y_{i})$. And
if $y_{i}$ does not appear in $f$, then we set
$\ord(f,y_{i})=-\infty$.
     The {\em order} of $f$ is defined to be $\max_{i}\,\ord(f,y_{i})$, that is,
     $\ord(f)=\max_{i}\,\ord(f,y_{i})$.

     A {\em ranking} $\mathscr{R}$ is a total order over $\Theta (\Y)$, which is compatible with
     the derivations over the alphabet:

     1) $\delta \theta y_{j} >\theta y_{j}$ for all derivatives $\theta y_{j}\in\Theta (\Y)$.

     2) $\theta_{1} y_{i} >\theta_{2} y_{j}$ $\Longrightarrow$ $\delta\theta_{1} y_{i} >\delta\theta_{2} y_{j}$
for $\theta_{1} y_{i}, \theta_{2} y_{j}\in \Theta (\Y)$.

     By convention, $1<\theta y_{j}$ for all $\theta y_{j}\in \Theta (\Y)$.

    Two important kinds of rankings are the following:

    1) {\em Elimination ranking}: \ $y_{i} > y_{j}$ $\Longrightarrow$ $\delta^{k}  y_{i} >\delta^{l}
    y_{j}$ for any $k,    l\geq 0$.

    2) {\em Orderly ranking}: \
     $k>l$  $\Longrightarrow$  $\delta^{k}  y_{i} >\delta^{l}
    y_{j}$,    for any $i,    j \in \{1,    2,    \ldots,    n\}$.

    Let $p$ be a differential polynomial in
    $\mathcal {F}\{\Y\}$ and $\mathscr{R}$  a ranking
    endowed on it.  The greatest derivative w.r.t.  $\mathscr{R}$ which  appears effectively in $p$ is called the {\em leader} of $p$,
    which will be denoted by $u_{p}$ or $\lead(p)$.  The
    two conditions mentioned above imply that the leader of $\theta
    p$ is $\theta u_{p}$ for $\theta\in\Theta$.  Let the degree of $p$  in $u_{p}$ be $d$.
    As a univariate polynomial in $u_{p}$, $p$ can be rewritten as
     $$p=I_{d} u_{p}^{d}+I_{d-1}u_{p}^{d-1}+\cdots+I_{0}.$$
    $I_{d}$ is called the {\em initial} of $p$ and is denoted by $\init_{p}$.
    The partial derivative of $p$ w.r.t. $u_{p}$ is called the {\em
separant} of $p$,
    which will be denoted by $\sep_{p}$.  Clearly,    $\sep_{p}$
is the initial of any   proper derivative of $p$.
The {\em rank} of $p$ is $u_{p}^{d}$, and is denoted by $\rk(p)$.
%For any two
%   differential polynomials $p$, $q$ in $\mathcal {F}\{\Y\}\backslash \mathcal {F}$,
%    $p$  is said to be of {\em lower rank} than  $q$ if either  $u_{p}<u_{q}$
%    or  $u_{p}=u_{q}=u$ and $\deg(p,u)<\deg(q,u)$.
%    By convention, any element of $\mathcal {F}$ is of lower rank than elements of $\mathcal{F}\{\Y\}\backslash\mathcal
%    {F}$.  We denote $p \preceq
%    q$ if and only if either $p$ is of lower rank than $q$ or they have
% the same rank.  Clearly,    $\preceq$ is a totally ordering of
%$\mathcal{F}\{\Y\}$.

    Let $p$ and $q$ be two differential polynomials and $u_{p}^{d}$  the rank of $p$.  $q$ is said to be
    {\em partially reduced} w.r.t. $p$ if no proper derivatives of $u_{p}$ appear
    in $q$.  $q$ is said to be {\em reduced} w.r.t. $p$ if $q$ is partially reduced
    w.r.t. $p$ and $\deg(q,u_{p})<d$.  Let $\mathcal {A}$ be a set of
    differential polynomials.    $\mathcal {A}$ is said to be an
    {\em auto-reduced set} if each polynomial of $\mathcal {A}$ is reduced
    w.r.t.  any other element of $\mathcal{A}$.
   %Clearly, a set consisting of a single element belonging to $\mathcal {F}$ is an auto-reduced set,
    % which will be called the trivial ones.
   Every auto-reduced set is  finite.

Let $\mathcal {A}=A_{1},A_{2},\ldots,A_{t}$ be an auto-reduced
   set with $\sep_{i}$ and $\init_{i}$ as the separant and initial of $A_{i}$, and $f$ be any  differential polynomial.
   Then there exists an
   algorithm, called Ritt's algorithm of reduction, which reduces
   $f$ w.r.t. $\mathcal {A}$ to a  polynomial $r$ that is
   reduced w.r.t. $\mathcal {A}$, satisfying the relation
   $$\prod_{i=1}^t\sep_{i}^{d_{i}}\init_{i}^{e_{i}} \cdot f \equiv
   r, \mod\, [\mathcal {A}],$$
   where $d_{i},e_{i}\,(i=1,2,\ldots,t)$ are nonnegative integers.
  The differential polynomial $r$ is called the {\em differential remainder} of $f$ w.r.t. $\A$.

Let $\mathcal {A}$ be an auto-reduced set. Denote $\H_{\mathcal
{A}}$ to be  the set of all the initials and
    separants of $\mathcal {A}$ and $\H_{\mathcal {A}}^\infty$ to be the minimal
    multiplicative set containing $\H_{\mathcal {A}}$.
    The {\em saturation ideal} of $\A$ is defined to be
    $$\sat(\A)=[\mathcal
   {A}]:H_{\mathcal {A}}^\infty = \{p: \exists h\in H_{\mathcal
{A}}^\infty, \,{\text s. t. }\, hp\in[A]\}.$$

 An auto-reduced set $\mathcal {C}$ contained in a differential polynomial set
 $\mathcal {S}$ is said to be a {\em characteristic set} of $\mathcal {S}$,
 if  $\mathcal {S}$ does not contain any nonzero element reduced w.r.t.
$\mathcal {C}$. A characteristic set $\mathcal{C}$ of an ideal
$\mathcal {J}$ reduces to zero all elements of $\mathcal {J}$. If
the ideal is prime, $\mathcal {C}$ reduces to zero only the elements
of $\mathcal {J}$ and  $\mathcal {J}=\sat(\C)$ (\cite[Lemma 2,
p.167]{kol}) is valid.

In terms of the characteristic set, the cardinal number of the
characteristic set of $\I$ is equal to the codimension of  $\I$,
that is $n-\dim(\I)$. When $\I$ is of codimension one, it has the
following property.
\begin{lemma}\cite[p.45]{ritt} \label{le-char-codim1}
Let $\I$ be a prime differential ideal  of codimension one in
$\ff\{\Y\}$. Then there exists an irreducible differential
polynomial $A$ such that $\I=\sat(\A)$ and $\{A\}$ is the
characteristic set of $\I$ w.r.t.  any ranking.
\end{lemma}

\section{Sparse differential resultant  for Laurent differential
polynomials}\label{sec-lauresultant}

In this section, the concepts of Laurent differential polynomials
and Laurent differentially essential systems are first introduced,
and then the sparse differential resultant for Laurent
differentially essential systems is defined.

\subsection{Laurent differential polynomial}\label{subsec-laurent}
Let $\F$ be an ordinary differential field with a derivation
operator $\delta$ and $\F\{\Y\}$ the ring of differential
polynomials in the differential indeterminates
$\Y=\{y_1,\ldots,y_n\}$. Let $\ee$ be a universal differential field
of $\ff$. For any element $e\in \ee$, $e^{[k]}$ is used to denote
the set $\{e^{(i)}:\,i=0,\ldots,k\}$.

The sparse differential  resultant is closely related with Laurent
differential polynomials, which will be defined below.
\begin{definition}
A Laurent differential monomial of order $s$ is a Laurent monomial
in variables $\Y^{[s]}=(y_i^{(k)})_{1\leq i\leq n;0\leq k\leq s}$.
More precisely, it has the form
$\prod_{i=1}^n\prod_{k=0}^s(y_i^{(k)})^{d_{ik}}$ where $d_{ik}$ are
integers which can be negative. A {\em Laurent differential
polynomial} is a finite linear combination of Laurent differential
monomials with coefficients from $\ee$.
\end{definition}

Clearly, the collections of all Laurent differential polynomials
form a commutative differential ring under the obvious sum, product
operations and the usual derivation operator $\delta$, where all
Laurent differential monomials are invertible.
We denote the differential ring of Laurent differential polynomials
with coefficients in $\mathcal {F}$ by $\mathcal
{F}\{y_1,y_1^{-1},\ldots,y_n,y_n^{-1}\}$, or simply by
$\ff\{\Y,\Y^{-1}\}$.

\begin{remark}
$\ff\{\Y,\Y^{-1}\}=\mathcal {F}\{y_1,y_1^{-1},\ldots,y_n,y_n^{-1}\}$
is only a notation for Laurent differential polynomial ring. It is
not equal to $\mathcal {F}[y_{i}^{(k)},(y_{i}^{-1})^{(k)}:k\geq0]$.
\end{remark}

Denote $\mathcal {S}$ to be the set of all differential ideals in
$\mathcal {F}\{\Y,\Y^{-1}\}$, which are finitely generated.  Let
$\mathbbm{m}$ be the set of all differential monomials in $\Y$ and
$\mathcal {T}$ the set of all differential ideals in $\ff\{\Y\}$,
each of which has the form
 $$[f_1,\ldots,f_r]:\mathbbm{m}=\{f\in\ff\{\Y\}\big|\, \exists\,
M\in\mathbbm{m}, \,\text{\rm s.t.}\,\,M\cdot f\in
[f_1,\ldots,f_r]\}$$
for arbitrary $f_i\in\ff\{\Y\}$. Now we give a one-to-one
correspondence between $\mathcal {S}$ and $\mathcal {T}$.

The maps $\phi:\,\mathcal {S}\longrightarrow \mathcal {T}$ and
$\psi:\,\mathcal {T}\longrightarrow \mathcal {S}$ are defined as
follows:
\begin{itemize}
\item Given any  $\I=[F_1,\ldots,F_s]\in \mathcal{S}$. Since each
$F_i\in\ff\{\Y,\Y^{-1}\}$, we can choose a vector
$(M_1,\ldots,M_s)\in\mathbbm{m}^s $ such that
$M_iF_i\in\ff\{\Y\}\,(i=1,\ldots,s)$. We then define
$\phi(\I)\stackrel{\triangle}{=}[M_1F_1,\ldots,M_sF_s]:\mathbbm{m}\subset\ff\{\Y\}$.

\item For any  $\mathcal {J}=[f_1,\ldots,f_r]:\mathbbm{m}\in\mathcal{T}$, define
$\psi(\mathcal{J})=[f_1,\ldots,f_r]$ in $\ff\{\Y,\Y^{-1}\}$.
\end{itemize}

\begin{lemma}\label{le-one-to-one}
The above maps $\phi$ and $\psi$ are well defined. Moreover,
$\phi\circ \psi=\text{\rm id}_{\mathcal{T}}$ and $\psi\circ
\phi=\text{\rm id}_{\mathcal{S}}$.
\end{lemma}
\proof $\psi$ is obviously well-defined. To show that $\phi$ is
well-defined, it suffices to show that given another
$(N_1,\ldots,N_s)\in\mathbbm{m}^s$ with
$N_iF_i\in\ff\{\Y\}\,(i=0,\ldots,n)$,
$[M_1F_1,\ldots,M_sF_s]:\mathbbm{m}=[N_1F_1,\ldots,N_sF_s]:\mathbbm{m}$
follows. It follows directly from the fact that
$N_iF_i\in[M_1F_1,\ldots,M_sF_s]:\mathbbm{m}$ and
$M_iF_i\in[N_1F_1,\ldots,N_sF_s]:\mathbbm{m}$.

For any $\I=[F_1,\ldots,F_s]\in \mathcal{S}$,
$\psi\circ\phi(\I)=\psi([M_1F_1,\ldots,M_sF_s]:\mathbbm{m})=[M_1F_1,\ldots,M_sF_s]=
\I\subset\ff\{\Y,\Y^{-1}\}$ where $M_iF_i\in\ff\{\Y\}$, since
Laurent differential monomials are invertible. So we have $\psi\circ
\phi=\text{\rm id}_{\mathcal{S}}$. And for any $\mathcal
{J}=[f_1,\ldots,f_r]:\mathbbm{m}\in\mathcal{T}$, $\phi\circ
\psi(\mathcal {J})=\phi([f_1,\ldots,f_r])=\mathcal {J}$. Thus,
$\phi\circ \psi=\text{\rm id}_{\mathcal{T}}$ follows. \qedd

From the above, for a finitely generated Laurent differential ideal
$\I=[F_1,\ldots,F_s]$, although  $\phi(\I)$ is unique, different
vectors $(M_1,\ldots,M_s)\in\mathbbm{m}^s$ can be chosen to give
different representations for $\phi(\I)$. Now the norm form for a
Laurent differential polynomial is introduced to fix the choice of
$(M_1,\ldots,M_s)\in\mathbbm{m}^s$ when we consider $\phi(\I)$.

\begin{definition}
For every  Laurent differential polynomial $F\in\ee\{\Y,\Y^{-1}\}$,
there exists a unique laurent differential monomial $M$ such that 1)
$M\cdot F\in\ee\{\Y\}$ and 2) for any Laurent differential monomial
$T$ with $T\cdot F\in\ee\{\Y\}$, $T\cdot F$ is divisible by $M\cdot
F$ as differential polynomials. This $M\cdot F$ is defined to be the
{\em norm form} of $F$, denoted by $F^{\text{N}}$. The order of
$F^{\text{N}}$ is defined to be the {\em effective order of $F$},
denoted by $\Eord(F)$. Clearly, $\Eord(F)\leq\ord(F)$. And the
degree of $F$ is defined to be the {\em  degree of $F^{\text{N}}$},
denoted by $\deg(F)$.
\end{definition}

In the following, we consider zeros for Laurent differential
polynomials.

\begin{definition}
Let $\ee^{\wedge}=\ee\backslash\{a\in\ee\big|\,\exists
k\in\mathbb{N}, \,\text{\rm s.t.}\, a^{(k)}=0 \}$. Let $F$ be a
Laurent differential polynomial in $\ff\{\Y,\Y^{-1}\}$. A point
$(a_1,\ldots,a_n)\in(\ee^{\wedge})^n$ is called a {\em
non-polynomial differential zero} of $F$  if $F(a_1,\ldots,a_n)=0.$
\end{definition}

%For an ideal  $\I$ in $\ff\{\Y,\Y^{-1}\}$, the zero set of $\I$ is
%the set of common non-polynomial differential zeros of all Laurent
%differential polynomials in $\I$.

It becomes apparent why  non-polynomial elements in $\ee^{\wedge}$
are considered as zeros of Laurent differential polynomials when
defining the zero set of an ideal. If $F\in\I$, then
$(y_i^{(k)})^{-1}F\in\I$ for any positive integer $k$, and in order
for $(y_i^{(k)})^{-1}F$ to be meaningful, we need to assume
$y_i^{(k)}\ne0$.
We will see later in Example \ref{ex-sp12}, how non-polynomial
solutions are naturally related with the sparse differential
resultant.

\subsection{Definition of sparse differential resultant} \label{sec-defresultant}

In this section, the definition of the sparse differential resultant
will be given.
Since the study of sparse differential resultants becomes more
transparent if we consider not individual differential polynomials
but differential polynomials with indeterminate coefficients, the
sparse differential resultant for Laurent differential polynomials
with differential indeterminate coefficients will be defined first.
Then the sparse differential resultant for a given Laurent
differential polynomial system with concrete coefficients is the
value which the resultant in the generic case assumes for the given
case.

Suppose $\mathcal
{A}_i=\{M_{i0},M_{i1},\ldots,M_{il_i}\}\,(i=0,1,\ldots,n)$ where
$M_{ik}=\prod_{j=1}^n\prod_{l=0}^{s_i}(y_j^{(l)})^{d_{ikjl}}\triangleq
(\Y^{[s_i]})^{\alpha_{ik}}$ is a Laurent differential monomial of
order $s_i$ with exponent vector
$\alpha_{ik}\in\mathbb{Z}^{n(s_i+1)}$ and for $k_1\neq k_2$,
$\alpha_{ik_1}\neq \alpha_{ik_2}$. Consider $n+1$ {\em generic
Laurent differential polynomials} defined over $\mathcal
{A}_i\,(i=0,1,\ldots,n)$:
\begin{equation} \label{eq-sparseLaurent}
\P_i=\sum\limits_{k=0}^{l_i}u_{ik} M_{ik}\,(i=0,\ldots,n),
\end{equation} where all the $u_{ik}$ are  differentially independent over $\Q$.
The set of exponent vectors $\SP_i =\{ \alpha_{ik}:\,
k=0,\ldots,l_i\}$ is called the {\em support} of $\P_i$. The number
$|\SP_i| = l_i +1$ is called the {\em size} of $\P_i$.  Note that
$s_i$ is the order of $\P_i$ and an exponent vector of $\P_i$
contains $n(s_i+1)$ elements.  Denote
\begin{equation} \label{eq-uu}
 \bu_i=(u_{i0},u_{i1},\ldots,u_{in})\,(i=0,\ldots,n) \hbox{ and }
   \bu=\{u_{ik}:i=0,\ldots,n;k=1,\ldots,l_i\}.
\end{equation}
 To avoid the triviality,  $l_i\geq1\,(i=0,\ldots,n)$ are
always assumed in this paper.

\begin{definition}\label{def-tdes} A set of Laurent differential
polynomials of form \bref{eq-sparseLaurent} is called a {\em Laurent
differentially essential system} if there exist
$k_i\,(i=0,\ldots,n)$ with $1\leq k_i\leq l_i$ such that
$\dtrdeg\,\qq\langle\frac{M_{0k_0}}{M_{00}},\frac{M_{1k_1}}{M_{10}},\ldots,\frac{M_{nk_n}}{M_{n0}}\rangle/\qq=n.$
%there exist pairs $(i_k,j_k)$\,
%$(k=1,\ldots,n)$ with $1\leq j_k\leq l_{i_k}$ and $i_{k_1}\neq
%i_{k_2}$ $(k_1\neq k_2)$
% such that
%$\frac{M_{i_1j_1}}{M_{i_10}},\ldots,\frac{M_{i_nj_n}}{M_{i_n0}}$ are
%differentially independent over $\Q$,   where $\qq$ is the field of
%rational numbers.
%
In this case, we also say that $\mathcal{A}_0,\ldots,\mathcal{A}_n$
or  $\SP_0,\ldots,$ $\SP_n$ form a Laurent differentially essential
system.
\end{definition}

Although  $M_{i0}$ are used as denominators to define differentially
essential system, the following lemma shows that the definition does
not depend on the choices of $M_{i0}$.

\begin{lemma} \label{le-defsparse}
The following two conditions are equivalent.
\begin{enumerate}
\item  There exist
$k_0,\ldots,k_n$ with $1\leq k_i\leq l_i$ such that
$\dtrdeg\,\qq\langle\frac{M_{0k_0}}{M_{00}},\ldots,\frac{M_{nk_n}}{M_{n0}}\rangle/\qq=n.$
\item  There exist pairs
$(k_i,j_i)\,(i=0,\ldots,n)$ with $k_i\neq j_i\in\{0,\ldots,l_i\}$
such that\newline $\dtrdeg\,\qq\langle\frac{M_{0k_0}}{M_{0j_0}},$
$\ldots,\frac{M_{nk_n}}{M_{nj_n}}\rangle/\qq=n.$

%\item There exist triples $(i_k,j_k,m_k)\,(k=1,\ldots,n)$ with
%$j_k\neq m_k\in\{0,1,\ldots,l_{i_k}\}$ and $i_{k_1}\neq
%i_{k_2}\,(k_1\neq k_2)$ such that
%$\frac{M_{i_kj_k}}{M_{i_km_k}}\,(k=1,\ldots,n)$ are differentially
%independent over $\Q$.
%\item
%There exist pairs $(i_k,j_k)$\, $(k=1,\ldots,n)$ with
%$j_k\in\{1,\ldots,l_{i_k}\}$ and $i_{k_1}\neq i_{k_2}\,(k_1\neq
%k_2)$ such that
%$\frac{M_{i_1j_1}}{M_{i_10}},\ldots,\frac{M_{i_nj_n}}{M_{i_n0}}$ are
%differentially independent over $\Q$.
\end{enumerate}

\end{lemma}

\proof 1)$\Longrightarrow$ 2) is trivial.

Now suppose 2) holds. Fix the $n+1$ pairs $(k_i,j_i)$, and without
loss of generality, suppose
$\frac{M_{1k_1}}{M_{1j_1}},\ldots,\frac{M_{nk_n}}{M_{nj_n}} $ are
differentially independent over $\qq$. We need to show 1) holds.
Suppose the contrary. Then we know that for any
$m_i\in\{1,\ldots,l_{i}\}$,
$\frac{M_{1m_1}}{M_{10}},\ldots,\frac{M_{nm_n}}{M_{n0}}$ are
differentially dependent over $\Q$. Since
$\frac{M_{ik_i}}{M_{ij_i}}=\frac{M_{ik_i}}{M_{i0}}\big/\frac{M_{ij_i}}{M_{i0}}$,
it follows that $\frac{M_{ik_i}}{M_{ij_i}}\,(i=1,\ldots,n)$ are
differentially dependent over $\Q$, which is a contradiction. \qedd

Let $[\P_0,\ldots,\P_n]$ be the differential ideal in
$\qq\{\Y,\Y^{-1};\bu_0,\ldots,\bu_n\}$ generated by $\P_i$. By
Lemma~\ref{le-one-to-one}, $[\P_0,\ldots,\P_n]$ correspondents to
$[\P_0^N,\ldots,\P_n^N]:\mathbbm{m}\subset\qq\{\Y;\bu_0,\ldots,\bu_n\}$
in a unique way. Moreover,
we have the following lemma.

\begin{lemma}\label{le-intersectideal}
$[\P_0,\ldots,\P_n]\cap\qq\{\bu_0,\ldots,\bu_n\}=([\P_0^N,\ldots,\P_n^N]:\mathbbm{m})\cap\qq\{\bu_0,\ldots,\bu_n\}$.
\end{lemma}
\proof Denote $\P_i^N=M_i\P_i\,(i=0,\ldots,n)$ where $M_i$ are
Laurent differential monomials. It is obvious that the right
elimination ideal is contained in the left one. For the other
direction, let $G$ be any element in the left ideal. Then there
exist $H_{ij}\in\qq\{\Y,\Y^{-1};\bu_0,$$\ldots,\bu_n\}$ such that
$G=\sum_{i,j}H_{ij}\P_i^{(j)}$. So
$G=\sum_{i,j}H_{ij}\big(\frac{\P_i^N}{M_i}\big)^{(j)}=\sum_{i,j}\widetilde{H}_{ij}\big(\P_i^N\big)^{(j)}$
with $\widetilde{H}_{ij}\in\qq\{\Y,\Y^{-1};\bu_0,\ldots,\bu_n\}$.
Thus, there exists an $M\in\mathbbm{m}$ such that
$MG\in[\P_0^N,\ldots,\P_n^N]$ and
$G\in([\P_0^N,\ldots,\P_n^N]:\mathbbm{m})\cap\qq\{\bu_0,\ldots,\bu_n\}$
follows. \qedd

In the whole paper, when talking about prime differential ideals, it
is assumed that they are distinct from the unit differential ideal.
The following result is the foundation for defining the sparse
differential
 resultant.
\begin{theorem} \label{th-Mcodim1}
Let $\P_0,\ldots,\P_n$ be Laurent differential polynomials defined
in (\ref{eq-sparseLaurent}). Then the following
assertions hold. \begin{itemize}
\item $([\P_0^N,\ldots,\P_n^N]:\mathbbm{m})$ is a prime
differential ideal in $\Q\{\Y,\bu_0,$ $\ldots,\bu_n\}$.
\item
$([\P_0^N,\ldots,\P_n^N]:\mathbbm{m})\cap\Q\{\bu_0,\ldots,\bu_n\}$
is of codimension 1 if and only if  $\P_0,\ldots,\P_n$ form a
Laurent differentially essential system.
\end{itemize}
\end{theorem}
\proof  Let $\eta=(\eta_1,\ldots,\eta_n)$ be a generic point of
$[0]$ over $\Q\langle\bu\rangle$, where $\bu$ is defined in
\bref{eq-uu}. Let
\begin{equation}\label{eq-zeta}
\zeta_i=-\sum_{k=1}^{l_i}u_{ik}\frac{M_{ik}(\eta)}{M_{i0}(\eta)}\,\,(i=0,1,\ldots,n).
\end{equation}
Then we claim that
$\theta=(\eta_1,\ldots,\eta_n;\zeta_0,u_{01},\ldots,u_{0l_0};\ldots;\zeta_n,u_{n1},\ldots,u_{nl_n})$
is a generic point of $([\P_0^N,\ldots,\P_n^N]:\mathbbm{m})$, which
follows that $([\P_0^N,\P_1^N,\ldots,\P_n^N]:\mathbbm{m})$ is a
prime differential ideal.

Denote $\P_i^N=M_i\P_i\,(i=0,\ldots,n)$ where where $M_i$ are
Laurent differential monomials. Clearly, $\P_i^N=M_i\P_i$ vanishes
at $\theta$\,$(i=0,\ldots,n)$. For any
 $f\in([\P_0^N,\P_1^N,\ldots,\P_n^N]:\mathbbm{m})$,
there exists an $M\in\mathbbm{m}$ such that
$Mf\in[\P_0^N,\P_1^N,\ldots,\P_n^N]$. It follows that $f(\theta)=0$.
Conversely, let $f$ be any differential polynomial in
$\Q\{\Y,\bu_0,\ldots,\bu_n\}$ satisfying $f(\theta)=0$. Clearly,
$\P_0^N,\P_1^N,\ldots,\P_n^N$ constitute  an autoreduced set with
$u_{i0}$ as leaders. Let $f_1$ be the differential remainder of $f$
w.r.t. this autoreduced set. Then $f_1$ is free from
$u_{i0}\,(i=0,\ldots,n)$ and there exist $k_i\geq0$ such that
$\prod_{i=0}^n(M_iM_{i0})^{k_i}\cdot f\equiv
f_1,\mod\,[\P_0^N,\P_1^N,\ldots,\P_n^N]$. Clearly, $f_1(\theta)=0$.
Since $f_1\in\Q\{\bu,\Y\}$, $f_1=0$. Thus,
$f\in[\P_0^N,\P_1^N,\ldots,\P_n^N]:\mathbbm{m}$. So
$[\P_0^N,\P_1^N,\ldots,\P_n^N]:\mathbbm{m}$ is a prime differential
ideal with $\theta$ as its generic point.

Consequently,
$([\P_0^N,\P_1^N,\ldots,\P_n^N]:\mathbbm{m})\cap\Q\{\bu_0,\ldots,\bu_n\}$
is a prime differential ideal with a generic point
$\zeta=(\zeta_0,u_{01},\ldots,$ $u_{0l_0};\ldots;$
$\zeta_n,u_{n1},\ldots,u_{nl_n})$. From \bref{eq-zeta}, it is clear
that $\dtrdeg\,\Q\langle \zeta\rangle/\Q\leq \sum_{i=0}^nl_i+n$. If
there exist pairs $(i_k,j_k)$\, $(k=1,\ldots,n)$ with $1\leq j_k\leq
l_{i_k}$ and $i_{k_1}\neq i_{k_2}$ $(k_1\neq k_2)$ such that
$\frac{M_{i_1j_1}}{M_{i_10}},\ldots,\frac{M_{i_nj_n}}{M_{i_n0}}$ are
differentially independent over $\Q$, then by
Lemma~\ref{lm-special}, $\zeta_{i_1},\ldots,\zeta_{i_n}$ are
differentially independent over $\Q\langle\bu\rangle$. It follows
that $\dtrdeg\,\Q\langle \zeta\rangle/\Q= \sum_{i=0}^nl_i+n$. Thus,
$([\P_0^N,\P_1^N,\ldots,\P_n^N]:\mathbbm{m})\cap\Q\{\bu_0,\ldots,\bu_n\}$
is of codimension 1.

Conversely, assume that
$([\P_0^N,\P_1^N,\ldots,\P_n^N]:\mathbbm{m})\cap\Q\{\bu_0,\ldots,\bu_n\}$
is of codimension 1. That is,  $\dtrdeg\,\Q\langle \zeta\rangle/\Q=
\sum_{i=0}^nl_i+n$. We want to show that there exist pairs
$(i_k,j_k)$\, $(k=1,\ldots,n)$ with $1\leq j_k\leq l_{i_k}$ and
$i_{k_1}\neq i_{k_2}$ $(k_1\neq k_2)$
 such that
$\frac{M_{i_1j_1}}{M_{i_10}},\ldots,\frac{M_{i_nj_n}}{M_{i_n0}}$ are
differentially independent over $\Q$. Suppose the contrary,
 i.e.,
$\frac{M_{i_1j_1}(\eta)}{M_{i_10}(\eta)},\ldots,\frac{M_{i_nj_n}(\eta)}{M_{i_n0}(\eta)}$
are differentially dependent for any $n$ different $i_k$ and
$j_k\in\{1,\ldots,l_{i_k}\}$. Since each $\zeta_{i_k}$ is a linear
combination of $\frac{M_{i_kj_k}(\eta)}{M_{i_k0}(\eta)}$
$(j_k=1,\ldots,l_{i_k})$, it follows that
$\zeta_{i_1},\ldots,\zeta_{i_n}$ are differentially dependent over
$\Q\langle\bu\rangle$. Thus, we have $\dtrdeg\,\Q\langle
\zeta\rangle/\Q< \sum_{i=0}^nl_i+n$, a contradiction to the
hypothesis.\qedd

Combining Lemma~\ref{le-intersectideal} and
Theorem~\ref{th-Mcodim1}, we have
\begin{cor}
$[\P_0,\P_1,\ldots,\P_n]\cap\Q\{\bu_0,\ldots,\bu_n\}$ is a prime
differential ideal of codimension one if and only if $\{\P_i:
i=0,\ldots,n\}$ is a Laurent differentially essential system.
\end{cor}

Now suppose $\{\P_0,\ldots,\P_n\}$ is a Laurent differentially
essential system. Denote the differential ideal
$[\P_0,\P_1,\ldots,\P_n]\cap\Q\{\bu_0,\ldots,\bu_n\}$ by $\I$. Since
$\I$ is of codimension one, by Lemma~\ref{le-char-codim1}, there
exists an irreducible differential polynomial $\SR(\bu;
u_{00},\ldots,u_{n0})=\SR(\bu_{0},\ldots,\bu_{n})$
$\in\Q\{\bu_{0},\ldots,\bu_{n}\}$ such that
 \begin{equation}\label{eq-lsres}
[\P_0,\P_1,\ldots,\P_n]\cap\Q\{\bu_0,\ldots,\bu_n\}=
  \sat(\SR)\end{equation}
where $\sat(\SR)$ is the {\em saturation ideal} of $\SR$. More
explicitly, $\sat(\SR)$ is the whole set of differential polynomials
having zero differential remainders  w.r.t. $\SR$ under any ranking
endowed on $\bu_{0},\ldots,\bu_{n}$.

Now  the definition of sparse differential resultant is given as
follows:
\begin{definition}\label{def-sparse}
$\SR(\bu_{0},\ldots,\bu_{n})\in\Q\{\bu_{0},\ldots,\bu_{n}\}$ in
\bref{eq-lsres} is defined to be the {\em sparse differential
resultant} of the Laurent differentially essential system
$\P_0,\ldots,\P_n$, denoted by
$\Res_{\mathcal{A}_0,\ldots,\mathcal{A}_n}$ or
$\Res_{\P_0,\ldots,\P_n}$. And when all the $\mathcal{A}_i$ are
equal to the same $\mathcal{A}$, we simply denote it by
$\Res_\mathcal{A}$.
\end{definition}

From the proof of Theorem~\ref{th-Mcodim1} and equation
\bref{eq-lsres}, $\SR$ has the following useful property.

\begin{cor}\label{cor-r1}
Let $\SR(\bu_0,\ldots,\bu_n)$ be the sparse differential resultant
of $\P_0,\P_1,\ldots,\P_n$. Then $\sat(\SR)\subset
\Q\{\bu_0,\ldots,\bu_n\}$ is a prime differential  ideal with a
generic zero  $(\bu;\zeta_0,\ldots,\zeta_n)$, where $\zeta_i$ are
defined in \bref{eq-zeta}.
\end{cor}

We give five examples which will be used throughout the paper.

\begin{example} \label{ex-1}
Let $n=2$ and $\P_i$ has the form
$$\P_i=u_{i0}y_1''+u_{i1}y_1'''+u_{i2}y_2'''\,(i=0,1,2).$$
It is easy to show that $y_1'''/y_1''$ and $y'''_2/y''_1$ are
differentially independent over $\Q$. Thus, $\P_0,\P_1,\P_2$ form a
Laurent differentially essential system. The sparse differential
resultant is \[\SR= \Res_{\P_0,\P_1,\P_2}=\left|\begin{array}{lll}
u_{00}&u_{01}&u_{02}\\
u_{10}&u_{11}&u_{12}\\
u_{20}&u_{21}&u_{22}
\end{array}  \right| .\]
 Pay attention to the fact
that $\SR$ does not belong to the differential ideal generated by
$\P_i$ in $\Q\{\Y;\bu_0,\ldots,\bu_n\}$ because each $\P_i$ is
homogenous in $y_1'',y_1''',y_2'''$ and $\SR$ does not involve $\Y$.
That is why we use the ideal
$([\P_0,\P_1,\P_2]:\mathbbm{m})\subset\qq\{\Y;\bu_0,\ldots,\bu_n\}$
rather than $[\P_0,\P_1,\P_2]\subset\qq\{\Y;\bu_0,\ldots,\bu_n\}$ in
Theorem~\ref{th-Mcodim1}. Of course, $\SR$ does belong to
$[\P_0,\ldots,\P_n]$ when regarded as a differential ideal of the
Laurent differential polynomial ring
$\qq\{\Y,\Y^{-1};\bu_0,\ldots,\bu_n\}$.
\end{example}

The following example shows that for a Laurent differentially
essential system, its sparse differential resultant may not involve
the coefficients of some $\P_i$.
\begin{example}\label{ex-2}
Let $n=2$ and $\P_i$ has the form
$$\P_0=u_{00}+u_{01}y_1y_1',\,
  \P_1=u_{10}+u_{11}y_1,\,
\P_2=u_{10}+u_{11}y'_2.$$
Clearly, $\P_0,\P_1,\P_2$ form a Laurent differentially essential
system. And the sparse differential resultant of $\P_0,\P_1,\P_2$ is
$$\SR=u_{01}u_{10}(u_{11}u_{10}'-u_{10}u_{11}')+u_{00}u_{11}^3,$$
which is free from the coefficients of $\P_2.$
\end{example}

\begin{example}\label{ex-3}
Let $\mathcal{A}_0=\{\textbf{1},y_1y_2\}$,
$\mathcal{A}_1=\{\textbf{1},y_1y'_2\}$ and
$\mathcal{A}_2=\{\textbf{1},y'_1y'_2\}$.    It is easy to verify
that $\mathcal{A}_0,\mathcal{A}_1,\mathcal{A}_2$ form a Laurent
differentially essential system. And
$\Res_{\mathcal{A}_0,\mathcal{A}_1,\mathcal{A}_2}=u_{10}u_{01}u_{21}u_{11}u'_{00}-u_{10}u_{00}u_{11}u_{21}u'_{01}
-u_{01}^2u_{21}u_{10}^2-u_{01}u_{00}u_{11}^2u_{20}.$

\end{example}
\begin{example}
Let $n=1$ and
$\mathcal{A}_0=\mathcal{A}_1=\{y_1^2,(y_1')^2,y_1y_1'\}$. Clearly,
$\mathcal{A}_0,\mathcal{A}_1$ form a Laurent differentially
essential system and
$\Res_{\mathcal{A}}=u_{11}^2u_{00}^2-2u_{01}u_{10}u_{11}u_{00}+u_{01}^2u_{10}^2-
u_{12}u_{02}u_{11}u_{00}-u_{12}u_{02}u_{01}u_{10}+u_{12}^2u_{01}u_{00}+u_{10}u_{11}u_{02}^2$.
\end{example}
\begin{example}\label{ex-z1z2-z2'}
Let $n=1$ and $\mathcal{A}_0=\mathcal{A}_1=\{y_1,y_1',y_1^2\}$.
Clearly, $\mathcal{A}_0,\mathcal{A}_1$ form a Laurent differentially
essential system and $\Res_{\mathcal{A}}=-u_{12}u_{01}u_{00}u_{10}
-u_{12}u_{01} ^2u'_{10}+u_{12}u_{01}u'_{11}u_{00} +
u_{12}u_{01}u_{11}u'_{00}-u_{11}u_{02}u_{00}u_{10}+u_{11}u_{02}u'_{10}u_{01}
+
u_{02}u_{01}u_{10}^2-u_{11}^2u_{02}u'_{00}+u_{11}u_{02}u'_{01}u_{10}
+u_{11}u_{00}^2u_{12}$ $ +
u_{11}^2u'_{02}u_{00}-u_{11}u'_{02}u_{01}u_{10}
-u_{11}u_{01}u'_{12}u_{00} + u_{01}^2u'_{12}u_{10}
-u_{11}u'_{01}u_{12}u_{00}-u'_{11}u_{02}u_{01}u_{10}. $
\end{example}

\begin{remark}
When all the $\mathcal{A}_i\,(i=0,\ldots,n)$ are sets of
differential monomials, unless explicitly mentioned, we always
consider $\P_i$ as Laurent differential polynomials. But when we
regard $\P_i$ as differential polynomials,
$\Res_{\mathcal{A}_0,\ldots,\mathcal{A}_n}$ is also called the
sparse differential resultant of the differential polynomials
$\P_i$. In this paper, sometimes we regard $\P_i$ as differential
polynomials where we will highlight it.
\end{remark}

We now define the sparse differential resultant for any set of
specific Laurent differential polynomials over a Laurent
differentially essential system.
For any finite set $\mathcal{A}$ of Laurent differential monomials,
denote by $\mathcal {L}(\mathcal {A})$ the set of Laurent
differential polynomials of the form $\sum_{M\in\mathcal{A}}a_{M}M$
where $a_{M}\in  \mathcal{E}$. Then  $\mathcal {L}(\mathcal {A})$
can be considered as the affine space $\mathcal{E}^l$ or the
projective space $\textbf{P}(l-1)$ over $\mathcal{E}$ where
$l=|\mathcal{A}|$.

\begin{definition}\label{def-sparse1}
Let $\mathcal
{A}_i=\{M_{i0},M_{i1},\ldots,M_{il_i}\}\,(i=0,1,\ldots,n)$ be finite
sets of Laurent differential monomials which form a Laurent
differentially essential system. Consider $n+1$ Laurent differential
polynomials
$(F_0,F_1,\ldots,F_n)\in\prod_{i=0}^n\mathcal{L}(\mathcal{A}_i)$.
The sparse differential resultant of $F_0,F_1,\ldots,F_n$, denoted
as $\Res_{F_0,\ldots,F_n}$, is obtained by replacing $\bu_i$ by the
corresponding coefficient vector of $F_i$ in
$\Res(\bu_{0},\ldots,\bu_{n})$ which is the sparse differential
resultant of the $n+1$ generic Laurent differential polynomials in
\bref{eq-sparseLaurent}.
\end{definition}

We will show in the next Section \ref{sec-sol} that the sparse
differential resultant $\Res_{F_0,\ldots,F_n}=0$ will approximately
measure whether or not the the over-determined equation system
$F_i=0\,(i=0,\ldots,n)$ have a common non-polynomial solution.

%Here, we give the following illustrative example.
%
%\begin{example} \label{ex-1-1}
%Continue from Example \ref{ex-1}.
%%
%When the coefficients $u_{i0}$ specialize to  $a_{i0}$ in $\E$, if
%$\overline{\P}_i=a_{i0}y_1''+a_{i1}y_1'''+a_{i2}y_2'''=0\,(i=0,1,2)$
%have non-polynomial solutions $(\xi_1,\xi_2)$ for $(y_1,y_2)$ then
%it is apparent that
%$\Res_{\overline{\P}_0,\overline{\P}_1,\overline{\P}_2}=0$.
%%
%On the other hand, if
%$\Res_{\overline{\P}_0,\overline{\P}_1,\overline{\P}_2}=0$ and
%$a_{ij}$ is generic enough, treating $\overline{\P}_i=0$ as linear
%equations in $y_1'', y_1''', y_2'''$, there exist rational functions
%$\phi_1, \phi_2$ in $a_{ij}$ such that $y_1''' = \phi_1 y_1''$,
%$y_2'''=\phi_2 y_1''$, from which non-polynomial solutions for $y_1$
%and $y_2$ can be found.
%\end{example}

\subsection{Necessary and sufficient condition for existence of
non-polynomial solutions }\label{sec-sol}%
In the algebraic case, the resultant gives a necessary and
sufficient condition for a system of homogenous polynomials to have
common solutions. We will show that this is also true for sparse
differential resultants in certain sense.

To be more precise, we first introduce some notations. Let
$\mathcal{A}_0,\ldots,\mathcal{A}_n$ be a Laurent differentially
essential system of monomial sets. Each element
$(F_0,\ldots,F_n)\in\mathcal{L}(\mathcal{A}_0)\times\cdots\times
\mathcal{L}(\mathcal{A}_n)$ can be represented by one and only one
point
$(\bv_0,\ldots,\bv_n)\in\ee^{l_0+1}\times\cdots\times\ee^{l_n+1}$
where $\bv_i=(v_{i0},v_{i1},\ldots,v_{il_i})$ is the coefficient
vector of $F_i$\footnote{Here, we can also consider the differential
projective space $\textbf{P}(l_i)$ over $\ee$}. Let $\mathcal
{Z}_0(\mathcal{A}_0,\ldots,\mathcal{A}_n)$ be the subset of
$\ee^{l_0+1}\times\cdots\times\ee^{l_n+1}$ consisting of points
$(\bv_0,\ldots,\bv_n)$ such that the corresponding
$F_i=0\,(i=0,\ldots,n)$ have non-polynomial common solutions. That
is,
\begin{eqnarray}\label{eq-zA} \quad&\quad&\mathcal
{Z}_0(\mathcal{A}_0,\ldots,\mathcal{A}_n)=\{(\bv_0,\ldots,\bv_n)\in\ee^{l_0+1}\times\cdots\times\ee^{l_n+1}:
F_0=\cdots=F_n=0 \,\,\text{have} \nonumber\\&\quad&\qquad\qquad
\qquad\qquad\text{ a common non-polynomial solution in}\,
(\ee^{\wedge })^n\}.
\end{eqnarray}
The following result shows that the vanishing of sparse differential
resultant gives a necessary condition for the existence of
non-polynomial solutions.

\begin{lemma} \label{le-necessarycondition}
${Z}_0(\mathcal{A}_0,\ldots,\mathcal{A}_n)\subseteq
\V\big(\sat(\Res_{\mathcal{A}_0,\ldots,\mathcal{A}_n})\big)$.
\end{lemma}
\proof Let $\P_0,\ldots,\P_n$ be a generic Laurent differentially
essential system corresponding to
$\mathcal{A}_0,\ldots,\mathcal{A}_n$ with coefficient vectors
$\bu_0,\ldots,\bu_n$. By (\ref{eq-lsres}),
$[\P_0,\P_1,\ldots,\P_n]\cap\Q\{\bu_0,\ldots,\bu_n\}=
  \sat(\Res_{\mathcal{A}_0,\ldots,\mathcal{A}_n})$. For
  any point $(\bv_{0},\ldots,\bv_{n})\in
  {Z}_0(\mathcal{A}_0,\ldots,\mathcal{A}_n)$,
   let $(\overline{\P}_{0},\ldots,\overline{\P}_{n})\in\mathcal{L}(\mathcal{A}_0)\times\cdots\times
\mathcal{L}(\mathcal{A}_n)$ be the differential polynomial system
represented by $(\bv_{0},\ldots,\bv_{n})$. Let $G$ be any
differential polynomial in
$\sat(\Res_{\mathcal{A}_0,\ldots,\mathcal{A}_n})$. Then
$G(\bv_0,\ldots,\bv_n)\in
[\overline{\P}_0,\ldots,\overline{\P}_n]\subset\ee\{\Y,\Y^{-1}\}$.
Since $\overline{\P}_{0},\ldots,\overline{\P}_{n}$ have a
non-polynomial common zero, $G(\bv_0,\ldots,\bv_n)$ should be zero.
Thus, $\sat(\Res_{\mathcal{A}_0,\ldots,\mathcal{A}_n})$ vanishes at
$(\bv_0,\ldots,\bv_n)$.
 \qedd

\begin{example}\label{ex-sp12}
Continue from Example~\ref{ex-1}. Suppose $\F=\Q(x)$ and $\delta =
\frac{d}{d x}$. In this example, we have
$\Res_{\P_0,\P_1,\P_2}\ne0$. But $y_1=c_{11}x+c_{10},
y_2=c_{22}x^2+c_{21}x+c_{20}$ consist of a non-zero solution of
$\P_0=\P_1=\P_2=0$ where $c_{ij}$ are distinct arbitrary constants.
This shows that Lemma \ref{le-necessarycondition} is not correct if
we do not consider non-polynomial solutions.
This example also shows why we need to consider non-polynomial
differential solutions, or equivalently why we consider Laurent
differential polynomials instead of usual differential polynomials.
\end{example}

Let $\overline{\mathcal {Z}_0(\mathcal{A}_0,\ldots,\mathcal{A}_n)}$
be the Kolchin differential closure of $\mathcal
{Z}_0(\mathcal{A}_0,\ldots,\mathcal{A}_n)$ in
$\ee^{l_0+1}\times\cdots\times\ee^{l_n+1}$. Then we have the
following theorem which gives another characterization for the
sparse differential resultant.

\begin{theorem}\label{th-nscond}
Suppose the Laurent differential monomial sets
$\mathcal{A}_i\,(i=0,\ldots,n)$ form a Laurent differentially
essential system. Then $\mathcal
{Z}(\mathcal{A}_0,\ldots,\mathcal{A}_n)=\V\big(\sat(\Res_{\mathcal{A}_0,\ldots,\mathcal{A}_n})\big)$.
\end{theorem}
\proof Firstly,  by Lemma~\ref{le-necessarycondition}, $\mathcal
{Z}_0(\mathcal{A}_0,\ldots,\mathcal{A}_n)\subseteq\V\big(\sat(\Res_{\mathcal{A}_0,\ldots,\mathcal{A}_n})\big)$.
So $\mathcal
{Z}(\mathcal{A}_0,\ldots,\mathcal{A}_n)=\overline{\mathcal
{Z}_0(\mathcal{A}_0,\ldots,\mathcal{A}_n)}\subseteq\V\big(\sat(\Res_{\mathcal{A}_0,\ldots,\mathcal{A}_n})\big)$.

For the other direction, follow the notations in the proof of
Theorem~\ref{th-Mcodim1}. By Theorem~\ref{th-Mcodim1},
$[\P_0^N,\ldots,\P_n^N]:\mathbbm{m}$ is a prime differential ideal
with a generic point $(\eta,\zeta)$ where
$\eta=(\eta_1,\ldots,\eta_n)$ is a generic point of $[0]$ over
$\qq\langle(u_{ik})_{i=0,\ldots,n;k\neq0}\rangle$ and
$\zeta=(\zeta_0,u_{01},\ldots,$ $u_{0l_0};\ldots;$
$\zeta_n,u_{n1},\ldots,u_{nl_n})$.  Let $(F_0,\ldots,F_n)\in
\mathcal{L}(\mathcal{A}_0)\times\cdots\times
\mathcal{L}(\mathcal{A}_n)$ be a set of Laurent differential
polynomials represented by $\zeta$. Clearly, $\eta$ is a
non-polynomial solution of $F_i=0$. Thus, $\zeta\in\mathcal
{Z}_0(\mathcal{A}_0,\ldots,\mathcal{A}_n)\subset\mathcal
{Z}(\mathcal{A}_0,\ldots,\mathcal{A}_n)$. By Corollary \ref{cor-r1},
 $\zeta$ is a generic point of
$\sat(\Res_{\mathcal{A}_0,\ldots,\mathcal{A}_n})$. It follows that
$\V\big(\sat(\Res_{\mathcal{A}_0,\ldots,\mathcal{A}_n})\big)\subseteq\mathcal
{Z}(\mathcal{A}_0,\ldots,\mathcal{A}_n)$. As a consequence,
$\V\big(\sat(\Res_{\mathcal{A}_0,\ldots,\mathcal{A}_n})\big)=\mathcal
{Z}(\mathcal{A}_0,\ldots,\mathcal{A}_n)$. \qedd

The above theorem shows that the sparse differential resultant gives
a sufficient and necessary condition for a differentially essential
system to have non-polynomial solutions over an open set of
$\prod_{i=0}^n \mathcal{L}(\mathcal{A}_i)$ in the sense of Kolchin
topology.
\vf With Theorem \ref{th-nscond},  property 3) of Theorem
\ref{th-i1} is proved.

\section{Criterion for Laurent differentially essential system in
terms of supports}

Let $\A_i\,(i=0,\ldots,n)$ be finite sets of  Laurent differential
monomials. According to Definition~\ref{def-tdes}, in order to check
whether they form a Laurent differentially essential system, we need
to check whether there exist  $M_{ik_i}, M_{ij_i}\in\A_i
(i=0,\ldots,n)$ such that $\dtrdeg\, \Q\langle M_{0k_0}/M_{0j_0},$
$\ldots,M_{nk_n}/M_{nj_n}\rangle/\Q=n$. This can be done with the
differential characteristic set method via symbolic computation
\cite{ritt,bc1,sit,gao}. In this section, a criterion will be given
to check whether a Laurent differential system is essential in terms
of their supports, which is conceptually and computationally simpler
than the naive approach based on the characteristic set method.

\subsection{Sets of Laurent differential monomials in reduced  and
T-shape forms}

In this section, two types of Laurent differential monomial sets are
introduced, whose differential transcendence degrees are easy to
compute.

Let $B_1,B_2,\ldots,B_m$ be $m$ Laurent differential monomials,
where $B_i = \prod_{j=1}^n\prod_{k\geq 0} (y_{j}^{(k)})^{d_{ijk}}$.
For each $j\in\{1,\ldots,n\}$, let $q_j =
\max_{i=1}^m\ord(B_i,y_j)$. Let $x_1,\ldots,x_n$ be new algebraic
indeterminates and
$$d_{ij} = \sum_{k=0}^{q_j} d_{ijk}x_j^{k}\, (i=1,\ldots,m ,j=1,\ldots,n)$$
univariate polynomials in $\mathbb{Z}[x_j]$ respectively. If
$\ord(B_i,y_j) = -\infty$, then $d_{ij}=0$ and we
 denote $\deg(d_{ij},x_j) = -\infty$.
The vector  $(d_{i1},d_{i2},$ $\ldots,d_{in})$ is called the {\em
symbolic support vector} of $B_i$.
The following $m\times n$ matrix
\[M=\left(\begin{array}{cccccccc}
d_{11} & \,d_{12} & \,\ldots & \,d_{1n} \\
d_{21} & \,d_{22} & \,\ldots & \,d_{2n} \\
&  & \ddots & \\
d_{m1} & \,d_{m2} & \,\ldots & \,d_{mn}
\end{array}\right) \]
is called the {\em symbolic support matrix} of $B_1,\ldots, B_m$.

Note that there is a one-to-one correspondence between Laurent
differential monomials and their symbolic support vectors, so we
will not distinguish these two concepts if there is no confusion.
The same is true for a set of Laurent differential monomials and its
symbolic support matrix.

\begin{definition}
A set of Laurent differential monomials $B_1, B_2, \ldots, B_m$ or
its symbolic support matrix $M$ is called {\em reduced} if for each
$i\leq \min(m,n)$, $-\infty \ne \ord(B_i,y_i)
> \ord(B_{i+k},y_i)$, or equivalently $-\infty\ne\deg(d_{ii},x_i)>\deg(d_{i+k,i},x_i)$, holds for all $k>0$.
\end{definition}

Note that a reduced symbolic support matrix is always of full rank
since the term $\prod_{i=1}^{\min(m,n)} x_i^{\ord(B_i,y_i)}$ will
 appear effectively in the determinant of the $\min(m,n)$-th
principal minor when expanded.

\begin{example}\label{exam-1}
Let $B_1 = y_{1}^2y''_{1}y'_{2}, B_2 =
y_{1}^3(y'_{2})^2y_{3}(y'_{3})^2, B_3 = y'_{1}y'_{3}$. Then $q_1 =
2, q_2 = 1, q_3 = 1$, and
 \[M=\left(\begin{array}{cccc}
x_1^2 + 2 &  x_2 & 0 \\
3 &  2x_2 & 2x_3+1 \\
x_1 &  0 & x_3
\end{array}\right) \]
is reduced.
\end{example}

Before giving the property of reduced symbolic support matrices, the
following simple result about the differential transcendence degree
will be proved.
\begin{lemma}\label{lm-dtr}
For $\eta_1,\eta_2$ in an extension field of $\qq$,
$\dtrdeg\,\qq\langle \eta_1^{a_1},\eta_1^{a_2}\eta_2
\rangle/\qq=\dtrdeg\,\qq\langle \eta_1,$ $\eta_2 \rangle/\qq $,
where $a_1,a_2$ are non-zero rational numbers.
\end{lemma}
\proof For any non-zero integer $p$, we have
\[
\begin{array}{lll}
\dtrdeg\, \qq\langle \eta_1,\eta_2 \rangle/\qq & = & \dtrdeg\,
\qq\langle \eta_1,\eta_2 \rangle/\qq\langle \eta_1^p,\eta_2 \rangle
+
\dtrdeg\, \qq\langle \eta_1^p,\eta_2 \rangle/\qq \\
& = &  \dtrdeg \,\qq\langle \eta_1^p,\eta_2 \rangle/\qq.
\end{array}
\]
So for each $a\in\qq\backslash\{0\}$, $\dtrdeg\, \qq\langle
\eta_1^a,\eta_2 \rangle/\qq =\dtrdeg\, \qq\langle \eta_1,\eta_2
\rangle/\qq $.
Let $a_i=p_i/q_i\,(i=1,2)$ where $p_i,q_i$ are non-zero integers.
Then,
\begin{eqnarray}
\dtrdeg\, \qq\langle \eta_1^{a_1},\eta_1^{a_2}\eta_2 \rangle/\qq & =
& \dtrdeg\, \qq\langle \eta_1^{1/q_2},\eta_1^{p_2/q_2}\eta_2 \rangle
/\qq  \nonumber\\
& = &  \dtrdeg \,\qq\langle \eta_1^{1/q_2}, \eta_2 \rangle
/\qq\,\,(\text{for $\qq\langle \eta_1^{1/q_2},\eta_1^{p_2/q_2}\eta_2
\rangle=\qq\langle \eta_1^{1/q_2}, \eta_2 \rangle $})
 \nonumber\\ & = & \dtrdeg \,\qq\langle \eta_1,\eta_2 \rangle/\qq.\nonumber
\end{eqnarray}
\qedd

\begin{theorem}\label{th-red}
Let $B_1, B_2, \ldots, B_m$ be $m$ reduced Laurent differential
monomials in $\Y$. Then $\dtrdeg\,\Q\langle B_1, B_2, \ldots,
B_m\rangle/\Q=\min(m,n)$.
\end{theorem}
\proof %By Lemma~\ref{lm-dtr}, without loss of generality, we assume
%that the symbolic supports of $B_i$ are polynomials with
%coefficients integers. We only need to prove the case of $m\leq n$
It suffices to prove the case $m = n$
%when $B_1, B_2, \ldots, B_m$ are reduced Laurent differential monomials
by the following two facts. In the case $m>n$, we need only to prove
that $B_1,\ldots, B_n$ are differentially independent. And in the
case $m<n$, we can treat $y_{m+1},\ldots,y_n$ as parameters, then
$B_1, B_2, \ldots, B_m$ are still reduced Laurent differential
monomials. So if we have proved the result for $m=n$, $\dtrdeg\,
\Q\langle B_1, B_2, \ldots, B_m\rangle/\Q \geq  \dtrdeg\, \Q\langle
y_{m+1},\ldots,y_n\rangle\langle B_1, B_2, \ldots,
B_m\rangle/\Q\langle y_{m+1},\ldots,y_n\rangle$ $ = m$ follows.

Since $B_1, B_2, \ldots, B_n$ are reduced, we have $o_i =
\ord(B_i,y_i)\geq 0$ for $i\leq n$.
In this proof, a Laurent differential monomial will be treated as an
algebraic Laurent monomial, or simply a monomial. Furthermore, the
lex order between two monomials induced by the following variable
order will be used.
\begin{eqnarray*}
 && \boxed{y_{1} > y'_{1}> \cdots > y_{1}^{(o_1-1)}}\\
 &>& \boxed{y_{2} > y'_{2}> \cdots > y_{2}^{(o_2-1)}}\\
 &>& \cdots\\
 &>& \boxed{y_{n} > y'_{n} > \cdots > y_{n}^{(o_n-1)} > y_{n}^{(o_n)} > y_{n}^{(o_n+1)} > \cdots}\\
 &>& \boxed{ y_{n-1}^{(o_{n-1})} > y_{n-1}^{(o_{n-1}+1)}> \cdots}\\
 &>& \cdots \\
 &>& \boxed{y_{1}^{(o_1)} > y_{1}^{(o_1+1)} > \cdots}.
 \end{eqnarray*}
Under this ordering, we claim that the leading monomial of $\delta^t
B_i\,( 1\leq i \leq n, t\in \NN)$  is $LM_{it} =
\frac{B_i*y_{i}^{(o_i+t)}}{y_{i}^{(o_i)}}$. Here by leading
monomial, we mean the monomial with the highest order appearing
effectively in a polynomial. Let $B_i = N_i
(y_{i}^{(o_i)})^{D_i}\,(1\leq i \leq n)$.
 %then the monomials
%of $\delta^t B_i$ is of the form $\delta^k N_i\delta^{t-k}
%y_{i,o_i}^{d_{i,i,o_i}}, 0\leq k \leq t$. If $k\ne 0$, then we have
%$\delta^k N_i < N_i$ and $\delta^k N_i\delta^{t-k}
%(y_{i}^{(o_i)})^{d_{i,i,o_i}} < N_i
%(y_{i}^{(o_i)})^{d_{i,i,o_i-1}}y_{i}^{(o_i+t)}$ for $1\leq i \leq
%m$.
If $N_i=1$, then the monomials of $\delta^t B_i$ is of the form
$\prod_{k=0}^t(y_{i}^{(o_i+k)})^{s_k}$, where $s_0,\ldots,s_t$ are
non-negative integers such that $\sum_{k=0}^t s_k = D_{i}$ and
$\sum_{k=1}^t ks_k = t$. Among these monomials, if $s_k>0$ for some
$1\leq k\leq t-1$, then $s_0$ is strictly less than $D_{i}-1$ and
$\prod_{k=0}^t(y_{i}^{(o_i+k)})^{s_k}<(y_{i}^{(o_i)})^{D_{i}-1}y_{i}^{(o_i+t)}=\frac{B_i*y_{i}^{(o_i+t)}}{y_{i}^{(o_i)}}$
follows. Hence, in the case $N_i=1$, the claim holds. Now suppose
$N_i\neq1$, then it is a product of variables with lex order larger
than $y_i^{(o_i)}$. Then $\delta^{t}B_i=\sum_{k=0}^t {t\choose
k}\delta^{k}N_i\delta^{t-k}(y_{i}^{(o_i)})^{D_{i}}$. If $k=0$, then
similar to the case $N_i=1$, we can show that the highest monomial
in $ N_i\delta^{t} (y_{i}^{(o_i)})^{D_{i}}$ is
$N_i(y_{i}^{(o_i)})^{D_{i}-1}y_{i}^{(o_i+t)}$. For each $k>0$,
$\delta^{k}N_i<N_i$ and
$\delta^{k}N_i\delta^{t-k}(y_{i}^{(o_i)})^{D_{i}}<N_i(y_{i}^{(o_i)})^{D_{i}-1}y_{i}^{(o_i+t)}=\frac{B_i*y_{i}^{(o_i+t)}}{y_{i}^{(o_i)}}$.
Hence, the leading monomial of $\delta^t B_i$ is
$N_i(y_{i}^{(o_i)})^{D_{i}-1}y_{i}^{(o_i+t)} =
\frac{B_i*y_{i}^{(o_i+t)}}{y_{i}^{(o_i)}}$.

We claim that these leading monomials $LM_{it} =
\frac{B_i*y_{i}^{(o_i+t)}}{y_i^{(o_i)}}\,( i=1,\ldots,m; t\geq 0)$
are algebraically independent over $\qq$. We prove this claim by
showing that the algebraic transcendence degree of these monomials
are the same as the number of monomials for any fixed $t$. Let  $Y_i
= [y_{i} , y_{i}' , \ldots , y_{i}^{(o_i-1)}] $, $Y_i^* = [
y_{i}^{(o_i+t+1)}, \ldots, y_{i}^{(q_i+t)}]$, $B_{it} = [B_i,
LM_{i1},\ldots, LM_{it}]$ for $1\leq i \leq n$.  We denote by
$BY_{i} = (y_{i}^{(o_i)})^{D_{i}}$, $BY_{it} = [
(y_{i}^{(o_i)})^{D_{i}}, (y_{i}^{(o_i)})^{D_{i}-1}y_{i}^{(o_i+1)},
\ldots, (y_{i}^{(o_i)})^{D_{i}-1}y_{i}^{(o_i+t)}]$ for $1\leq i \leq
n$. Then, by Lemma~\ref{lm-dtr}, we have { \[
\begin{array}{lll}
n(t+1) &\geq & \trdeg\,\Q(B_{1t},B_{2t},\ldots,B_{nt})/\Q \\
 &\geq & \trdeg\,\Q_1(B_{1t},B_{2t},\ldots,B_{nt})/\qq_1 \\
&=& \trdeg\,\Q_1(BY_{1t},BY_{2t},\ldots,BY_{nt})/\qq_1\\
&=& n(t+1)
\end{array}
\]}
where $\qq_1=\Q(Y_1,\ldots,Y_n, Y_1^*, \ldots, Y_n^*)$.
Hence, this claim is proved.

Now, we prove that $B_1,\ldots,B_n$ are differentially independent
over $\qq$. Suppose the contrary, then there exists a nonzero
differential polynomial $P\in\qq\{z_1,\ldots,z_n\}$ such that
$P(B_1,\ldots,B_n) = 0$. Let $P = \sum_k c_k P_k $, where $P_k$ is a
monomial and $c_k\in\Q\backslash\{0\}$. Then, the leading monomial
of $P_k(B_1,\ldots,B_n)$ is a product of $LM_{it}\, (i=1,\ldots,n;
t\geq 0)$. We denote this product by $LMP_k$, then $LMP_k \neq
LMP_j$ for $k\neq j$ since these $LM_{it}$ are algebraically
independent. But there exists one and only one product which has the
highest order, which can not be eliminated by the others, which
means that $P(B_1,\ldots,B_n)\neq 0$, a contradiction. \qedd

\begin{figure}[ht]
%\begin{minipage}{0.9\textwidth}
\centering{\includegraphics[width=6.0cm,angle=0]{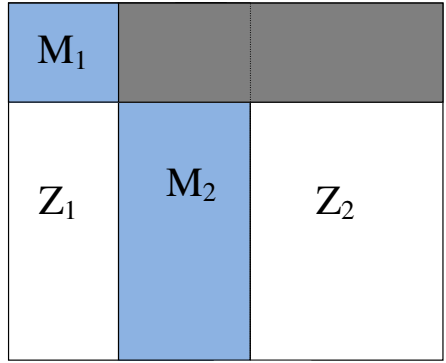}}
%\end{minipage}
\caption{A T-shape Matrix} \label{fig-T}
\end{figure}

In general, we cannot reduce a symbolic support matrix to a reduced
one. But, in the next section, we will show that any symbolic
support matrix can be reduced to T-shape to be defined below.

\begin{definition}
A set of Laurent differential monomials $B_1,\ldots,B_m$ or their
symbolic support matrix $M$  is said to be in {\em T-shape} with
index $(i,j)$, if there exist $1\leq i\leq \min(m,n), 0\leq j \leq
\min(m,n)-i$ such that
all elements except those in the first $i$ rows and the
$i+1,\ldots,(i+j)$-th columns of $M$ are zeros and the sub-matrix
consisting of the first $i+j$ columns of $M$ is reduced.
The zero sub-matrix $(Z_1,Z_2)$ in Figure \ref{fig-T} is called the
{\em zero sub-matrix} of $M$.
\end{definition}

In Figure~\ref{fig-T}, an illustrative form of a matrix in T-shape
is given, where the sub-matrices $M_1$ and $M_2$ of the matrix are
reduced ones. It is easy to see that $M_1$ must be an $i\times i$
square matrix.
Since the first $i+j$ columns of a T-shape matrix $M$ is a reduced
sub-matrix, we have
\begin{lemma}\label{lm-tshape}
The rank of a T-shape matrix with index $(i,j)$ equals to $i+j$.
Furthermore, a T-shape matrix is reduced if and only if it is of
full rank, that is, $i+j=\min{(m,n)}$.
\end{lemma}

For a zero matrix $S$ with $k$ rows and $l$ columns whose elements
are zeros, we define its {\em $0$-rank} to be $k+l$. A T-shape
matrix $M$ is not of full rank if and only if $i+j < \min(m,n)$. As
a consequence, we have
\begin{lemma}\label{lm-tshape2}
A T-shape matrix of index $(i,j)$ is not of full rank if and only if
its zero sub-matrix is an $(m-i)\times(n-j)$ zero matrix with
$0$-rank $m+n-i-j \geq \max(m,n)+1$.
\end{lemma}

%
%In the case that the T-shape matrix $M$ is not of full rank, that
%is, $i+j < \min(m,n)$, its zero sub-matrix is an $(m-i)\times(n-j)$
%zero matrix with $0$-rank $m+n-i-j \geq \max(m,n)+1$.

The differential transcendence degree of $m$  Laurent differential
monomials in T-shape can be easily determined, as shown by the
following result.
\begin{theorem}\label{th-T}
Let $B_1,\ldots,B_m$ be $m$  Laurent differential monomials and $M$
their symbolic support matrix which is in T-shape with index
$(i,j)$. Then $\dtrdeg\,$ $ \Q\langle B_1, B_2, \ldots,
B_m\rangle/\Q=\rank(M)=i+j$.
\end{theorem}
\proof Since $M$ is a T-shape matrix with index $(i,j)$, by Lemma
\ref{lm-tshape}, the rank of $M$ is $i+j$.

Deleting the zero columns of the symbolic support matrix of
$B_{i+1},\ldots,B_{m}$, we can get a reduced matrix. By
Theorem~\ref{th-red}, we have~$\dtrdeg\,\Q\langle
B_{i+1},\ldots,B_{m}\rangle/Q = j$. Since the symbolic support
matrix of~$B_1,\ldots,B_{i}$ is also a reduced one, by
Theorem~\ref{th-red}, we have $\dtrdeg\,\Q\langle
B_1,\ldots,B_{i}\rangle/\Q$ $ = i$. Hence,
\[
\begin{array}{lll}
\dtrdeg\,\qq\langle B_1,\ldots,B_m\rangle/\Q & = & \dtrdeg\,\Q\langle B_1,\ldots,B_m\rangle/\Q\langle B_{i+1},\ldots,B_{m}\rangle\\& & +  \dtrdeg\,\Q\langle B_{i+1},\ldots,B_{m}\rangle/\Q \\
                          & \leq & \dtrdeg\,\Q\langle B_1,\ldots,B_{i}\rangle/\Q + j \\
                          & = & i+j.
\end{array}
\]

On the other hand, if we treat $y_{i+1},\ldots, y_{i+j}$ and their
derivatives as parameters, the symbolic support matrix
of~$B_1,\ldots,B_{i}$ is also a reduced one and the rank of this
matrix is $i$.  By Theorem~\ref{th-red},
we have~$\dtrdeg\,\Q\langle y_{i+1},\ldots,y_{i+j}\rangle\langle
B_1,\ldots,B_{i}\rangle/\Q\langle y_{i+1},\ldots,y_{i+j}\rangle =
i$. Since $B_{i+1},\ldots,B_m$ are monomials in
$y_{i+1},\ldots,y_{i+j}$ (see Figure \ref{fig-T}), $\Q\langle
B_{i+1},\ldots,B_{m}\rangle\subset\Q\langle
y_{i+1},\ldots,y_{i+j}\rangle$. Hence,
\[
\begin{array}{lll}
\dtrdeg\,\qq\langle B_1,\ldots,B_m\rangle/\Q & = & \dtrdeg\,\Q\langle B_1,\ldots,B_m\rangle/\Q\langle B_{i+1},\ldots,B_{m}\rangle\\ & & + \dtrdeg\,\Q\langle B_{i+1},\ldots,B_{m}\rangle/\Q \\
                          & \geq & \dtrdeg\,\Q\langle y_{i+1},\ldots,y_{i+j}\rangle\langle B_1,\ldots,B_{i}\rangle /\Q\langle y_{i+1},\ldots,y_{i+j}\rangle + j \\
                          & = & i+j.
\end{array}
\]
Thus, $\dtrdeg\,\qq\langle B_1,\ldots,B_m\rangle/\Q=\rk(M)=i+j$.
\qedd

\subsection{An algorithm to reduce Laurent differential monomials to T-shape}

In this section,  an algorithm is given to reduce any set of Laurent
differential monomials to a set of Laurent differential monomials in
T-shape, which has the same differential transcendence degree with
the original one.

First, we will define the transformations that will be used to
reduce any symbolic support matrix to a T-shape one.
A {\em $\Q$-elementary transformation} for a matrix $M$ consists of
two types of matrix row operations and one type of  matrix column
operations. To be more precise, Type 1 operations consist of
interchanging two rows of $M$; Type 2 operations consist of adding a
rational number multiple of one row to another; and Type 3
operations consist of interchanging  two columns.

Let $B_1,\ldots,B_m$ be Laurent differential monomials and $M$ their
symbolic support matrix. Then $\Q$-elementary transformations of $M$
correspond to certain transformations of the monomials. Indeed,
interchanging the $i$-th and the $j$-th rows  of $M$ means
interchanging $B_i$ and $B_j$, and interchanging the $i$-th and the
$j$-th columns of $M$ means interchanging $y_i$ and $y_j$ in
$B_1,\ldots,B_m$(or in the variable order). Multiplying the $i$-th
row of $M$ by a rational number $r$ and adding the result to the
$j$-th row means changing $B_j$ to $B_i^rB_j$.

\begin{lemma}\label{lem-1}
Let $B_1,\ldots,B_m$ be Laurent differential monomials and
$C_1,\ldots,C_m$ obtained by a series of $\Q$-elementary
transformations from $B_1,\ldots,B_m$. Then $\dtrdeg\,\Q\langle B_1,
\ldots, B_m\rangle/\Q=\dtrdeg\, \Q\langle C_1,$ $\ldots,
C_m\rangle/\Q$.
\end{lemma}
\proof It is a direct consequence of Lemma~\ref{lm-dtr}.\qedd

Now, an algorithm {\bf RDM}($M$) will be given to reduce a given
symbolic support matrix to a T-shape matrix by a series of
$\Q$-elementary transformations. We sketch the algorithm below. Note
that we still denote by $M$ the matrix obtained by $\Q$-elementary
transformations from $M$. We assume that $m \le n$ and hence
$p=\max(m,n)=n$. The case $m > n$ can be shown similarly.

Let $N$ be a sub-matrix of $M$. Then the {\em complementary matrix}
of $N$ in $M$ is the sub-matrix of $M$ from which all the rows and
columns associated with $N$ have been removed.

The algorithm consists of three major steps.
In the first step, a procedure similar to the Gauss elimination will
be used to construct a reduced square sub-matrix $R$ of $M$ such
that the complementary matrix of $R$ in $M$ is a zero matrix.
Precisely, choose a column of $M$, say the first column, which
contains at least one non-zero element. Then, choose an element, say
$d_{11}$, of this column, which has the largest degree among all
elements in the same column.
If there exists a $d_{i1}, i>1$ such that $\deg(d_{i1}) =
\deg(d_{11})$, then replace $d_{ij}$ by $ d_{ij} -
\frac{a_i}{a_1}d_{1j} $ for $j=1,\ldots,n$, where $a_i,a_1$ are the
leading coefficients of $d_{i1},d_{11}$ respectively. This is a
$\Q$-elementary transformation of Type 2.
Repeat the above procedure until the first column is in reduced
form, that is $\deg(d_{i1}) < \deg(d_{11})$ for $i=2,\ldots,n$.
Consider the lower-right $(m-1)\times (n-1)$ sub-matrix $N$ of $M$
and repeat the above procedure for $N$. In this way, we will obtain
a reduced square matrix whose complementary matrix is a zero matrix
$Z$ in the lower-right corner of $M$.

In the second step, a recursive procedure is used to construct a
reduced form of $M$.
Let the zero matrix $Z$ obtained above be an $i\times j$ matrix.
Denote $r = i+j$ to be the $0$-rank of it.
If $j = n$, the last $i$ rows of $M$ are zero rows. Delete the last
$i$ rows from $M$, then we have a strictly smaller matrix, which can
be treated recursively.

\begin{figure}[ht]
\begin{minipage}{0.33\textwidth}
\centering{\includegraphics[width=4.0cm,angle=0]{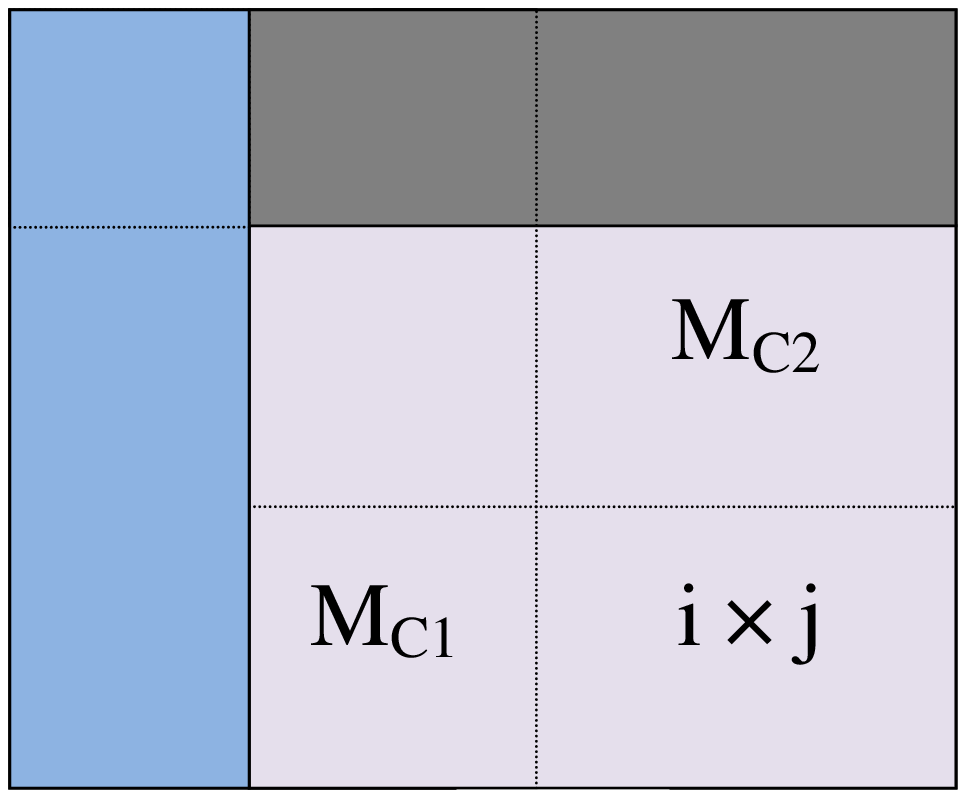}}\\
{~$(a)$~Sub-matrices for $m<n$}
\end{minipage} \hfill
\begin{minipage}{0.33\textwidth}
\centering{\includegraphics[width=3.0cm,angle=0]{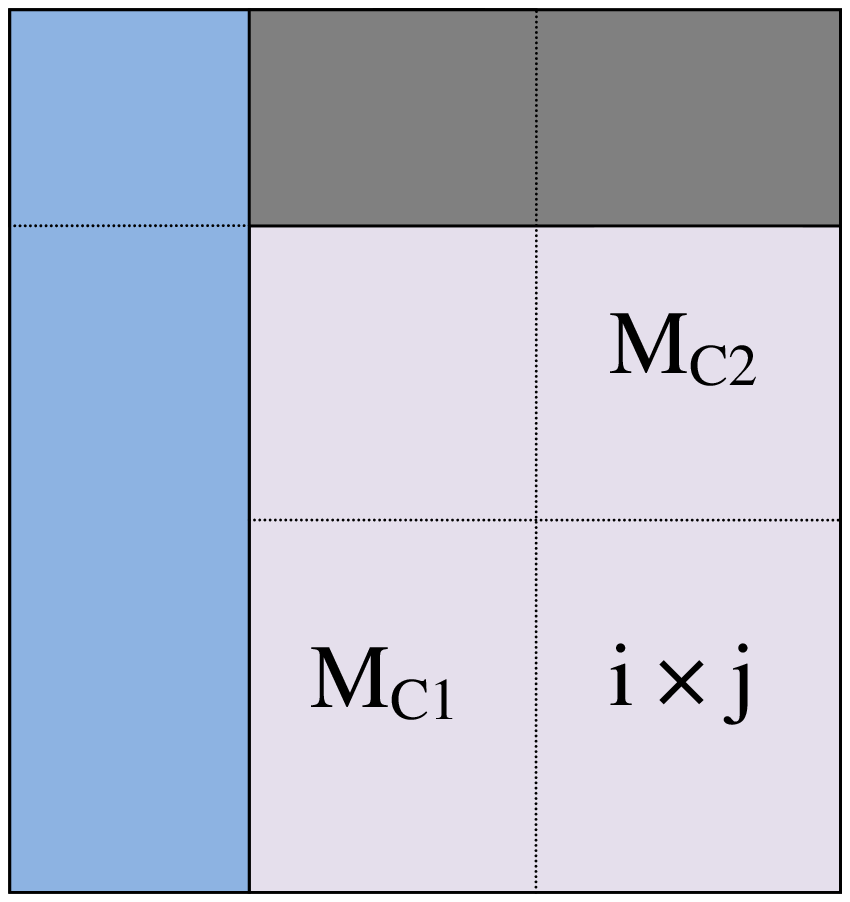}}\\
{~$(b)$~Sub-matrices for $m\ge n$}
\end{minipage}
 \hfill
 \begin{minipage}{0.32\textwidth}
\centering{\includegraphics[width=4.0cm,angle=0]{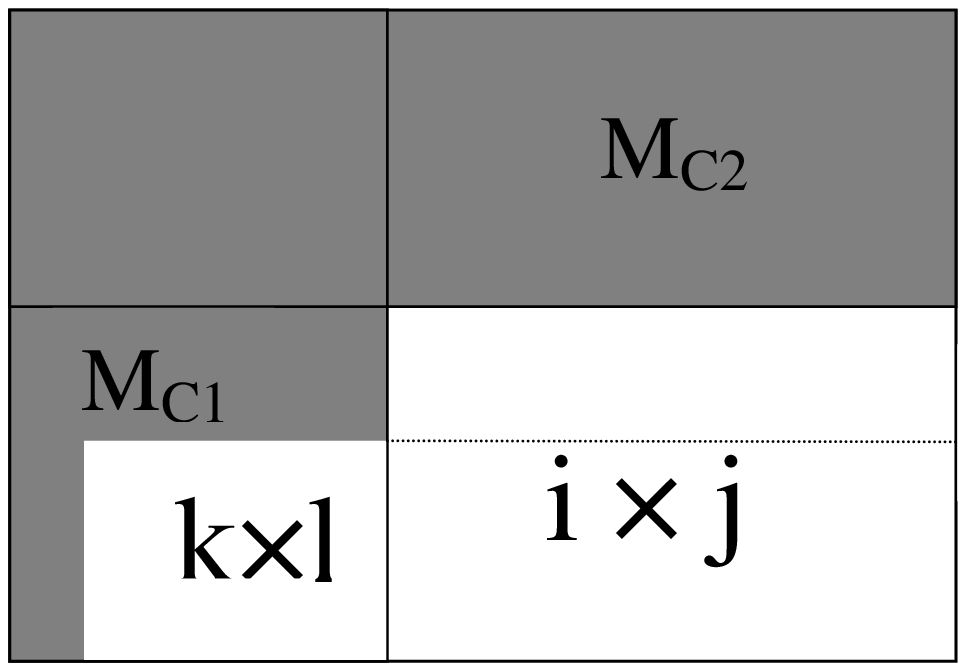}}\\
{~$(c)$~Two zero sub-matrices }
\end{minipage}
 \hfill
\begin{minipage}{0.33\textwidth}
\centering{\includegraphics[width=4.0cm,angle=0]{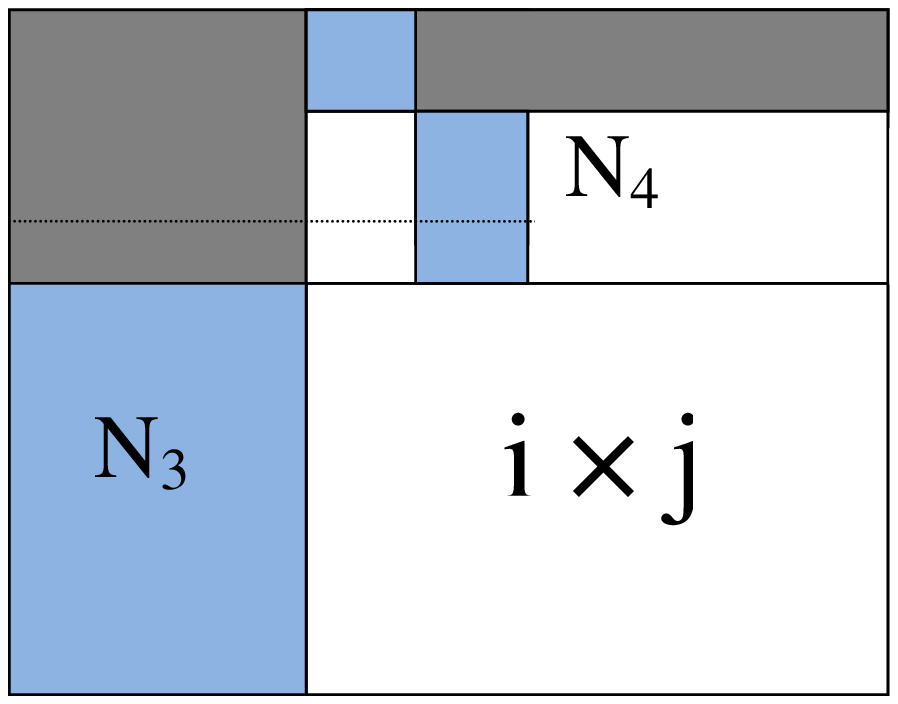}}\\
{~$(d)$~ $N_3$ and $N_4$}
\end{minipage}
%\vfill
\begin{minipage}{0.32\textwidth}
\centering{\includegraphics[width=4.0cm,angle=0]{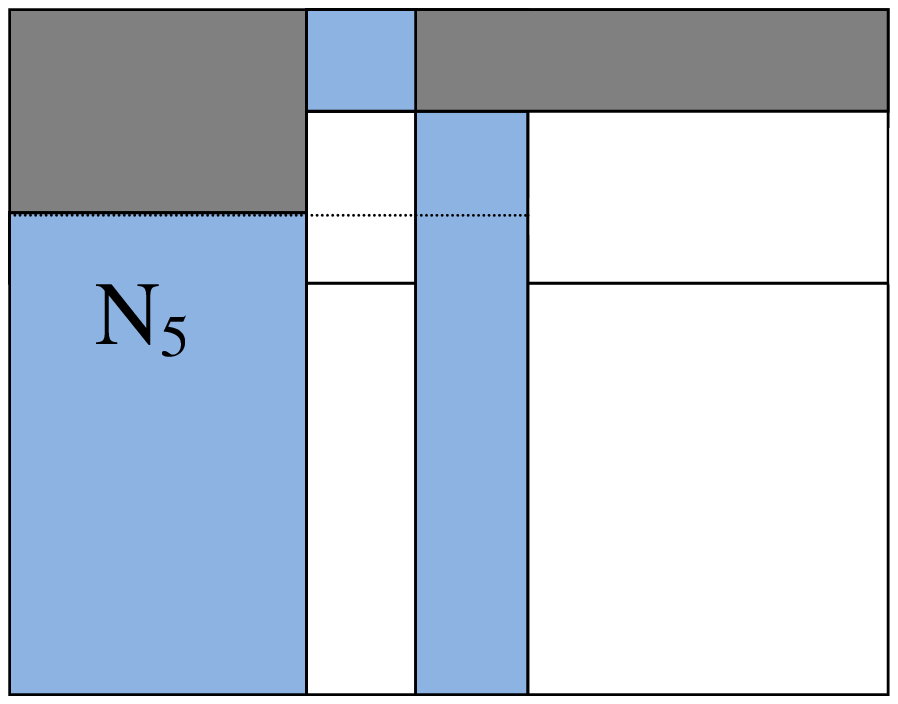}}\\
{~$(e)$~Compute $N_5$}
\end{minipage}
 \hfill
\begin{minipage}{0.33\textwidth}
\centering{\includegraphics[width=4.0cm,angle=0]{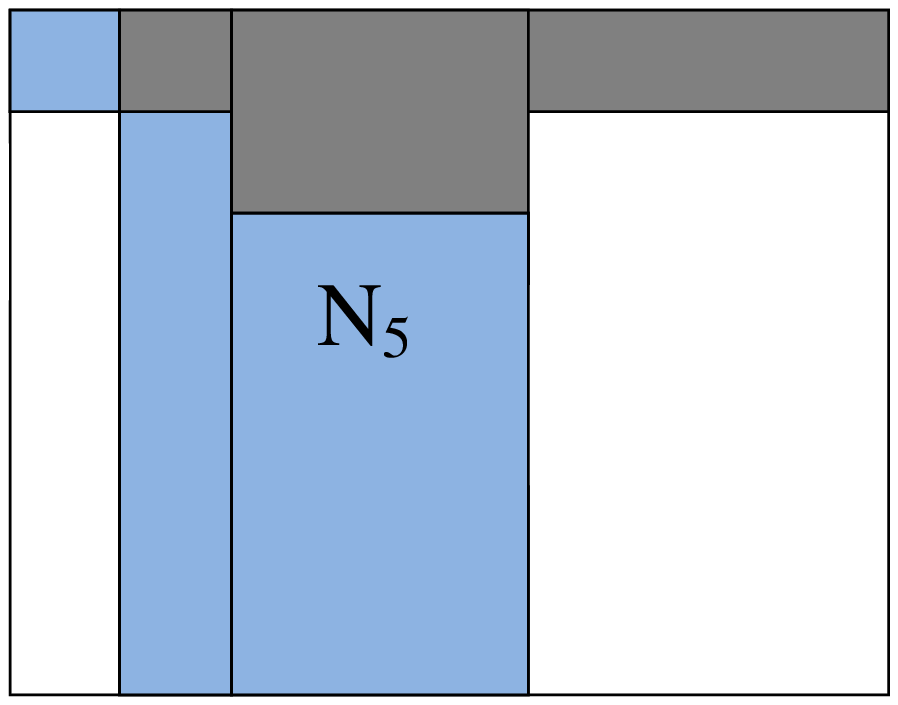}}\\
{~$(f)$~Interchanging rows }
\end{minipage}
% the white parts are zero matrix and the blue parts are reduced matrices
\caption{ Matrix forms in Algorithm~\ref{algo-red}, the blue parts
are reduced ones} \label{pic}
\end{figure}

If $r \geq n+1$, $M$  cannot be of full row rank, which will be
considered in step three.
Otherwise, let $M_C$ be the lower-right $(m+r-\max(m,n))\times
(n+r-\max(m,n))=(m+r-n)\times r$ sub-matrix of $M$, $M_{C1}$ the
lower-left $i\times (n+i-\max(m,n))=i\times i$ sub-matrix of $M_C$,
and $M_{C2}$ the upper-right $(m+j-\max(m,n))\times j=(m+j-n)\times
j$ sub-matrix of $M_C$. In Figure~\ref{pic}(a,b), $M_C$ is
represented by the pink area. Here, $M_C$ is chosen to be the
minimal $(m-q)\times(n-q)$ sub-matrix of $M$ at the lower-right
corner, which may have full rank.

Let $N_1=${\bf RDM}($M_{C1}$) and $N_2=${\bf RDM}($M_{C2}$). Note
that the $\Q$-elementary transformations of these sub-procedures are
for the whole rows and columns of $M$. By doing so, the sub-matrix
consisting of the first $n-r$ columns of $M$ remains to be a reduced
one.

If $N_1$ and $N_2$ are reduced matrices, we can obtain a reduced
matrix for $M$ by a suitable column interchanging.
Otherwise, either $N_1$ or $N_2$ is not of full rank. Assume $N_{1}$
is not of full rank. Then merging the zero sub-matrix of $N_1$ and
$Z$, we obtain a zero matrix with 0-rank larger than that of $Z$
(Figure \ref{pic}(c)). Repeat the second step for $M$ with this new
zero sub-matrix.

In the third step, $M$ contains a ``large" zero sub-matrix and a
T-shape matrix of $M$ can be constructed directly as follows.
Let the zero matrix $Z$ at the lower-right corner of $M$ be an
$i\times j$ matrix and $r = i+j$.
Let $M_{C3}$ be the lower-left $i\times (n-j)$ sub-matrix of $M$ and
$N_3=$ {\bf RDM}($M_{C3}$). In this case, $M_{C3}$ has more rows
than columns. We can assume that $N_3$ is of full column rank.
Otherwise, a sub-matrix of $N_3$ can be used as $N_3$.

Let $M_{C4}$ be the upper-right $(m-i)\times j$ sub-matrix of $M$,
$N_4=$ {\bf RDM}($M_{C4}$), and $s=\rank(N_4)$ (see
Figure~\ref{pic}(d)).
If $N_4$ is of full row rank, then by suitable column
interchangings, we can obtain a T-shape matrix.
Otherwise, let the lower-left $(m-s)\times(n-j)$ sub-matrix of $M$
be $M_{C5}$, and $N_5=$ {\bf RDM}($M_{C5}$), which is a  reduced
matrix with full column rank, see Figure~\ref{pic}(e). Now, by
suitable column interchangings, we can obtain a T-shape matrix (see
Figure~\ref{pic}(f)).

The idea of the algorithm is as follows. Try to use the first step
to construct a reduced matrix. If the first step fails to do so, use
the second step to change the matrix so that it contains a larger
zero sub-matrix after each iteration. The procedure will end until
either a T-shape matrix is obtained or the matrix has a zero
sub-matrix with size larger than $\max(m,n)+1$, in which case a
T-shape matrix can be obtained directly.

We now use the following example to illustrate the first two steps
of the algorithm.

\begin{example}\label{exam-2}
Let $B_1 = y_1y'_1y'''_2 y_3y'_3 $,
$B_2 = y_1^3(y'_1)^2 y''_2(y'''_2)^2 y_3^3(y'_3)^2 $,
$B_3 = y_1^2(y'_1)^3$ $y'_2(y'''_2)^3$ $y_3^3(y'_3)^3$ . Then, the
symbolic support matrix is
 \[ M=\left(\begin{array}{cccc}
x_1+1   &   x_2^3      &  x_3+1\\
2x_1+3  &  2x_2^3+x_2^2 & 2x_3+3 \\
3x_1+2  &  3x_2^3+x_2 & 3x_3+3
\end{array}\right). \]
We will use this matrix to illustrate the algorithm.
\[  M
\stackrel{(a)}{\Longrightarrow }
  \left( \begin{array}{cccc}
x_1 + 1 &  x_2^3 & x_3+1 \\
1 & \, x_2^2 & \, 1 \\
-1 & \, x_2  & \, 0
\end{array}\right)
\stackrel{(b)}{\Longrightarrow }\left( \begin{array}{cccc}
x_1+1  & x_3+1 &  x_2^3 \\
1 & \, 1 &\, x_2^2\\
-1 & \, 0 &\, x_2
\end{array}\right).
\]
The matrix after $\stackrel{(a)}{\Longrightarrow }$ is obtained with
the first step of the algorithm. We first use $d_{11}=x_1+1$ to
reduce the degrees of $2x_1+2$ and $3x_1+3$ with $\Q$-elementary
transformations of Type~2. Since  $x_2^2$ is of greater degree than
$x_2$, nothing needs to do. Finally, we obtain a $1\times 1$ zero
matrix at the lower-right corner at the end of step 1.

Now, goto the second step of the algorithm. We have $r=2 <
\max(m,n)+1=4$.  $M_C$ is the lower-right $2\times 2$ sub-matrix of
$M$, $M_{C1} = (x_2)$, and $M_{C2}=(1)$.

Since both $M_{C1}$ and $M_{C2}$ are reduced, we interchange the
second and third columns of $M$ to obtain the final matrix after
$\stackrel{(b)}{\Longrightarrow }$, which is reduced. The
corresponding monomials are $D_1 =y_1y'_1 y'''_2y_3y'_3 $,
$D_2 =y_1 y''_2y_3$,
and $D_3 =y'_2/y_1$.
It is of T-shape under the variable order $y_1 > y_3> y_2$.
\end{example}

We use the following example to illustrate the third step of the
algorithm.

\begin{example}\label{exam-3}
Let $B_1 = y'''_1y'''_2 y'_3y_4y_5^2$,
$B_2 = y''_1y'''_2y'_3y''_3y_4y_5^2$,
$B_3 = y'_1y_3y'_3$,
$B_4 = y'_1$,
$B_5 = y_1^2$.  Then, the symbolic support matrix is $M$ given
below.
%
% \[M=\left(\begin{array}{cccccc}
%x_1^3 &  x_2^3 & x_3 & 1 & 2\\
%x_1^2 &  x_2^3 & x_3^2+x_3 & 1 & 2\\
%x_1   &  0     & x_3+1 & 0 & 0 \\
%x_1   &  0     & 0 & 0 & 0 \\
%2     &  0     & 0 & 0 & 0 \\
%\end{array}\right). \]
%We will use this matrix to illustrate the algorithm.
%We give the major steps to reduce $M$ to T-shape below.
%
\[ M =\left(\begin{array}{cccccc}
x_1^3 &  x_2^3 & x_3 & 1 & 2\\
x_1^2 &  x_2^3 & x_3^2+x_3 & 1 & 2\\
x_1   &  0     & x_3+1 & 0 & 0 \\
x_1   &  0     & 0 & 0 & 0 \\
2     &  0     & 0 & 0 & 0 \\
\end{array}\right)
\stackrel{(c)}{\Longrightarrow} \left(\begin{array}{cccccc}
x_1^3        &  x_2^3 & x_3   & 1 & 2\\
-x_1^3+x_1^2 &  0     & x_3^2 & 0 & 0\\
x_1          &  0     & x_3+1 & 0 & 0 \\
x_1          &  0     & 0 & 0 & 0 \\
2            &  0     & 0 & 0 & 0 \\
\end{array}\right)\]\[
\stackrel{(d)}{\Longrightarrow }
 \left(\begin{array}{cccccc}
x_1^3        &  x_2^3 & x_3    & 1 & 2\\
-x_1^3+x_1^2 &  0     & x_3^2  & 0 & 0\\
x_1          &  0     & x_3+1  & 0 & 0 \\
0            &  0     & -x_3-1 & 0 & 0 \\
2            &  0     & 0      & 0 & 0 \\
\end{array}\right)
\stackrel{(e)}{\Longrightarrow } \left(\begin{array}{cccccc}
 x_2^3 & x_3 &  x_1^3           & 1 & 2\\
 0    & x_3^2  & -x_1^3+x_1^2 & 0 & 0\\
   0          & x_3+1  & x_1 &  0  & 0 \\
0             & -x_3-1   &  0  & 0 & 0 \\
0            &  0     & 2     & 0 & 0 \\
\end{array}\right)
%\Longrightarrow \left(\begin{array}{cccccc}
% x_1^3 & x_2    &  x_3^3           & 1 & 2\\
% 0    & x_2^2  & -x_3^3+x_3^2 & 0 & 0\\
%   0  & x_2+1  & x_3 &  0  & 0 \\
%0    & -x_2-1   &  0  & 0 & 0 \\
%0      &  0     & 2     & 0 & 0 \\
%\end{array}\right).
\]
For step $1$ of the algorithm, we do nothing to $M$ and the zero
matrix $Z$ obtained at the end of this step is a $2\times 2$ zero
sub-matrix at the lower-right corner of $M$.
In step 2, $M_C$ is set to be the lower-right $4\times 4$ sub-matrix
of $M$, $M_{C1}=\left(\begin{array}{cc}
 0 & 0 \\
 0 & 0 \\
\end{array}\right)$,
and $M_{C2}=\left(\begin{array}{cc}
 1 & 2 \\
 0 & 0 \\
\end{array}\right)$.

Merging $Z$ and $M_{C1}$, we obtain a $2\times 4$ zero sub-matrix at
the lower-right corner of $M$. Up to now, $M$ is not changed. Then,
step~3 of the algorithm is applied.

In step 3, we have
$M_{C3}=\left(\begin{array}{c}x_1\\2\\\end{array}\right)$.
%$M_{C3}=(x_1,\, 2)^{\text{T}}$
%
Since $M_{C3}$ is reduced and of full rank, we execute case 1 by
setting $M_{C4}= \left(\begin{array}{cccc}
 x_2^3 & x_3 & 1 & 2\\
 x_2^3 & x_3^2+x_3 & 1 & 2\\
 0     & x_3+1 & 0 & 0 \\
\end{array}\right)$ and $N_4 =$ {\bf RDM}($M_{C4}$) which is a T-shape matrix with
index $(1,1)$ and is not of full rank. Now, $M$ becomes the matrix
after $\stackrel{(c)}{\Longrightarrow}$, which contains $N_4$.
Since $N_4$ is not of full rank, let
%$M_{C5}=\left(\begin{array}{c}x_1\\x_1\\2\\\end{array}\right)$
$M_{C5}=(x_1,\,x_1,\,2)^{\text{T}}$ and compute $N_5 =$ {\bf
RDM}($M_{C5}$). Now $M$ becomes the matrix after
$\stackrel{(d)}{\Longrightarrow}$. We interchange the first column
and the $2,3$-th columns of $M$ to obtain the final matrix which is
in T-shape with index $(1,2)$.

The corresponding monomials are $D_1 = y'''_1y'''_2 y'_3y_4y_5^2$,
$D_2 = y''_1y''_3/y'''_1$,
$D_3 = y'_1y_3y'_3$,
$D_4 = 1/(y_3y'_3)$,
$D_5 = y_1^2$. It is of T-shape under the variable order $y_2> y_3>
y_1>y_4>y_5$.

%$C_1 = y_1^{(3)} y_2^{(1)}y_3^{(3)}y_4y_5^2$,
%%
%$C_2 = y_2^{(2)}y_3^{(2)}/y_3^{(3)}$,
%%
%$C_3 = y_2y_2^{(1)}y_3^{(1)}$
%%
%$C_4 = 1/(y_2y_2^{(1)})$
%%
%$C_5 = y_1^2$.
\end{example}

\begin{algorithm}[H]\caption{\bf --- RDM($M$)}\label{algo-red}
 {\bf{Input}}: Laurent differential monomials $B_1,\ldots, B_m$ in $\Y$ or
             their symbolic support  matrix \\
             \SPC\SPC\,\,\, $M=(d_{ij})_{1\le i \le m, 1\le j \le n} $.\\
{\bf{Output}}: A T-shape matrix which is obtained from
     $M$ by $\Q$-elementary transformations.\\
{\bf{ Initial:}} Let $s = 1$, $p=\max(m,n)$.%
% and $t=$ number of columns of $M$.
%columns of
%$M$.

1.   While $ s \leq \min(m,n)$ do \\
\SPC 1.1 If for any $j,l \geq s$, $\deg(d_{jl}) = -\infty $, let $i=m-s+1,j=n-s+1$ and go to Step~2.\\
\SPC 1.2 Select $j,l \geq s$ such that $-\infty\ne \deg(d_{jl})
 \geq \deg(d_{il})$ for any $i\geq s$. Interchange the $j$-th\\
 \SPC \SPC row  and the $s$-th row, the $l$-th column and the $s$-th column of $M$. Using $d_{ss}$ to do \\
  \SPC\SPC $\Q$-elementary transformations such that $\deg(d_{ss})> \deg(d_{is})$ for $i>s$.\\
\SPC 1.3 $s=s+1$. \\
2. Let $r=i+j$ be the $0$-rank of the $i\times j$ zero
sub-matrix in the lower-right side of $M$.\\
\SPC 2.1 If $M$ is already a T-shape matrix, return $M$.\\
\SPC 2.2 If $j = n$, delete the last $i$ rows from $M$, and let
$N$={\bf RDM($M$)}.\\
\SPC\SPC  Then add $i$ rows of zeros at the bottom of $N$ and return this matrix.\\
%\SPC\SPC\SPC {\bf RDM($M'$)}. Adding $i$ zero rows at the bottom of {\bf RDM($M'$)}, return this matrix.\\
%
\SPC 2.3 If $r \geq p+1$, go to Step~3.\\
\SPC 2.4
 Let $M_C$ be the lower-right $(m+r-p)\times (n+r-p)$ sub-matrix of $M$.\\
\SPC\SPC Let the lower-left $i\times (n+i-p)$ sub-matrix of $M_C$ be
$M_{C1}$ and the \\
\SPC\SPC  upper-right $(m+j-p)\times j$ sub-matrix of $M_C$ be $M_{C2}$. (see (a, b) of Fig. \ref{pic}) \\
%
%\SPC\SPC\SPC \,\, Call {\bf RDM($M_{C1}$)} and {\bf RDM($M_{C2}$)}.\\
%
\SPC 2.5 Let $N_1=$ {\bf RDM($M_{C1}$)} and $N_2=$ {\bf RDM($M_{C2}$)}.\\
\SPC 2.6 If $N_1,N_2$ are reduced matrices, interchange the $p-r+1$
    to  $n-j$ columns and the\\
\SPC\SPC   $n-j+1$ to  $n+m-p$ columns of $M$, return the obtained T-shape (reduced) matrix.  \\
\SPC 2.7 If the $k\times l$ zero sub-matrix $Z_1$ of $N_1$ has
$0$-rank $k+l\geq \max(i,n+i-p)+1=i+1$,\\
\SPC\SPC combine $Z_1$ and
the $i\times j$ zero matrix to obtain a $k\times (l+j)$ zero matrix with 0-rank\\
\SPC\SPC $k+l+j > i+j$ (see (c) of Fig.~\ref{pic}). Let $i=k, j=l+j$, go to Step~2. \\
\SPC 2.8 Else, the $k\times l$ zero sub-matrix $Z_2$ of $N_2$ has
$0$-rank $k+l\geq \max(m+j-p,j)+1=j+1$,\\
\SPC\SPC combine $Z_2$ and
the $i\times j$ zero matrix to obtain a $(k+i)\times l$ zero matrix with 0-rank\\
\SPC\SPC $k+l+i > i+j$. Let $i=k+i, j=l$, go to Step~2. \\
%
%\SPC 3.3 If $N_2$ contains a $k\times l$ zero matrix with $0$-rank
%$k+l\geq \max(m+j-\max(m,n),j)+1$. Combining this $k\times l$ zero
%matrix
%and the $i\times j$ matrix, we get a \\
%%
%\SPC\SPC$(k+i)\times l$ zero
%matrix with $0$-rank $k+l+i > i+j$, go to Step~2. \\
%
3. Let  $M_{C3}$ be the lower-left $i\times (n-j)$ sub-matrix of $M$
 and $N_3=$ {\bf RDM($M_{C3}$)} with index \SPC $(k,l)$.\\
\SPC 3.1 If $l=0$, delete the last $i-k$ rows from $M$, let
$N$={\bf RDM($M$)}, add $i-k$ zero \\
\SPC\SPC\,\,\, rows at
the bottom of $N$ and return this matrix.\\
 \SPC 3.2 If $N_3$ is  not of full rank,
       put the  $k+1,\ldots,(k+l)$-th columns as the first $l$ columns\\
 \SPC\SPC   of $M$. Let $i=i-k$, $j=n-l$,
    $N_3$ the lower-left $i\times (n-j)$ sub-matrix of $M$.\\
\SPC 3.3 Now $N_3$ is of full column rank. Let the upper-right
$(m-i)\times j$ sub-matrix of $M$\\
\SPC\SPC be $M_{C4}$,  $N_4=$ {\bf RDM($M_{C4}$)}, and
$s=\rank(N_4)$.\\
\SPC 3.4 Let the lower-left $(m-s)\times(n-j)$ sub-matrix of $M$ be
$M_{C5}$ and $N_5=$ {\bf
RDM($M_{C5}$)}.\\
\SPC\SPC   Interchange the first $n-j$ columns and the $n-j+1$ to $n-j+s$ columns of $M$, and \\
\SPC\SPC   return the obtained T-shape
matrix. (See (d,e,f) of Fig.~\ref{pic}.) \\
%%
%\SPC\SPC interchanging to obtain a T-shape matrix. Return this matrix. \\
%

\vskip 1pt
\noindent In this algorithm, the $\Q$-elementary transformations in
 {\bf RDM($M_{Ci}$)}$\,(i=1,\ldots,5)$  are also for the whole  $m\times n$ matrix. And in
each step the new $m\times n$ matrix  obtained after doing
$\Q$-elementary transformations    is also denoted by $M$.
\end{algorithm}

The following theorem proves what we claimed before.
\begin{theorem}\label{th-algo}
The symbolic support matrix of any Laurent differential monomials
$B_1,$ $ \ldots,$ $B_m$ can be reduced to a T-shape matrix by a
finite number of $\Q$-elementary transformations.
\end{theorem}
\proof
%
%It suffices to show that Algorithm~\ref{algo-red} is correct and
%will terminate in a finite number of steps.
%
%To be more precise, we claim that if $M$ is of full rank, then ${\bf
%RDM(M)}$ returns a reduced matrix, while in the contrary case,
%${\bf RDM(M)}$ returns a T-shape matrix, and the zero sub-matrix of
%the T-shape matrix has $0$-rank no less than $\max(m,n)+1$.
%
We assume that $m \le n$ and hence $p=\max(m,n)=n$. The case $m > n$
can be proved similarly.

We prove the theorem by induction on the size of the matrix $M$,
that is, $m+n$. One can easily verify that the claim is true when
$m+n = 2,3,4$. Assume it holds for $m+n \leq s-1$, we consider the
case $m+n=s$.

If a T-shape matrix is obtained in Step 1, then the theorem is
proved. Otherwise, let $Z$ be the $i\times j$ zero matrix obtained
in this step. Since the complementary matrix of $Z$ in $M$ is a
square matrix, the $0$-rank of $Z$ is larger than
$\max(m,n)-\min(m,n)+1$.

In Step~2.2, $M$ contains zero rows. By deleting these zero rows,
the size of $M$ is decreased by one at least. By induction, the
algorithm is valid.

In Step~2.3, from $r \geq \max(m,n)+1$, we have $r=i+j \ge n+1$ and
$i > n-j$. Then the $i\times (n-j)$ left-lower sub-matrix of $M$ has
more rows than columns. As a consequence, and $M$ cannot be of full
rank.

In Step~2.4, $M_C$ is chosen as the minimal sub-matrix of $M$ such
that it is of type $(m-q)\times(n-q)$ which may have full row rank.
This implies that $M_{C1}$ must be an $i\times i$ square matrix, and
hence $q=n-r$ and $M_C$ is an $(m+r-n)\times r$ matrix.
Since the complementary matrix of $Z$ in $M$ is a square matrix, we
have $j \ge j-i=n-m$. Hence $m+r-n\ge i$ and $M_C$ contains $Z$ as a
sub-matrix for the first loop, and this is always true since $Z$ is
from $M_C$ and the size of $M_C$ is increasing for each loop.

In Step~2.5, by the induction hypothesis, $N_1=$ {\bf RDM($M_{C1}$)}
and $N_2=$ {\bf RDM($M_{C2}$)} can be computed. Moreover, the
lower-left $m\times (n-r)$ sub-matrix of $M$ is always a reduced one
although the $\Q$-elementary transformations are for the whole rows
of $M$.

In Step 2.6, $N_{1}$ and $N_{2}$ are reduced with full rank. The
algorithm terminates and returns a reduced matrix by suitable column
interchanging given in the algorithm.

In Step 2.7,  $N_1$ is not of full rank. Then by Lemma
\ref{lm-tshape2}, the $k\times l$ zero sub-matrix of $N_1$ has
$0$-rank $k+l \geq \max(i,n+i-\max(m,n))+1=i+1$. The $i\times j$
zero sub-matrix $Z$ and this $k\times l$ zero sub-matrix form a
$k\times (l+j)$ zero-matrix, with $0$-rank $k+j+l \geq i+j+1$
(Figure~\ref{pic}(c)). Step 2.8 can be considered similarly.
Since in each loop of Step 2, the $0$-rank of the zero-matrix $Z$ of
$M$ increases strictly, this loop will terminate.

Step~3 treats the case when $M$ is not of full rank. Note that
$M_{C3}$ has more rows than columns. Step~3.1 is correct due to the
induction hypothesis.

For Step~3.2, since $N_3$ is not of full rank and does not contain
zero rows, we have $l>0$ and $i>k$. These conditions make the
constructions given in the algorithm possible.
%
%Put the $k+1,\ldots,(k+l)$-th columns of $M$ as the first $l$
%columns of $M$. Let $i=i-k, j=n-l$ and $N_3$ the lower-left $i\times
%(n-j)=i\times l$ sub-matrix of $M$, which is a reduced matrix.

In Step~3.3, $N_3$ is an $i\times (n-j)$ reduced matrix with full
column rank and the lower-right $i\times j$ sub-matrix of $M$ is a
zero matrix. Due to this condition, the remaining steps are clearly
valid.
Also note that  $M_{C5}$ is obtained from $N_3$ by adding several
more rows. Then $M_{C5}$ is also of full column rank and hence $N_5$
is a reduced matrix of full column rank. \qedd

\begin{theorem}\label{th-rank}
The differential transcendence degree of the Laurent differential
monomials $B_1,\ldots, B_m$ over $\qq$ equals to the rank of their
symbolic support matrix.
\end{theorem}
\proof By Lemma~\ref{lem-1}, $\Q$-elementary transformations keep
the differential transcendence degree. The result follows from
Theorems~\ref{th-T} and~\ref{th-algo}. \qedd

Theorem \ref{th-rank} can be used to check whether the Laurent
polynomial system \bref{eq-sparseLaurent} is differentially
essential as shown by the following result.
\begin{cor}\label{cor-rk00}
The Laurent differential system \bref{eq-sparseLaurent} is Laurent
differentially essential if and only if there exist
$M_{ij_i}\,(i=0,\ldots,n)$ with $1\leq j_i\leq l_{i}$ such that the
symbolic support matrix of the Laurent differential monomials
$M_{0j_0}/M_{00},\ldots,M_{nj_n}/M_{n0}$ is of rank $n$.
\end{cor}

By Corollary 3.4 of \cite{emiris2005} , the complexity to compute
the determinant of a sub-matrix $M_s$ of $M$ with size $k\times k$
is bounded by $O(k^{k+2}L\gamma^{\frac{2}{k+3}}\Delta)$, where $L =
\log ||M_s||$, $\gamma$ denotes the number of arithmetic operations
required for multiplying a scalar vector by the matrix $M_s$, and
$\Delta$ is the degree bound of $M_s$. So, the complexity to compute
the rank of $M$ is  single exponential at most.

\begin{remark}\label{rem-lde}
A practical way to check whether the Laurent differential system
\bref{eq-sparseLaurent} is Laurent differentially essential is given
below.
\begin{itemize}
\item Choose $n+1$ monomials $M_{ij_i}\,(i=0,\ldots,n)$ with $1\leq
j_i\leq l_{i}$.

\item Use Algorithm \ref{algo-red} to reduce the symbolic support matrix of
 $M_{0j_0}/M_{00},\ldots,M_{nj_n}/M_{n0}$ to a T-shape matrix $M$.

\item Use Theorem \ref{th-T} to check whether the rank of $M$ is $n$.

\item If the rank of $M$ is $n$, then the system is essential.
Otherwise, we need to choose another set of $n+1$ monomials and
repeat the procedure.
\end{itemize}
\end{remark}
The number of possible choices for the $n+1$ monomials is
$\prod_{i=0}^n l_i$, which is very large. But, the procedure is
efficient for two reasons. Firstly, Algorithm \ref{algo-red} is very
efficient, since we are essentially doing numerical computation
instead of symbolic ones. Secondly, the probability for $n+1$
Laurent monomials to have differential transcendence degree $n$ is
very high. As a consequence, we do not need to repeat the procedure
for many choices of $n+1$ monomials.

\vskip10pt By  Corollary \ref{cor-rk00}, property 3) of Theorem
\ref{th-i0} is proved.

\subsection{Differential transcendence degree of
generic Laurent differential polynomials}

Algorithm~\ref{algo-red} and Theorem \ref{th-rank} show how to
reduce the computation of the differential transcendence degree of a
set of Laurent differential monomials to the computation of the rank
of their symbolic support matrix. In this section, this result will
be extended to compute the differential transcendence degree of a
set of generic Laurent differential polynomials.

Consider $m$ generic Laurent differential polynomials
\begin{equation} \label{eq-sparseLaurent1}
\P_i= u_{i0}M_{i0} + \sum\limits_{k=1}^{l_i}u_{ik}
M_{ik}\,(i=1,\ldots,m),
\end{equation} where all the $u_{ik}$ are differentially independent over $\Q$.
Let $\beta_{ik}$ be the symbolic support vector of $M_{ik}/M_{i0}$.
Then the vector $w_i = \sum_{k=0}^{l_i} u_{ik}\beta_{ik}$ is called
the {\em symbolic support vector} of $\P_i$, and the matrix $M_\P$
with $w_1,\ldots,w_m$ as its rows is called the {\em symbolic
support matrix} of $\P_1,\ldots,\P_m$. Then, we have the following
results.

\begin{lemma}\label{lem-mmm}
Let $M_{k_1,\ldots,k_m}$ be the symbolic support matrix of the
Laurent differential monomials
$(M_{1k_1}/M_{10},\ldots,M_{mk_m}/M_{m0})$. Then $\rank(M_\P) =
\max_{ 1\leq k_i \leq l_i} \rank(M_{k_1,\ldots,k_m})$.
\end{lemma}
\proof Let  the rank of $M_\P$ be $r$. Without loss of generality,
we assume that the $r\times r$ leading principal sub-matrix of
$M_\P$, say $M_{\P,r}$, is of full rank. By the properties of
determinant, $\det(M_{\P,r}) = \sum\limits_{k_1=1}^{
l_1}\cdots\sum\limits_{k_r=1}^{ l_r} \prod_{i=1}^r
u_{ik_i}D(k_1,\ldots,k_r)$ where $D(k_1,\ldots,k_r)$ is the
determinant of the $r\times r$ leading principal sub-matrix  of
$M_{k_1,\ldots,k_m}$. So $\det(M_{\P,r})\neq0$ if and only if there
exist $k_1,\ldots,k_r$ such that $D(k_1,\ldots,k_r)\neq0$.
 Hence,  the rank of $M_{k_1,\ldots,k_m}$ is no less than the rank of
$M_\P$.
On the other hand, let $s= \max_{ 1\leq k_i \leq l_i}
\rank(M_{k_1,\ldots,k_m})$. Without loss of generality, we assume
$D(k_1,\ldots,k_s)\ne 0$, then, $\det(M_{\P,s})\ne 0$. Hence, $s$ is
no greater than the rank of $M_\P$. \qedd

The following result is interesting in that it reduces the
computation of differential transcendence degree for a set of
generic differential polynomials to the computation of the rank of a
matrix, which is analogue to the similar result for linear
equations.
\begin{theorem}\label{th-rank11}
%For $\P_i$ defined in \bref{eq-sparseLaurent1},
$\dtrdeg\,\Q\langle\cup_{i=1}^m\bu_i\rangle\langle
\P_1/M_{10},\ldots,
\P_m/M_{m0}\rangle/\Q\langle\cup_{i=1}^m\bu_i\rangle $ $=
\rank(M_\P)$, where $\bu_i$ $=(u_{i0},\ldots,u_{il_i})$.
\end{theorem}
\proof By Lemma~\ref{lm-special}, the differential transcendence
degree of  $\P_1/M_{10},\ldots, \P_m/M_{m0}$ is no less than the
maximal differential transcendence degree of
$M_{1k_1}/M_{10},\ldots,M_{mk_m}/M_{m0}$.

On the other hand, the differential transcendence degree will not
increase by linear combinations since $\dtrdeg\Q\langle \lambda
\rangle\langle
a_1+\lambda\bar{a_1},a_2,\ldots,a_k\rangle/\Q\langle\lambda\rangle
\leq \max(\dtrdeg\Q\langle a_1,a_2\ldots,a_k\rangle/\Q,$ $
\dtrdeg\Q\langle\bar{a_1},a_2,\ldots,a_k\rangle/\Q)$ for any
differential polynomial $a_i\,( 1\leq i \leq k)$ and $ \bar{a_1}$.
So, the differential transcendence degree of  $\P_1/M_{10},\ldots,
\P_m/M_{m0}$ over $\Q\langle\cup_{i=1}^m\bu_i \rangle$ is no greater
than the maximal differential transcendence degree of
$M_{1k_1}/M_{10},\ldots,M_{mk_m}/M_{m0}$.

So, we have  $\dtrdeg\,\Q\langle
\cup_{i=1}^m\bu_i\rangle\langle\P_1/M_{10},\ldots,
\P_m/M_{m0}\rangle/\Q\langle \cup_{i=1}^m\bu_i\rangle=\max_{
k_1,\ldots,k_m}\dtrdeg\,$ $\qq\langle M_{1k_1}/M_{10},$
$\ldots,M_{mk_m}/M_{m0}\rangle/\qq$. By Theorem~\ref{th-rank} and
Lemma~\ref{lem-mmm}, the differential transcendence degree of
$\P_1/M_{10},\ldots, \P_m/M_{m0}$ equals to the rank of $M_\P$.\qedd

By Theorem \ref{th-rank11}, we have the following criterion for
system \bref{eq-sparseLaurent} to be differentially essential.
\begin{cor}\label{cor-rk11}
The Laurent differential system \bref{eq-sparseLaurent} is Laurent
differentially essential if and only if $\rank(M_\P)=n$.
\end{cor}

The difference between Corollary \ref{cor-rk00} and Corollary
\ref{cor-rk11} is that, in the later case we need only to compute
the rank of a single matrix whose elements are multivariate
polynomials in $\sum_{i=0}^n (l_i+1)+n$ variables, while in the
former case we have to compute the ranks of $\prod_{i=0}^n l_i$
matrices whose elements are univariate polynomials in $n$ separate
variables. One also can replace $u_{ik_i}$ by $v_i^{k_i}$ in $M_\P$,
where $v_i$ is a new variable, then the elements of $M_\P$ will be
multivariate polynomials in $2n+1$ variables.
%
%In practice, we believe using Corollary \ref{cor-rk00} is more
%efficient because any $n$ Laurent differential monomials are likely
%to be differentially independent.

\vskip 10pt In the rest of this section, properties for the
elimination ideal
 \begin{equation}\label{eq-iu1}
 \I_u=([\P_1^N,\ldots,\P_m^N]:\mathbbm{m})\cap\Q\{\bu_1,\ldots,\bu_m\}
 \end{equation}
will be studied, where  $\P_i$ are defined in
\bref{eq-sparseLaurent1} and $\bu_i=(u_{i0},\ldots,u_{il_i})$. These
results will lead to deeper understandings for the sparse
differential resultant.

\begin{theorem}\label{th-cod}
Let $\I_u$ be defined in \bref{eq-iu1}. Then $\I_u$ is a
differential prime ideal with codimension $m-\rank(M_{\P})$.
\end{theorem}
\proof
Let $\eta=(\eta_1,\ldots,\eta_n)$ be a generic point of $[0]$ over
$\Q\langle\bu\rangle$, where
$\bu=\{u_{ik}:i=1,\ldots,m;k=1,\ldots,l_i\}$ and
\begin{equation}\label{eq-zetaw}
\zeta_i=-\sum_{k=1}^{l_i}u_{ik}\frac{M_{ik}(\eta)}{M_{i0}(\eta)}\,\,(i=1,\ldots,m).
\end{equation} Similar to the proof of Theorem~\ref{th-Mcodim1}, we
can show that
$\theta=(\eta_1,\ldots,\eta_n;\zeta_1,u_{11},\ldots,u_{1l_1};\ldots;\zeta_m,$
$u_{m1},\ldots,u_{ml_m})$ is a generic point of
$[\P_1^N,\ldots,\P_m^N]:\mathbbm{m}$, which follows that
$[\P_1^N,\ldots,\P_m^N]:\mathbbm{m}$ is a prime differential ideal
in $\Q\{\Y,\bu_1,$ $\ldots,\bu_m\}$. As a consequence, $\I_u$ is a
prime differential ideal.
Since $\zeta_1,\ldots,\zeta_m$ are free of $u_{i0}\,
(i=1,\ldots,m)$, by Theorem~\ref{th-rank11},
\begin{equation*}
\begin{array}{lll}
 & & \dtrdeg\,\Q\langle\bu\rangle\langle\zeta_1,\ldots,\zeta_m\rangle/\Q\langle\bu\rangle   \\
 & & = \dtrdeg\,\Q\langle\bu_1,\ldots,\bu_{m}\rangle\langle\zeta_1,\ldots,\zeta_m\rangle/\Q\langle\bu_1,\ldots,\bu_{m}\rangle   \\
 & & = \dtrdeg\,\Q\langle\bu_1,\ldots,\bu_{m}\rangle
      \langle\frac{\P_1(\eta)}{M_{10}(\eta)},\ldots,\frac{\P_m(\eta)}{M_{m0}(\eta)}\rangle/\Q\langle\bu_1,\ldots,\bu_{m}\rangle   \\
 & & = \rank(M_\P).
\end{array}
\end{equation*}
Hence, the codimension of $\I_u$
%$\I_u=([\P_1^N,\ldots,\P_m^N]:\mathbbm{m})\cap\Q\{\bu_1,\ldots,\bu_m\}$
is $m-\rank(M_{\P})$.\qedd

In the following, two applications of Theorem \ref{th-cod} will be
given.
The first application is to identify certain $\P_i$ such that their
coefficients will not occur in the sparse differential resultant.
This will lead to simplification in the computation of the
resultant.
Let $\TT\subset\{0,1,\ldots,n\}$. We denote by $\P_\TT$ the Laurent
differential polynomial set consisting of $\P_i\,( i\in\TT)$, and
$M_{\P_\TT}$ its symbolic support matrix. For a subset
$\TT\subset\{0,1,\ldots,n\}$, if $\card(\TT) = \rank(M_{\P_\TT})$,
then $\P_\TT$, or $\{\mathcal{A}_i:\,i\in\TT\}$, is called a {\em
differentially independent set}.
\begin{definition}
Let $\TT\subset\{0,1,\ldots,n\}$. Then we say $\TT$ or $\P_\TT$ is
{\em rank essential} if the following conditions hold: (1)
$\card(\TT) - \rank(M_{\P_\TT}) = 1$ and (2) $\card(\JJ) =
\rank(M_{\P_\JJ})$ for each proper subset $\JJ$ of $\TT$.
\end{definition}

Note that rank essential system is the differential analogue of
essential system introduced in \cite{sturmfels2}. Using this
definition, we have the following property, which is similar to
Corollary 1.1 in \cite{sturmfels2}.
\begin{theorem}\label{th-rankessential}
If $\{\P_0,\ldots,\P_n\}$ is a Laurent differentially essential
system, then for any $\TT\subset\{0, 1,\ldots, n \}$, $\card(\TT) -
\rank(M_{\P_\TT}) \leq 1$ and there exists a unique $\TT$ which is
rank essential. In this case, the sparse differential resultant of
$\P_0,\ldots,\P_n$ involves only the coefficients of
$\P_i\,(i\in\TT)$.
\end{theorem}
\proof Since $n = \rank(M_{\P}) \leq \rank(M_{\P_\TT}) + \card(\P) -
\card(\P_\TT) = n+1 + \rank(M_{\P_\TT}) - \card(\TT)$, we have
$\card(\TT) - \rank(M_{\P_\TT}) \leq 1$.
Since $\card(\TT) - \rank(M_{\P_\TT})\ge 0$, for any $\TT$, either
$\card(\TT) - \rank(M_{\P_\TT})=0$ or $\card(\TT) -
\rank(M_{\P_\TT})=1$. From this fact, it is easy to check the
existence of a rank essential set $\TT$.
For the uniqueness, we assume that there exist two subsets
$\TT_1,\TT_2\subset\{1,\ldots,m\}$ which are rank essential. Then,
we have
\begin{equation*}\begin{array}{lll}
\rank(M_{\P_{\TT_1\cup\TT_2}}) &\leq& \rank(M_{\P_{\TT_1}}) + \rank(M_{\P_{\TT_2}}) - \rank(M_{\P_{\TT_1\cap\TT_2}})\\
&= & \card(\TT_1) - 1 + \card(\TT_2) -1 - \card(\TT_1\cap\TT_2) =
\card(\TT_1\cup\TT_2) - 2,
\end{array}
\end{equation*}
%-li I think the first relation may not be =, but should be \leq.
which means that $M_{\P}$ is not of full rank, a contradiction.

Let $\TT$ be a rank essential set. By Theorem \ref{th-cod},
$[\P_i]_{i\in\TT}\cap\Q\{\bu_i\}_{i\in\TT}$ is of codimension one,
which means that the sparse differential resultant of
$\P_0,\ldots,\P_n$ involves of coefficients of $\P_i\,({i\in\TT})$
only. \qedd

Using this property, one can determine which polynomial is needed
for computing the sparse differential resultant, which  will
eventually reduce the computation complexity.

\begin{example}\label{ex-2_1}
Continue from Example \ref{ex-2}. $\{\P_0,\P_1\}$ is a rank
essential sub-system since they involve $y_1$ only.
\end{example}

A more interesting example is given below.
\begin{example}\label{ex-9}
Let $\P$ be a Laurent differential polynomial system where
\begin{equation*}
\begin{array}{llll}
\P_0 &=&  u_{00}y_1y_2+ u_{01}y_3 \\
\P_1 &=&  u_{10}y_1y_2+ u_{11}y_3y_3' \\
\P_2 &=&  u_{20}y_1y_2+ u_{21}y_3' \\
\P_3 &=&  u_{30}y_1^{(o)}+ u_{31}y_2^{(o)} + u_{32}y_3^{(o)} \\
\end{array}
\end{equation*}
where $o$ is a very large positive integer. It is easy to show that
$\P$ is Laurent differentially essential and $\widetilde{\P}=\{
\P_0, \P_1, \P_2 \}$ is the rank-essential sub-system. Note that all
$y_1,y_2,y_3$ are in $\widetilde{\P}$. $\widetilde{\P}$ is rank
essential because $y_1y_2$ can be treated as one variable.
\end{example}

The second application is to prove the dimension conjecture for a
class of generic differential polynomials.
The {\em differential dimension conjecture} proposed by Ritt
\cite[p.178]{ritt} claims that the dimension of any component of $m$
differential polynomial equations in $n\ge m$ variables is no less
than $n-m$.
In \cite{gao}, the dimension conjecture is proved for quasi-generic
differential polynomials. The following theorem proves the
conjecture for a larger class of differential polynomials.

\begin{theorem}\label{dim-conj}
Let $\P_i= u_{i0} + \sum\limits_{k=1}^{l_i}u_{ik}
M_{ik}\,(i=1,\ldots,m;\,m\le n)$ be generic differential polynomials
in $n$ differential indeterminates $\Y$ and
$\bu_i=(u_{i0},\ldots,u_{il_i})$.
Then
$[\P_1,\ldots,\P_m]\subset\Q\langle\bu_1,\ldots,\bu_m\rangle\{\Y \}$
is either the unit ideal or a  prime differential ideal of dimension
$n-m$.
\end{theorem}
\proof Use the notations introduced in the proof of Theorem
\ref{th-cod} with $M_{i0}=1$.
Let $\I_0 = [\P_1,\ldots,\P_m]\subset\Q\{\bu_1,\ldots,\bu_m, \Y \}$
and $\I_1 =
[\P_1,\ldots,\P_m]\subset\Q\langle\bu_1,\ldots,\bu_m\rangle\{\Y \}$.
Since $\P_i$ contains a non-vanishing degree zero term $u_{i0}$, it
is clear that $[\P_1,\ldots,\P_m]:\mathbbm{m} = \I_0$.

From the proof of Theorem \ref{th-cod}, $\I_0$ is a prime
differential ideal with $\theta=(\eta_1,\ldots,\eta_n;\zeta_1,$ $
u_{11},\ldots,u_{1l_1};\ldots;\zeta_m,u_{m1},\ldots,u_{ml_m})$ as a
generic point.
Note that $\rank(M_\P) \le m$ and two cases will be considered.
If $\rank(M_\P) < m$, by Theorem~\ref{th-cod},
$\I_u=[\P_1,\ldots,\P_m]\cap\Q\{\bu_1,\ldots,\bu_m\}$ is of
codimension $m-\rank(M_\P)>0$, which means that $\I_1$ is the unit
ideal in $\Q\langle\bu_1,\ldots,\bu_m\rangle\{\Y \}$.
If $\rank(M_\P) = m$, by the proof of Theorem~\ref{th-cod},
$\dtrdeg\,\Q\langle\bu\rangle\langle\zeta_1,\ldots,\zeta_m\rangle/\Q\langle\bu\rangle
= m$ and $\I_u = [0]$ follows.
Since $\I_0 = \I_1 \cap \Q\{\bu_1,\ldots,\bu_m, \Y \}$ and $\I_0$ is
prime, it is easy to see that $\I_1$ is also a differential prime
ideal in $\Q\langle\bu_1,\ldots,\bu_m\rangle\{\Y \}$.
Moreover, we have
\begin{equation*}
\begin{array}{lll}
 n&=& \dtrdeg\,\Q\langle\bu\rangle\langle\eta_1,\ldots,\eta_n,\zeta_1,\ldots,\zeta_m\rangle/\Q\langle\bu\rangle   \\
 & =& \dtrdeg\,\Q\langle\bu\rangle\langle\eta_1,\ldots,\eta_n,\zeta_1,\ldots,\zeta_m\rangle/\Q\langle\bu,\zeta_1,
 \ldots,\zeta_m\rangle \\
 &\quad &  + \dtrdeg\,\Q\langle\bu,\zeta_1,\ldots,\zeta_m\rangle/\Q\langle\bu\rangle   \\
 & =& \dtrdeg\,\Q\langle\bu\rangle\langle\eta_1,\ldots,\eta_n,\zeta_1,\ldots,\zeta_m\rangle/\Q\langle\bu,\zeta_1,
 \ldots,\zeta_m\rangle + m.
\end{array}
\end{equation*}
Hence, $\dtrdeg\,\Q\langle\bu,\zeta_1,
 \ldots,\zeta_m\rangle\langle\eta_1,\ldots,\eta_n\rangle/\Q\langle\bu,\zeta_1,
 \ldots,\zeta_m\rangle=n-m$. Without loss of generality, suppose $\eta_1,\ldots,\eta_{n-m}$ are differentially independent over
  $\Q\langle\bu,\zeta_1,
 \ldots,\zeta_m\rangle$.
Since $\I_0 = \I_1 \cap \Q\{\bu_1,\ldots,\bu_m, \Y \}$,
$\{y_1,\ldots,y_{n-m}\}$ is a parametric set of $\I_1$.
 Thus,
$[\P_1,\ldots,\P_m]\subset\Q\langle\bu_1,\ldots,\bu_m\rangle\{\Y \}$
is of dimension $n-m$. \qedd

\vskip10pt By Theorem \ref{th-rank11},  Theorem \ref{th-cod}, and
Corollary \ref{cor-rk11}, properties  1) and 2) of Theorem
\ref{th-i0} are proved.

\section{Basic properties of sparse differential resultant}
In this section, we will prove some basic properties for the sparse
differential resultant $\SR(\bu_0,\ldots,\bu_n)$.
%
%Firstly, we will give bounds for  $\ord(\SR,\bu_i)$ and show that $\SR$
%is differentially homogeneous in each $\bu_i$. Secondly, we will
%show that on a Kolchin open set of
%$\prod_{i=0}^n\mathcal{L}(\mathcal{A}_i)$, $\Res_{F_0,\ldots,F_n}=0$
%is a sufficient and necessary condition for the Laurent differential
%polynomials $F_0,\ldots,F_n$ to have common non-polynomial
%solutions. Thirdly, the concept of differential toric varieties is
%introduced and its connection with the differential sparse resultant
%is studied. Finally, a Poisson-type factorization formula for $\SR$ is
%proved.

\subsection{Sparse differential resultant is differentially homogeneous }
Following Kolchin \cite{kol4}, we now introduce the concept of
differentially homogenous polynomials.
\begin{definition} \label{d-homogenous}
A differential polynomial $p \in \mathcal {F}\{y_{0},\ldots,y_{n}\}$
is called differentially homogenous of degree $m$ if for a new
differential indeterminate $\lambda$, we have $p(\lambda
y_{0},\lambda y_{1}\ldots,\lambda
y_{n})=\lambda^{m}p(y_{0},y_{1},\ldots,y_{n}) $.
\end{definition}

The differential analogue of Euler's theorem related to homogenous
polynomials is valid.
\begin{theorem}\cite{kol4} \label{th-dhomo}\,
 $f \in \mathcal{F}\{y_{0},y_{1},\ldots,y_{n}\}$ is differentially
homogenous of degree $m$ if and only if \newline \[\sum_{j=0}^{n}
\sum_{k \in \mathbb{N}} {k+r \choose r} y_{j}^{(k)} \frac{\partial
f(y_{0},\ldots,y_{n})}{\partial y_{j}^{(k+r)} } = \left\{
\begin{array}{ccc} mf & & r = 0 \\ 0 & & r \neq 0 \\ \end{array} \right.\]
\end{theorem}

Sparse differential resultants have the following property.
\begin{theorem}\label{th-homo}
The sparse differential resultant is differentially homogenous in
each $\bu_i$ which is the coefficient set of  $\P_i$.
\end{theorem}
\proof Suppose $\ord(\SR,\bu_i)=h_i\geq0$. Follow the notations used
in Theorem~\ref{th-Mcodim1}. By Corollary \ref{cor-r1},
$\SR(\bu;\zeta_0,\ldots,\zeta_n)=0$. Differentiating this identity
w.r.t. $u_{ij}^{(k)}\,(j=1,\ldots,l_i)$ respectively, we have
{\scriptsize\[
\begin{array}{l} \overline{\frac{\partial
\SR}{\partial u_{i j}}}+\overline{\frac{\partial \SR}{\partial u_{i
0}}}\big(-\frac{M_{ij}(\eta)}{M_{i0}(\eta)}\big)+\overline{\frac{\partial
\SR}{\partial u'_{i
0}}}\big(-[\frac{M_{ij}(\eta)}{M_{i0}(\eta)}]'\big)
+\overline{\frac{\partial \SR}{\partial u_{i
0}''}}\big(-[\frac{M_{ij}(\eta)}{M_{i0}(\eta)}]''\big)+\cdots+
 \overline{\frac{\partial \SR}{\partial u_{i
0}^{(h_i)}}}\big(-{h_i \choose 0}[\frac{M_{ij}(\eta)}{M_{i0}(\eta)}]^{(h_i)}\big)=0\,\qquad(0*)\\
\overline{\frac{\partial \SR}{\partial u_{i
j}'}}+\qquad0\qquad\qquad\quad+\overline{\frac{\partial
\SR}{\partial u'_{i
0}}}\big(-\frac{M_{ij}(\eta)}{M_{i0}(\eta)}\big)+\overline{\frac{\partial
\SR}{\partial u_{i0 }''}}\big(-{2 \choose
1}[\frac{M_{ij}(\eta)}{M_{i0}(\eta)}]'\big)
+\cdots+ \frac{\partial \SR}{\partial u_{i0 }^{(h_i)}}\big(-{h_i \choose 1}[\frac{M_{ij}(\eta)}{M_{i0}(\eta)}]^{(h_i-1)}\big)=0\,\,\,\,(1*)\\
\overline{\frac{\partial \SR}{\partial u_{i
j}''}}+\qquad0\qquad+\qquad\qquad0\qquad\qquad\qquad+\overline{\frac{\partial
\SR}{\partial u_{i0 }''}}\big(-{2 \choose
2}\frac{M_{ij}(\eta)}{M_{i0}(\eta)}\big)+\cdots+\overline{\frac{\partial
\SR}{\partial u_{i0
}^{(h_i)}}}\big(-{h_i \choose 2}[\frac{M_{ij}(\eta)}{M_{i0}(\eta)}]^{(h_i-2)}\big)=0\,\quad(2*)\\
\dotfill \\
\overline{\frac{\partial \SR}{\partial u_{i
j}^{(h_i)}}}+\quad0\qquad+\qquad\qquad0\qquad\qquad\qquad+\qquad\quad0\qquad\qquad\qquad+\cdots+
\overline{\frac{\partial \SR}{\partial u_{i 0}^{(h_i)}}}\big(-{h_i \choose h_i }[\frac{M_{ij}(\eta)}{M_{i0}(\eta)}]^{(0)}\big)=0\,\,\,\,\quad(h_i*)\\
\end{array}
\] }
In the above equations,    $\overline{\frac{\partial \SR}{\partial
u_{i j}^{(k)}}}$ $(k=0,\ldots,h_i; j=0,\ldots,l_i)$ are  obtained by
replacing $u_{i0}$ by $\zeta_{i}\,(i=0, 1,    \ldots,    n)$ in each
$\frac{\partial \SR}{\partial u_{i j}^{(k)}}$ respectively.

Now, let us consider  $\sum_{j=0}^{l_i} \sum_{k\geq 0} {k+r \choose
k}u_{i j}^{(k)}\frac{\partial \SR}{\partial u_{i j}^{(k+r)}}$. Of
course, it needs only to consider $r\leq h_i$.
%
% In the case $r=0$, adding $(0*)\times u_{i j}
%+(1*)\times u_{i j}'+\cdots+(h_i*)\times u_{i j}^{(h_i)}$  for $j$
%from 1 to $l_i$, we obtain $$\sum_{j=1}^{l_i} u_{i
%j}\overline{\frac{\partial \SR}{\partial u_{i j}}}+\sum_{j=1}^{l_i}
%u_{i j}' \overline{\frac{\partial \SR}{\partial u_{i
%j}'}}+\cdots+\sum_{j=1}^{l_i} u_{i j}^{(h_i)}
%\overline{\frac{\partial \SR}{\partial u_{i j}^{(h_i)}}}+\zeta_{i}
%\frac{\partial \SR}{\partial \zeta_{i }}+ \zeta_{i}' \frac{\partial
%R}{\partial \zeta_{i }'}+\cdots+\zeta_{i}^{(h_i)} \frac{\partial
%R}{\partial \zeta_{i }^{(h_i)}}=0.$$
% So  the differential polynomial $\sum\limits^{l_i}_{j=0} u_{i j}\frac{\partial \SR}{\partial
%u_{i j}}+\sum\limits^{l_i}_{j=0} u_{i j}' \frac{\partial \SR}{\partial
%u_{i j}'}+\sum\limits^{l_i}_{j=0} u_{i j}'' \frac{\partial
%R}{\partial u_{i j}''}+\cdots+\sum\limits^{l_i}_{j=0} u_{i
%j}^{(h_i)} \frac{\partial \SR}{\partial u_{i j}^{(h_i)}} $ vanishes at
%$(u_{00},\ldots,u_{n0})=(\zeta_{0},\ldots,\zeta_{n})$. So it can be
%divisible by $\SR$, i.e. $\sum\limits_{j=0}^{l_i}
%\sum\limits_{k=0}^{h_i} u_{i j}^{(k)}$ $ \frac{\partial \SR}{\partial
%u_{i j}^{(k)}}$ $=m\cdot R$ for some $m\in\mathbb{Z}$.
%
%
%In  the case $r \neq 0$,
For each $r\leq h_i$ and each $j\in\{1,\ldots,l_i\}$, {\small \[
\begin{array}{l} 0=(r*)\times {r \choose r} u_{i j} +(r+1*)\times
{r+1 \choose r} u_{i
j}'+\cdots+(h_i*)\times {h_i \choose r} u_{i j}^{(h_i-r)}\\
 = {r \choose r}u_{i j}\overline{\frac{\partial \SR}{\partial u_{i
j}^{(r)}}}+{r+1 \choose r}u_{i j}'\overline{\frac{\partial
\SR}{\partial u_{i j}^{(r+1)}}}+\cdots+{h_i \choose r}u_{i
j}^{(h_i-r)}\overline{\frac{\partial \SR}{\partial u_{i
j}^{(h_i)}}}+\overline{\frac{\partial \SR}{\partial
u_{i0}^{(r)}}}\Big(-u_{i j} \frac{M_{ij}(\eta)}{M_{i0}(\eta)}\Big)
\\+\overline{\frac{\partial \SR}{\partial u_{i0}^{(r+1)}}}\Big(-{r+1 \choose
r}u_{i j} [\frac{M_{ij}(\eta)}{M_{i0}(\eta)}]'-{r+1 \choose r}u_{i
j}'\frac{M_{ij}(\eta)}{M_{i0}(\eta)}\Big) + \cdots\\
+\overline{\frac{\partial \SR}{\partial u_{i0}^{(h_i)}}}\Big(-{h_i
\choose r}u_{i j} [\frac{M_{ij}(\eta)}{M_{i0}(\eta)}]^{(h_i-r)}
-{r+1 \choose r}{h_i \choose r+1}u_{i j}'
[\frac{M_{ij}(\eta)}{M_{i0}(\eta)}]^{(h_i-r-1)}-\cdots -{h_i \choose
r}{h_i \choose h_i}u_{i j}^{(h_i-r)} \frac{M_{ij}(\eta)}{M_{i0}(\eta)}\Big) \\
= {r\choose r}u_{i j}\overline{\frac{\partial \SR}{\partial u_{i
j}^{(r)}}}+{r+1 \choose r}u_{i j}'\overline{\frac{\partial \SR}{
\partial u_{i j}^{(r+1)}}}+\cdots+{h_i \choose r}u_{i
j}^{(h_i-r)}\overline{\frac{\partial \SR}{\partial u_{i j}^{(h_i)}}}
+{r \choose r}\overline{\frac{\partial \SR}{\partial
u_{i0}^{(r)}}}\big(-u_{i j} \frac{M_{ij}(\eta)}{M_{i0}(\eta)}\big) \\
\qquad+{r+1 \choose r}\overline{\frac{\partial \SR}{\partial
u_{i0}^{(r+1)}}}\Big(-u_{i j}
\frac{M_{ij}(\eta)}{M_{i0}(\eta)}\Big)'+\cdots+{h_i \choose
r}\overline{\frac{\partial \SR}{\partial u_{ i0}^{(h_i)}}}\Big(-u_{i
j}\frac{M_{ij}(\eta)}{M_{i0}(\eta)}\Big)^{(h_i-r)} \\
\end{array}\] } It follows that $\sum_{j=1}^{l_i} {r \choose r}u_{i
j}\overline{\frac{\partial \SR}{\partial u_{i
j}^{(r)}}}+\sum_{j=1}^{l_i} {r+1 \choose r}u_{i
j}'\overline{\frac{\partial \SR}{\partial u_{i
j}^{(r+1)}}}+\cdots+\sum_{j=1}^{l_i} {h_i \choose r}u_{i
j}^{(h_i-r)}\overline{\frac{\partial \SR}{\partial u_{i
j}^{(h_i)}}}+{r \choose r}\zeta_{i}\overline{\frac{\partial
\SR}{\partial u_{i0}^{(r)}}}$ $+{r+1 \choose r}$
$\zeta_{i}'\overline{\frac{\partial \SR}{\partial
u_{i0}^{(r+1)}}}+\cdots+{h_i \choose
r}\zeta_{i}^{(h_i-r)}\overline{\frac{\partial \SR}{\partial
u_{i0}^{(h_i)}}}=0$.

By Corollary \ref{cor-r1}, $G=\sum_{k\geq0}\sum_{j=0}^{l_i} {r+k
\choose r}u_{i j}^{(k)}\frac{\partial \SR}{\partial u_{i
j}^{(r+k)}}\in \sat(\SR)$. Since $\ord(G)\leq \ord(\SR)$, $G$ can be
divisible by $\SR$. In the case $r=0$, $\sum\limits_{j=0}^{l_i}
\sum\limits_{k=0}^{h_i} u_{i j}^{(k)}$ $ \frac{\partial
\SR}{\partial u_{i j}^{(k)}}$ $=m\cdot R$ for some $m\in\mathbb{Z}$.
While in the case  $r>0$, if $G\neq 0$, it can not be divisible by
$\SR$. Thus, in this case,  $G$ must be identically zero. From the
above, we conclude that
\[ \sum_{j=0}^{l_i} \sum_{k\geq 0} {k+r \choose r}u_{i
j}^{(k)}\frac{\partial \SR}{\partial u_{i j}^{(k+r)}}=\left\{
\begin{array}{ccc} 0&\quad& r\neq 0 \\ mR&\quad& r=0
\end{array} \right.\]
By Theorem~\ref{th-dhomo}, $\SR(\bu_0,\ldots,\bu_n)$ is
differentially homogenous in each $\bu_i$ and the theorem is
obtained. \qedd

With Theorem \ref{th-homo}, property 1) of Theorem \ref{th-i1} is
proved.

\subsection{Order bound in terms of Jacobi number}

In this section, we will give an order bound for the sparse
differential resultant in terms of the Jacobi number of the given
system.

Consider a generic Laurent differentially essential system
$\{\P_0,\ldots,\P_n\}$ defined in \bref{eq-sparseLaurent} with
$\bu_i=(u_{i0},u_{i1},\ldots,u_{il_i})$ being the
coefficient vector of $\P_i\,(i=0,\ldots,n).$ %That is, $$
%\P_i=\sum\limits_{k=0}^{l_i}u_{ik} M_{ik}\,(i=0,\ldots,n),$$ where
%$u_{ik}$ are differentially independent over $\qq$.
%Suppose the norm
%form of $\P_i\,(i=0,\ldots,n)$ has the following form:
%$$\P_i^{N}=\sum\limits_{k=0}^{l_i}u_{ik} N_{ik}\,(i=0,\ldots,n)$$
%where $u_{ik}$ are differentially independent over $\qq$.
Suppose $\SR$ is the sparse differential resultant of
$\P_0,\ldots,\P_n$. Denote $\ord(\SR,\bu_i)$ to be the maximal order
of $\SR$ in $u_{ik}\,(k=0,\ldots,l_i)$, that is,
$\ord(\SR,\bu_i)=\max_{k}\ord(\SR,u_{ik})$. If $\bu_i$ does not
occur in $\SR$, as shown in Example \ref{ex-2}, then set
$\ord(\SR,\bu_i)=-\infty$. Firstly, we have the following result.
\begin{lemma}
For each $i$, if $\ord(\SR,\bu_i)=h_i\geq0$, then
$\ord(\SR,u_{ik})=h_i\,(k=0,\ldots,l_i)$.
\end{lemma}
\proof  Firstly, we claim that $\ord(\SR,u_{i0})=h_i$. For if not,
suppose $\ord(\SR,u_{ik})=h_i\geq0$ for some $k\neq0$. Then by
differentiating $\SR(\bu;\zeta_0,\ldots,\zeta_n)=0$ w.r.t.
$u_{ik}^{(h_i)}$, we have $\frac{\partial \SR}{\partial
u_{ik}^{(h_i)}}(\bu;\zeta_0,\ldots,\zeta_n)=0$, where $\zeta_i$ are
defined in \bref{eq-zeta}. By Corollary \ref{cor-r1}, we have
$\frac{\partial \SR}{\partial u_{ik}^{(h_i)}}\in\sat(\SR)$, a
contradiction. Thus, $\ord(\SR,u_{i0})=h_i$.
For each $k\neq0$, $\ord(\SR,u_{ik})\leq h_i$. If $\ord(\SR,u_{ik})<
h_i$, differentiate $\SR(\bu;\zeta_0,\ldots,\zeta_n)=0$ w.r.t.
$u_{ik}^{(h_i)}$, we have $\frac{\partial \SR}{\partial
u_{i0}^{(h_i)}}(\bu;\zeta_0,\ldots,\zeta_n)\cdot(-\frac{M_{ik}(\eta)}{M_{i0}(\eta)})=0$.
So $\frac{\partial \SR}{\partial
u_{i0}^{(h_i)}}(\bu;\zeta_0,\ldots,\zeta_n)=0$ and $\frac{\partial
R}{\partial u_{i0}^{(h_i)}}\in\sat(\SR)$, a contradiction. Thus, for
each $k=0,\ldots,l_i$, $\ord(\SR,u_{ik})=h_i$. \qedd

Let $A=(a_{ij})$ be an $n\times n$ matrix where $a_{ij}$ is an
integer or $-\infty$. A {\em diagonal sum} of $A$ is any sum
$a_{1i_1}+a_{2i_2}+\cdots+a_{ni_n}$ with $i_1,\ldots,i_n$ a
permutation of $1,\ldots,n$. If $A$ is an $m\times n$ matrix with
$M=\min\{m,n\}$, then a diagonal sum of $A$ is a diagonal sum of any
$M\times M$ submatrix of $A$. The {\em Jacobi number}  of a matrix
$A$ is the maximal diagonal sum of $A$, denoted by $\Jac(A)$.

%Two $m\times n$ matrices $A$ and $A^\star$ are called $I$-equivalent
%if $A^\star$ can be obtained from $A$ by interchanging rows and
%columns.
%

Let $\ord(\P_i^N,y_j)=e_{ij}\,(i=0,\ldots,n;j=1,\ldots,n)$ and
$\Eord(\P_i)=\ord(\P_i^{N})=e_i$. We call the $(n+1)\times n$ matrix
$A=(e_{ij})$ the {\em order matrix} of $\P_0,\ldots,\P_n$. By
$A_{\hat{i}}$, we mean the submatrix of $A$ obtained by deleting the
$(i+1)$-th row from $A$. We use $\P$ to denote the set
$\{\P_0^{N},\ldots,\P_n^N\}$ and by $\P_{\hat{i}}$, we mean the set
$\P\backslash\{\P_i^N\}$. We call $J_i=\Jac(A_{\hat{i}})$ the {\em
Jacobi number} of the system $\P_{\hat{i}}$, also denoted by
$\Jac(\P_{\hat{i}})$. Before giving an order bound for sparse
differential resultant in terms of the Jacobi numbers, we first give
several lemmas.

Given a vector
$\overrightarrow{K}=(k_0,k_1,\ldots,k_n)\in\mathbb{Z}_{\geq0}^{n+1}$,
we can obtain a prolongation of $\P$:
\begin{equation}\label{eq-pk1}
 \P^{[\overrightarrow{K}]} = \bigcup_{i=0}^n(\P_i^N)^{[k_i]}.\end{equation}
Let $t_j=\max \{e_{0j}+k_0, e_{1j}+k_1, \ldots, e_{nj}+k_n)$. Then
$\P^{[\overrightarrow{K}]}$ is contained in $\Q[\buk,\YK]$, where
$\buk=\cup_{i=0}^n \bu_i^{[k_i]}$ and $\YK = \cup_{j=1}^n
y_j^{[t_j]}$.

Denote $\nu(\PK)$ to be the number of $\Y$ and their derivatives
appearing effectively in $ \PK$. In order to derive a differential
relation among $\bu_{i}\,(i=0,\ldots,n)$ from $\PK$, a sufficient
condition is
\begin{equation}\label{eq-psv1}|\PK|\geq \nu(\PK)+1.\end{equation}
Note that $ \nu(\PK)\leq|\YK|= \sum_{j=1}^{n} (t_j+1) =
\sum_{j=1}^{n} \max(e_{0j}+k_0, e_{1j}+k_1, \ldots, e_{nj}+k_n)+n$.
Thus, if $|\PK|\geq\YK+1$, or equivalently,
\begin{equation}\label{constraint} k_0 + k_1 + \cdots + k_n \geq
\sum\limits_{j=1}^{n} \max(e_{0j}+k_0, e_{1j}+k_1, \ldots,
e_{nj}+k_n)
\end{equation}
is satisfied,  then so is the inequality \bref{eq-psv1}.

\begin{lemma}\label{le-order-cons}
Let $\P$ be a Laurent differentially  essential system and
$\overrightarrow{K}=(k_0,k_1,\ldots,k_n)\in\mathbb{Z}_{\geq0}^{n+1}$
a vector satisfying \bref{constraint}. Then $\ord(\SR,\bu_i)\le k_i$
for each $i=0,\ldots,n$.
\end{lemma}
\proof Denote $\mathbbm{m}^{[\overrightarrow{K}]}$ to be the set of
all monomials in variables $\YK$. Let
$\CI=(\P^{[\overrightarrow{K}]}):\mathbbm{m}^{[\overrightarrow{K}]}$
be an   ideal in the polynomial ring $\qq[\YK,\buk]$. Denote
$U=\buk\backslash\cup_{i=0}^nu_{i0}^{[k_i]}$. Assume
$\P_i^{N}=\sum_{k=0}^{l_i}u_{ik} N_{ik}\,(i=0,\ldots,n)$. Let
$\zeta_{il}=-(\sum_{k=1}^{l_i}u_{ik}N_{ik}/N_{i0})^{(l)}$ for
$i=0,1,\ldots,n;l=0,1,\ldots,k_i$. Denote
$\zeta=(U,\zeta_{0k_0},\ldots,\zeta_{00},\ldots,\zeta_{nk_n},\ldots,\zeta_{n0})$.
It is easy to show that $(\YK,\zeta)$ is a generic point of $\CI$.
 Indeed, on the one hand, each polynomial in $\CI$ vanishes at
$(\YK,\zeta)$. On the other hand, if $f$ is an arbitrary polynomial
in $\qq[\YK,\buk]$ such that $f(\YK,\zeta)=0$,  substitute
$u_{i0}^{(l)}=\big((\P_i^{N}-\sum_{k=1}^{l_i}u_{ik}N_{ik})/N_{i0}\big)^{(l)}$
into $f$, then we have
$\prod_{i=0}^nN_{i0}^{a_i}f\,\equiv\,f_1,\mod\,(\P^{[\overrightarrow{K}]}),$
where $f_1\in\qq[\YK,U]$. Clearly, $f_1=0$ and $f\in\CI$ follows.

Let $\CI_1=\CI\cap\qq[\buk]$. Then $\CI_1$ is a prime ideal with
$\zeta$ as its generic point. Since $\qq(\zeta)\subset\qq(\YK,U)$,
$\codim(\CI_1)=|U|+\sum_{i=0}^n(k_i+1)-\trdeg\,\qq(\zeta)/\qq\geq
|U|+|\P^{[\overrightarrow{K}]}|-\trdeg\,\qq(\YK,U)/\qq=|\P^{[\overrightarrow{K}}|-|\YK|\geq1$.
Thus, $\CI_1\neq (0)$. Suppose $f$ is any nonzero polynomial in
$\CI_1$. Clearly, $\ord(f,\bu_{i})\leq k_i$. Since
$\CI_1\subset[\P_0^N,\ldots,\P_n^N]:\mathbbm{m}\cap\qq\{\bu_0,\ldots,\bu_n\}=\sat(\SR)$,
$f\in\sat(\SR)$. Note that $\SR$ is a characteristic set of
$\sat(\SR)$ w.r.t. any ranking by Lemma~\ref{le-char-codim1}. Thus,
$\ord(\SR,\bu_{i})\leq\ord(f,\bu_i)\leq k_i$. \qedd

\begin{lemma}\label{le-jac-cons}
Let $\P$ be a Laurent differentially essential system and $J_i\geq0$
for each $i=0,\ldots,n$. Then $k_i=J_i\,(i=0,\ldots,n)$ satisfy
~\bref{constraint} in the equality case.
\end{lemma}
\proof Let $A = (e_{ij})$ be the $(n+1)\times n$ order matrix of
$\P$, where $e_{ij} = \ord(\P_i^{N},y_j)$.
Without loss of generality, suppose $J_{0} = e_{11} + e_{22} +
\cdots + e_{nn}$.

Firstly, we will show that for each $k\ne 1$, $e_{11}+J_1 \geq
e_{k1}+J_k$. Since $J_k$ is the Jacobi number of $\P_{\hat{k}}$ and
$k\ne1$, $J_k$ has a summand of the form $e_{1p_1}$. Consider the
longest sequence of summands in $J_k$ in the following form:
 $$T_0=e_{1p_1}+e_{p_1p_2}+ \cdots+ e_{p_{m-1}p_m}$$
 and suppose $J_k = T_0 + T_1$.
 Since $J_k$ is a diagonal sum, $p_i\ne p_j$ for $1< i <
j$. For otherwise, $J_k$ contains $e_{p_{i-1}p_i}$ and
$e_{p_{j-1}p_i}$ as summands, a contradiction.
Also note that $p_i\ne 0$ for $1\leq i \leq m$. Now we claim that
$p_m$ is either equal to $1$ or equal to $k$. Indeed, if $p_m=1$ or
$p_m=k$, $T_0$ cannot be any longer and these two cases may happen.
But if $p_m\ne1$ and $p_m\neq k$, then we can add another summand
$e_{p_{m}p_{m+1}}$ to $T_0$, which contradicts to the fact that
$T_0$ is the longest one.
Now three cases are considered.

\vf Case 1)\, If $p_1=1$, $J_k = e_{11}+T_1$ and $e_{k1}+J_k =
e_{11} + e_{k1} + T_1$. Since $e_{k1} + T_1$ is a diagonal sum of
$\P_{\hat{1}}$, $e_{k1} + T_1 \leq J_1$. Thus, $e_{11}+J_1 \geq
e_{k1}+J_k$.

\vf Case 2)\, If $p_m=1$ for $m>1$, $T_0 =
e_{1p_1}+e_{p_1p_2}+\cdots +e_{p_{m-1}1}$. Since $J_{0} =
e_{11}+\cdots + e_{nn}$, $T_0 \le e_{11} + e_{p_1p_1} + \cdots +
e_{p_{m-1}p_{m-1}}$. For otherwise, since $p_i\neq0$,
$T_0+\sum_{k\in\{2,\ldots,n\}\backslash\{p_1,\ldots,p_{m-1}\}}e_{kk}$
is a diagonal sum of $\P_{\hat{0}}$ which is greater than $J_0$.
Then $e_{k1}+J_k = e_{k1}+T_0 + T_1 \le  e_{k1}+e_{11} + e_{p_1p_1}
+ \cdots + e_{p_{m-1}p_{m-1}} + T_1\le e_{11}+J_1$, where the last
inequality follows from the fact that $e_{k1}+e_{p_1p_1} + \cdots +
e_{p_{m-1}p_{m-1}} + T_1$ is a diagonal sum of $\P_{\hat{1}}$.

\vf Case 3)\, If $p_m=k$, $T_0 = e_{1p_1}+e_{p_1p_2}+\cdots +
e_{p_{m-1}k}$.
Then, similar to case 2), we can show that
$e_{k1}+e_{1p_1}+e_{p_1p_2}+\cdots + e_{p_{m-1}k}\leq
e_{11}+e_{kk}+e_{p_1p_1}+\cdots+e_{p_{m-1}p_{m-1}}$. Thus,
\begin{equation*}
\begin{array}{lll}
e_{k1}+ J_k &=& e_{k1}+e_{1p_1}+e_{p_1p_2}+\cdots + e_{p_{m-1}k} + T_1 \\
&\leq &   e_{kk} + e_{11} + e_{p_1p_1} + \cdots + e_{p_{m-1}p_{m-1}} + T_1 \\
& \leq & e_{11} + J_1.
\end{array}
\end{equation*}

Similarly, we can prove that for each $j$, $e_{jj}+J_j\geq
e_{kj}+J_k$ with $0\leq k\leq n$. So
\begin{equation*}
\begin{array}{lll}
\sum\limits_{j=1}^{n} \max(e_{0j}+J_0,\cdots,e_{nj}+J_{n}) &=& e_{11}+J_1 + e_{22}+J_2 + \cdots + e_{nn}+J_{n} \\
 & = & J_0 + J_1 +\ldots + J_{n}.
\end{array}
\end{equation*}

\qedd

%\begin{theorem}\label{th-order-jacb1}
%Let $\P$ be a Laurent differentially essential system and $J_i\geq0$
%for each $i=0,\ldots,n$. Then $\ord(\SR,\bu_i)\leq
%J_i\,(i=0,\ldots,n)$.
%\end{theorem}

\begin{cor}\label{th-order-jacb1}
Let $\P$ be a Laurent differentially essential system and $J_i\geq0$
for each $i=0,\ldots,n$. Then $\ord(\SR,\bu_i)\leq
J_i\,(i=0,\ldots,n)$.
\end{cor}

\proof It is a direct consequence of Lemma~\ref{le-order-cons} and
Lemma~\ref{le-jac-cons}. \qedd

The above theorem shows that when all the Jacobi numbers are not
less that $0$, then Jacobi numbers are order bounds for the sparse
differential resultant. In the following, we deal with the remaining
case when some $J_i=-\infty$.
To this end, two more lemmas are needed.

\begin{lemma}\cite{cohn,lando}\label{lem-lando}
Let $A$ be an $m\times n$ matrix whose entries are $0$'s and $1$'s.
Let $\Jac(A)=J<\min\{m,n\}$. Then $A$ contains an $a\times b$ zero
sub-matrix with $a+b=m+n-J$.
\end{lemma}

\begin{lemma}\label{lm-essord}
Let $\P$ be a Laurent differentially essential system with the
following $(n+1)\times n$ order matrix
\[
\mbox{\bf $A$}=\left(\begin{array}{cc}
A_{11} & \, (-\infty)_{r\times t} \\
A_{21} & \,A_{22}
\end{array}\right),
\]
where $r+t\geq n+1$. Then $r+t=n+1$ and $\Jac(A_{22})\geq0$.
Moreover, when regarded as differential polynomials in
$y_1,\ldots,y_{r-1}$, $\{\P_{0},\ldots,\P_{r-1}\}$ is a Laurent
differentially essential system.
\end{lemma}
\proof  From the structure of $A$, it follows that the symbolic
support matrix of $\P$ has the following form:
\[\mbox{\bf $M_\P$}=\left(\begin{array}{cc}
B_{11} & \, 0_{r\times t} \\
B_{21} & \,B_{22}
\end{array}\right). \]
Since $\P$ is Laurent differentially essential, by
Corollary~\ref{cor-rk11}, $\rk(M_\P)=n$. As $\rank(M_\P)\leq\rk(
B_{11})+\rk\big((B_{21}\,\,\,B_{22})\big)$, $n \le (n-t)+(n+1-r) =
2n+1-(r+t)$.
 Thus, $r+t\le n+1$,  and
$r+t = n+1$ follows. Since the above inequality becomes equality,
$B_{11}$ has full column rank. As a consequence, $\rank(M_\P) =
\rank(B_{11})+\rank(B_{22})$. Hence, $B_{22}$ is a $t\times t$
nonsingular  matrix. Regarding $\P_{0},\ldots,\P_{r-1}$ as
differential polynomials in $y_1,\ldots,y_{r-1}$, then $B_{11}$ is
the symbolic support matrix of $\{\P_{0},\ldots,\P_{r-1}\}$ which is
of full rank. Thus, $\{\P_{0},\ldots,\P_{r-1}\}$ is a Laurent
differentially essential system.

It remains to show that $\Jac(A_{22})\ge0$. Suppose the contrary,
i.e. $\Jac(A_{22})=-\infty$.  Let $\bar{A}_{22}$ be a $t\times t$
matrix obtained from $A_{22}$ by replacing $-\infty$ by  $0$ and
replacing all other elements  in $A_{22}$ by $1$'s. Then
$\Jac(\bar{A}_{22})<t$, and by Lemma~\ref{lem-lando}, $\bar{A}_{12}$
contains an $a\times b$ zero submatrix with
$a+b=2t-\Jac(\bar{A}_{22})\geq t+1$. By interchanging rows and
interchanging columns when necessary, suppose such a zero submatrix
is in the upper-right corner of $\bar{A}_{22}$. Then
\[\mbox{\bf$A_{22}$}=\left(\begin{array}{cc }
C_{11} & \,  (-\infty)_{a\times b}\\
C_{21} & \,C_{22}
\end{array}\right), \]
where $a+b\geq t+1$. Thus,
\[\mbox{\bf $B_{22}$}=\left(\begin{array}{cc }
D_{11} & \,  0_{a\times b} \\
D_{21} & \,D_{22}
\end{array}\right), \]
which is singular for $a+b\ge t+1$, a contradiction. Thus,
$\Jac(A_{22})\geq0.$\qedd

Now, we are ready to prove the main result of this section.

\begin{theorem}\label{th-ord-jacobi2}
Let $\P$ be a Laurent differentially essential system and $\SR$ the
sparse differential resultant of $\P$. Then
\[\ord(\SR,\bu_i)=\left\{\begin{array}{lll}
-\infty&& \text{if}\quad\,J_i = -\infty,\\
h_i\leq J_i&& \text{if}\quad\,J_i \geq0.\end{array}\right.\]
\end{theorem}
\proof Corollary~\ref{th-order-jacb1} proves the case when
$J_i\geq0$ for each $i$.
 Now suppose there exists at least one $i$ such that $J_i =
-\infty$. Without loss of generality, we assume $J_n = -\infty$ and
let $A_n = (e_{ij})_{0\le i\le n-1; 1\le j\le n}$ be the order
matrix of $\P_{\hat{n}}$. By Lemma~\ref{lem-lando} and similarly as
the procedures in the proof of Lemma~\ref{lm-essord}, we can assume
that $A_n$ is of the following form
\[\mbox{\bf $A_n$}=\left(\begin{array}{cc}
A_{11} & \,(-\infty)_{r\times t} \\
\bar{A}_{21} & \,\bar{A}_{22}
\end{array}\right), \]
where $r+t\geq n+1$. Then the order matrix of $\P$ is equal to
 \[\mbox{\bf $A$}=\left(\begin{array}{cc}
A_{11} & \,(-\infty)_{r\times t} \\
A_{21} & \,A_{22}
\end{array}\right). \]

 Since $\P$ is  Laurent differentially essential, by
Lemma~\ref{lm-essord}, $r+t = n+1$ and $\Jac(A_{22})\ge 0$.
Moreover, considered as differential polynomials in
$y_1,\ldots,y_{r-1}$, $\widetilde{\PS} = \{p_{0},\ldots,p_{r-1}\}$
is Laurent differentially essential and $A_{11}$ is its order
matrix.
Let $\widetilde{J}_i=\Jac((A_{11})_{\hat{i}})$. By applying the
above procedure when necessary, we can suppose that
$\widetilde{J}_i\geq0$ for each $i=0,\ldots,r-1$.
Since
$[\P]\cap\qq\{\bu_0,\ldots,\bu_n\}=[\widetilde{\P}]\cap\qq\{\bu_0,\ldots,\bu_{r-1}\}=\sat(\SR)$,
$\SR$ is also the sparse differential resultant of the system
$\widetilde{\P}$ and $\bu_r,\ldots,\bu_{n}$ will not occur in $\SR$.
By Corollary~\ref{th-order-jacb1}, $\ord(\SR,\bu_{i})\leq
\widetilde{J_i}$. Since $J_{i} = \Jac(A_{22})+
\widetilde{J_i}\geq\widetilde{J_i}$ for $0\le i\le r-1$,
$\ord(\SR,\bu_i)\leq J_{i}$ for $0\leq i\leq r-1$ and
$\ord(\SR,\bu_i)=-\infty$ for $i=r,\ldots,n.$ \qedd

\begin{cor}
Let $\P$ be rank essential. Then $J_i\ge 0$ for $i=0,\ldots,n$ and
 $\ord(\SR,\bu_i) \le J_i$.
\end{cor}
\proof From the proof of Theorem \ref{th-ord-jacobi2}, if
$J_i=-\infty$ for some $i$, then $\P$ contains a proper
differentially essential subsystem, which contradicts to Theorem
\ref{th-rankessential}. Therefore, $J_i\ge 0$ for $i=0,\ldots,n$.
\qedd

The following example shows that  in spite of $J_i\ge 0$,
$\ord(\SR,\bu_{i}) = -\infty $ may happen.
\begin{example}
Let $\P = \{ \P_0, \P_1, \P_2,\P_3 \}$ be a Laurent differential
polynomial system where
\begin{equation*}
\begin{array}{llll}
\P_0 &=& u_{00}+ u_{01}y_1y_1'y_2y_2'' \\
\P_1 &=&  u_{10}+ u_{11}y_1y_1'y_2y_2'' \\
\P_2 &=&  u_{20}+ u_{21}y_1 +u_{22}y_2 \\
\P_3 &=&  u_{30}+ u_{31}y_1' + u_{32}y_3. \\
\end{array}
\end{equation*}
Then, the corresponding order matrix is
\[\mbox{\bf $A$}=\left(\begin{array}{cccccccc}
1 & \, 2 & \, -\infty \\
1 & \, 2 & \, -\infty \\
0 & \, 0 & \, -\infty \\
1 & \, -\infty & \, 0
\end{array}\right). \]
It is easy to show that $\P$ is  Laurent differentially essential
and $\{ \P_0, \P_1 \}$ is the rank-essential sub-system. Here
$\SR=u_{00}u_{11}-u_{01}u_{10}$. Clearly, $\ord(\SR,\bu_0) =
\ord(\SR,\bu_1)= 0$ and $\ord(\SR,\bu_2) = \ord(\SR,\bu_3)=
-\infty$, but $J_0 = 2, J_1 = 2, J_3 = 3, J_4 = -\infty$.
Also note that in this example, the sub-matrix $A_{11}$ in the proof
of Theorem \ref{th-ord-jacobi2} corresponds to
$\left(\begin{array}{cc}
1 &  2  \\
1 & 2  \\
0 &  0
\end{array}\right).$
\end{example}

We conclude this section by giving two improved order bounds based
on the Jacobi bound given in Theorem~\ref{th-ord-jacobi2}.

For each $j\in\{1,\ldots,n\}$, let
$\underline{o}_j=\min\{k\in\mathbb{N}| \,\exists\, i\, \text{s.t.}\,
\deg(\P_i^N,y_j^{(k)})>0\}$. In other words, $\underline{o}_j$ is
the smallest number such that $y_j^{(\underline{o}_j)}$ occurs in
$\{\P_0^N,\ldots,\P_n^{N}\}$. Let $B=(e_{ij}-\underline{o}_j)$ be an
$(n+1)\times n$ matrix. We call $\bar{J}_i=\Jac(B_{\hat{i}})$ the
{\em modified Jacobi number} of the system $\P_{\hat{i}}$. Denote
$\underline{\gamma}=\sum_{j=1}^n\underline{o}_j$. Clearly,
$\bar{J}_i=J_i-\underline{\gamma}.$ Then we have the following
result.

\begin{theorem} \label{th-jacobi-order3}
Let $\P$ be a Laurent differentially essential system and $\SR$ the
sparse differential resultant of $\P$. Then
\[\ord(\SR,\bu_i)=\left\{\begin{array}{lll}
-\infty&& \text{if}\quad\,J_i = -\infty,\\
h_i\leq J_i-\underline{\gamma}&& \text{if}\quad\,J_i
\geq0.\end{array}\right.\]
\end{theorem}
\proof Follow the notations given above. Let $\hat{\P}_i$ be
obtained from $\P_i$ by replacing $y_j^{(k)}$ by
$y_j^{(k-\underline{o}_j)}\,$ $(j=1,\ldots,n;k\geq \underline{o}_j)$
in $\P_i\,(i=0,\ldots,n)$ and denote
$\hat{\P}=\{\hat{\P}_0,\ldots,\hat{\P}_n\}$.  Since
\[M_\P=M_{\hat{\P}}\cdot
\left(\begin{array}{cccc} x_1^{\underline{o}_1}& & & \\
& x_2^{\underline{o}_2}& \multicolumn{2}{c}{\raisebox{0.5ex}[0pt]{\Huge0}}\,\, \\ & & \ddots& \\
\multicolumn{2}{c}{\raisebox{1.0ex}[0pt]{\Huge0}}& &
x_n^{\underline{o}_n}
\end{array}\right),
\]
it follows that $\rk(M_{\hat{\P}})=\rk(M_{\P})=n.$ Thus,
$\CI=[\hat{\P}]\cap\qq\{\bu_0,\ldots,\bu_n\}$ is a prime
differential ideal of codimension $1$. We claim that
$\CI=\sat(\SR)$. Suppose $\P_i=u_{i0}M_{i0}+T_i$ and
$\hat{\P}_i=u_{i0}\hat{M}_{i0}+\hat{T}_i$. Let $\zeta_i=-T_i/M_{i0}$
and $\theta_i=-\hat{T}_i/\hat{M}_{i0}$. Denote
$\bu=\cup_{i=0}^n\bu_i\backslash\{u_{i0}\}$. Then
$\zeta=(\bu,\zeta_0,\ldots,\zeta_n)$ is a generic point of
$\sat(\SR)$ and $\theta=(\bu, \theta_0,\ldots,\theta_n)$ is a
generic point of $\CI$. For any differential polynomial
$G\in\sat(\SR)$,
$G(\zeta)=0=(\sum\phi(\Y)F_\phi(\bu))/(\prod_{i=1}^n M_{i0}^{a_i})$
where $\phi(\Y)$ are distinct differential monomials in $\Y$. Then
$F_\phi(\bu)\equiv0$ for each $\phi$. Thus,
$G(\theta)=(\sum\hat{\phi}(\Y)F_\phi(\bu))/(\prod_{i=1}^n
\hat{M}_{i0}^{a_i})=0$ and $G\in\CI$ follows. So
$\sat(\SR)\subseteq\CI$. In the similar way, we can show that
$\CI\subseteq\sat(\SR)$. Hence, $\SR$ is the sparse differential
resultant of $\hat{\P}$. Since
$\Jac(\hat{\P}_{\hat{i}})=\Jac(\P_{\hat{i}})-\underline{\gamma}$, by
Theorem~\ref{th-ord-jacobi2}, the theorem is proved. \qedd

\begin{remark}\label{re-ordercmparing}
Let $\overrightarrow{K}=(e-e_0,e-e_1,\ldots,e-e_n)$ where
$e=\sum_{i=0}^n e_i$. Clearly,
$|\P^{[\overrightarrow{K}]}|=ne+n+1=|\Y^{[e]}|+1 \ge |\YK|+1$. Then
by Lemma~\ref{le-order-cons}, $\deg(\SR,\bu_i) \leq e-e_i\leq
s-s_i$. Here $s_i$ is the the order of $\P_i\,(i=0,\ldots,n)$.
If $L_i = e-e_i-\gamma(\P)$ where $\gamma(\P)=\sum_{j=1}^n
(\underline{o}_j + \overline{e}_j)$ and $\overline{e}_j = \min_i
\{e_i-\ord(\P^N_i,y_j) | \ord(\P^N_i,y_j)\ne -\infty\}$. By
\cite{sonia-arkiv}, $(L_0,\ldots,L_n)$ also consists of a  solution
to \bref{constraint}. Then $\deg(\SR,\bu_i)\le L_i$.
One can easily check that $\bar{J}_i\le L_i \le e-e_i$ for each $i$,
and the modified Jacobi bound is better than the other two bounds as
shown by the following example.
\end{remark}

\begin{example}
Let $A=(e_{ij})_{0\le i\le n, 1\le j\le n}$ be the  order matrix of
a system $\P$:
\[\mbox{\bf $A$}=\left(\begin{array}{cccccccc}
5 & \, -\infty & \, 0 \\
5 & \, 0 & \, -\infty \\
0 & \, 3 & \, 5 \\
5 & \, 2 & \, -\infty
\end{array}\right). \]
Then $\{ J_0, J_1, J_2, J_3\} = \{12,12,7,10\}$, $\{ L_0, L_1, L_2,
L_3\} = \{13,13,13,13\}$, $\{ e-e_0, e-e_1, e-e_2,e-e_3\} =
\{15,15,$ $15,15\}$. This shows that the modified Jacobi bound could
be strictly less than the other two bounds.
\end{example}

Now, we assume that $\P$ is a Laurent differentially essential
system which is not rank essential. Let $\SR$ be the sparse
differential resultant of $\P$. We will give a better order bound
for $\SR$.
By Theorem \ref{th-rankessential}, $\P$ contains a unique rank
essential sub-system $\P_{\TT}$. Without loss of generality, suppose
$\TT = \{ 0,\ldots, r\}$ with $r< n$.
Let $A_\TT$ be the order matrix of $\P_\TT$ and for $i=0,\ldots,r$,
let $A_{\TT\hat{i}}$ be the matrix obtained from $A_\TT$ by deleting
the $(i+1)$-th row. Note that $A_{\TT\hat{i}}$ is an $r\times n$
matrix.
Then we have the following result.
\begin{theorem}\label{th-jb12}
With the above notations, we have
\[\ord(\SR,\bu_i)=\left\{\begin{array}{lll}
h_i\leq \Jac(A_{\TT\hat{i}})&&i=0,\ldots,r,\\
-\infty&&i=r+1,\ldots,n.\\\end{array}\right.\]
\end{theorem}
\proof It suffices to show that $\ord(\SR,\bu_i) \le
\Jac(A_{\TT\hat{i}})$ for $i=0,\ldots,r$.
%
%Let $A_\TT$ be the order matrix of $\P_\TT$.
%
Let $\L_i = u_{i0} + \sum_{j=1}^n u_{ij}y_j$ for $i=r+1,\ldots,n$.
Since $\P_{\TT}$ is rank essential, there exist
$\frac{M_{ik_i}}{M_{i0}}\,( i= 1,\ldots,r)$ such that their symbolic
support matrix $B$ is of full rank. Without loss of generality, we
assume that the $r$-th principal submatrix of $B$ is of full rank.
Consider a new Laurent differential polynomial system
$\widetilde{\mathbb{P}} = \P_{\TT} \cup\{\L_{r+1},\ldots,\L_{n}\}$.
This system is also Laurent differentially essential since the
symbolic support matrix of $\frac{M_{1k_1}}{M_{10}},\ldots,
\frac{M_{rk_r}}{M_{r0}},y_{r+1},\ldots,y_n$ is of full rank. And
$\SR$ is also the sparse differential resultant of
$\widetilde{\mathbb{P}}$, for $\P_\TT$ is the rank-essential
subsystem of $\widetilde{\mathbb{P}}$.
The order vector of $\L_i$ is $(0,\ldots,0)$ for $i=r+1,\ldots,n$.
So $\Jac(\widetilde{\mathbb{P}}_{\hat{i}})=\Jac(A_{\TT\hat{i}})$ for
$i=0,\ldots,r$.
By Theorem \ref{th-ord-jacobi2}, $\ord(\SR,\bu_i) \le
\Jac((A_{\TT\hat{i}})$ for $i=0,\ldots,r$.\qedd

\begin{example}
%Let $\P$ be a Laurent differential polynomial system where
%\begin{equation*}
%\begin{array}{llll}
%\P_0 &=&  u_{00}y_1y_2+ u_{01}y_3 \\
%\P_1 &=&  u_{10}y_1y_2+ u_{11}y_3y_3' \\
%\P_2 &=&  u_{20}y_1y_2+ u_{21}y_3' \\
%\P_3 &=&  u_{30}y_1^{(o)}+ u_{31}y_2^{(o)} + u_{32}y_3^{(o)} \\
%\end{array}
%\end{equation*}
%where $o$ is a very large positive integer.
%
Continue from Example \ref{ex-9}. The corresponding order matrix is
\[\mbox{\bf $A$}=\left(\begin{array}{cccccccc}
0 & \, 0 & \, 0 \\
0 & \, 0 & \, 1 \\
0 & \, 0 & \, 1 \\
o & \, o&\, o
\end{array}\right). \]
%It is easy to show that $\P$ is Laurent differentially essential and
%$\{ \P_0, \P_1, \P_2 \}$ is the rank-essential sub-system.
%
Here
$\SR=u_{01}u_{10}((u_{21}u_{10})'u_{20}u_{11}-u_{21}u_{10}(u_{20}u_{11})')
- u_{01}u_{10}u_{20}^2u_{11}^2 $.
Clearly, $\ord(\SR,\bu_0) =0,  \ord(\SR,\bu_1)= \ord(\SR,\bu_2)=1$,
and $\ord(\SR,\bu_3) = -\infty$. But $J_0 = J_1 =  J_2 = o+1, J_3 =
1$, and $\ord(\SR,\bu_i)\ll J_i$ for $i=0,1,2$.
If using Theorem \ref{th-jb12}, then $A_\TT$ consists of the first
three rows of $A$ and Jacobi numbers for $A_\TT$ are $1,1,1$
respectively, which give much better bounds for the sparse
differential resultant.
\end{example}

With Theorem~\ref{th-ord-jacobi2}, property 2) of Theorem
\ref{th-i1} is proved.

\subsection{Differential toric variety and sparse differential resultant}
\label{sec-toric}

In this section, we will introduce the concept of differential toric
variety and establish its relation with the sparse differential
resultant.

We will deal with the special case when all the $\mathcal{A}_i$
coincide with each other, i.e.,
$\mathcal{A}_0=\cdots=\mathcal{A}_n=\mathcal{A}$. In this case,
$\mathcal{A}$ is said to be Laurent differentially essential when
$\mathcal{A}_0,\ldots,\mathcal{A}_n$ form a Laurent differentially
essential system.
Let
$\mathcal{A}=\{M_0=(\Y^{[o]})^{\alpha_{0}},M_1=(\Y^{[o]})^{\alpha_{1}},\ldots,M_l=(\Y^{[o]})^{\alpha_{l}}\}$
be Laurent differentially essential where
$\alpha_k\in\mathbb{Z}^{n(o+1)}$. Then by Definition~\ref{def-tdes},
$l\geq n$ and there exist indices $k_1,\ldots,k_n\in\{1,\ldots,l\}$
such that
$\frac{(\Y^{[o]})^{\alpha_{k_1}}}{(\Y^{[o]})^{\alpha_{0}}},\ldots,\frac{(\Y^{[o]})^{\alpha_{k_n}}}{(\Y^{[o]})^{\alpha_{0}}}$
are differentially independent over $\qq$.
Let
$$\P_i=u_{i0}M_0+u_{i1}M_1+\cdots+u_{il}M_l\,(i=0,\ldots,n)$$
be $n+1$ generic Laurent differential polynomials w.r.t $\A$.

Consider the following map
$$\phi_{\mathcal{A}}:\,(\ee^\wedge )^n\,\longrightarrow\,\textbf{P}(l)$$ defined by
\begin{equation} \label{eq-toric-map}
\phi_{\mathcal{A}}(\xi_1,\ldots,\xi_n)=((\xi^{[o]})^{\alpha_{0}},(\xi^{[o]})^{\alpha_{1}},\ldots,(\xi^{[o]})^{\alpha_{l}})
\end{equation}
where $\textbf{P}(l)$ is the $l$-dimensional differential projective
space over $\ee$ %\cite{kol74,li1}
 and $\xi=(\xi_1,\ldots,\xi_n)$
$\in (\ee^\wedge )^n$. Note that
$((\xi^{[o]})^{\alpha_{0}},(\xi^{[o]})^{\alpha_{1}},\ldots,(\xi^{[o]})^{\alpha_{l}})$
is never the zero vector since $\xi_i\in\ee^\wedge$ for all $i$.
Thus $\phi_{\mathcal{A}}$ is well defined on all of $ (\ee^\wedge
)^n$, though the image of $\phi_{\mathcal{A}}$ need not be a
differential projective variety of $\textbf{P}(l)$. Now we give the
definition of differential toric variety.
\begin{definition}\label{def-toricvariety}
The Kolchin projective differential closure of the image of
$\phi_{\mathcal{A}}$ is defined to be the {\em differential toric
variety w.r.t. $\mathcal{A}$}, denoted by $X_\mathcal{A}$. That is,
$X_\mathcal{A}=\overline{\phi_{\mathcal{A}}\big((\ee^\wedge
)^n\big)}$.
\end{definition}

 Then we have the following theorem.
\begin{theorem} \label{th-XA}
$X_\mathcal{A}$ is an irreducible projective differential variety
over $\qq$ of dimension $n$.
\end{theorem}
\proof Denote $\P_i^N=\sum_{k=0}^lu_{ik}N_{k}\,(i=0,\ldots,n)$ where
$N_{k}\in\mathbbm{m}$. Clearly, $M_k/M_0=N_k/N_0\,(k=1,\ldots,l)$.
Let
$\J=[N_0z_1-N_1z_0,\ldots,N_0z_l-N_lz_0]:\mathbbm{m}\in\qq\{\Y;z_0,z_1,\ldots,z_l\},$
where $\mathbbm{m}$ denotes the set of all differential monomials in
$\Y$. Let $\eta=(\eta_1,\ldots,\eta_n)$ be a generic point of $[0]$
over $\qq$ and $v$ a differential indeterminate over
$\qq\langle\eta\rangle$. Let
$\theta=(v,\frac{N_1(\eta)}{N_0(\eta)}v,\ldots,\frac{N_l(\eta)}{N_0(\eta)}v)$.
We claim that $(\eta;\theta)$ is a generic point of $\J$ which
follows that $\J$ is a prime differential ideal. Indeed, on the one
hand, since each $N_0z_i-N_iz_0\,(i=1,\ldots,l)$ vanishes at
$(\eta;\theta)$ and $\eta$ annuls none of the elements of
$\mathbbm{m}$, $(\eta;\theta)$ is a common zero of $\J$. On the
other hand, for any $f\in\qq\{\Y;z_0,z_1,\ldots,z_l\}$ which
vanishes at $(\eta;\theta)$, let $f_1$ be the differential remainder
of $f$ w.r.t. $N_0z_i-N_iz_0\,(i=1,\ldots,l)$ under the elimination
ranking $z_1\succ \ldots\succ z_l\succ z_0\succ\Y$. Then
$f_1\in\qq\{\Y;z_0\}$ satisfies that
$N_0^af\equiv\,f_1\,\mod\,[N_0z_1-N_1z_0,N_0z_2-N_2z_0,\ldots,N_0z_l-N_lz_0]$.
Since $f(\eta;\theta)=0$, $f_1(\eta_1,\ldots,\eta_n,v)=0$, and
$f_1=0$ follows. Thus, $f\in\,\J$ and the claim is proved.

Let $\J_1=\J\cap\qq\{z_0,z_1,\ldots,z_l\}$. Then $\J_1$ is a prime
differential ideal with a generic point $\theta$. Denote
$\bz=(z_0,z_1,\ldots,z_l)$. For any $f\in\J_1:\bz$, since
$z_0f\in\J_1$, $z_0f$ vanishes at $\theta$ and $f(\theta)=0$
follows. So $f\in\J_1$, and it follows that  $\J_1:\bz=\J_1$.
%Suppose $\xi=(\xi_0,\ldots,\xi_l)\in\ee^{l+1}$ is a zero of $\J_1$,
%then for $s\in\ee\backslash\{0\}$, we will show that
%$(s\xi_0,\ldots,s\xi_l)$ is also a zero of $\J_1$.
And for any $f\in\J_1\subset\J$ and any differential indeterminate
$\lambda$ over $\Q\langle\eta,v\rangle$, let
$f(\lambda\bz)=\sum\phi(\lambda)f_\phi(\bz)$ where $\phi(\lambda)$
are distinct differential monomials in $\lambda$ and
$f_\phi(\bz)\in\Q\{\bz\}$. Then
$f(\lambda\theta)=0=\sum\phi(\lambda)f_\phi(\theta)$. So each
$f_\phi(\theta)=0$ and $f_\phi\in\J_1$ follows. Thus,
$f(\lambda\bz)\in\Q\{\lambda\}\J_1$. By
Definition~\ref{def-diffhomo}, $\J_1$ is a differentially homogenous
differential ideal. Then $V=\V(\J_1)$ is an irreducible projective
differential variety in $\textbf{P}(l)$. Since $\theta$ is a generic
point of $V$,
$\dim(V)=\dtrdeg\,\qq\langle\frac{N_1(\eta)}{N_0(\eta)},\ldots,\frac{N_l(\eta)}{N_0(\eta)}\rangle/\qq=n$.
If we can show  $X_\mathcal{A}=V$, then it follows that
$X_\mathcal{A}$ is an irreducible  projective differential variety
of dimension $n$.

For any point $\xi\in(\ee^\wedge )^n$, it is clear that
$(\xi;N_0(\xi),N_1(\xi),\ldots,N_l(\xi))$ is a differential zero of
$\J$ and consequently
$(N_0(\xi),N_1(\xi),\ldots,N_l(\xi))\in\V(\J_1)=V$. So
$\phi(\xi)=(N_0(\xi),N_1(\xi),\ldots,$ $N_l(\xi))\in V$. Thus
$\phi_{\mathcal{A}}\big((\ee^\wedge )^n\big)\subseteq V$ and
$X_{\mathcal{A}}=\overline{\phi_{\mathcal{A}}\big((\ee^\wedge
)^n\big)} \subseteq V$ follows. Conversely, since
$\phi(\eta)=(1,\frac{N_1(\eta)}{N_0(\eta)},\ldots,\frac{N_l(\eta)}{N_0(\eta)})\in
X_\mathcal{A}$ is the generic point of $V$, $V\subseteq
X_\mathcal{A}$. Thus, $V= X_\mathcal{A}$. \qedd

Now, suppose $z_0,\ldots,z_l$ are the homogenous coordinates of
$\textbf{P}(l)$. Let
$$\L_i=u_{i0}z_0+u_{i1}z_1+\cdots+u_{il}z_l\,(i=0,\ldots,n)$$
be generic differential hyperplanes in $\textbf{P}(l)$.  Then,
clearly, $\P_i=\L_i\circ\phi_{\mathcal{A}}$. In the following, we
will explore the close relation between $\Res_{\mathcal{A}}$ and the
differential Chow form of $X_\mathcal{A}$. Before doing so, we first
recall the concept of projective differential Chow form
(\cite{li1}).

  Let $V$ be an irreducible
projective differential variety  of dimension $d$ over $\qq$ with
 a generic point $\xi=(\xi_0,\xi_1,\ldots,\xi_l)$. Suppose $\xi_0\neq0$. Let
$\L_i=-\sum_{k=0}^lu_{ik}z_k\,(i=0,\ldots,d)$ be $d+1$ generic
projective differential hyperplanes. Denote
$\zeta_i=-\sum_{k=1}^lu_{ik}\xi_0^{-1}\xi_k\,(i=0,\ldots,d)$ and
$\bu_i=(u_{i0},\ldots,u_{il})$. Then we showed in \cite{li1} that
the prime ideal  $\mathbb{I}(\zeta_0,\ldots,\zeta_d)$ over
$\qq\langle\cup_i\bu_i\backslash \{u_{i0}\}\rangle$
%with$(\zeta_0,\ldots,\zeta_d)$ as a generic point
is of codimension one. That is, there exists an irreducible
differential polynomial $F\in\qq\{\bu_0,\ldots,\bu_d\}$ such that
$\mathbb{I}(\zeta_0,u_{01},\ldots,u_{0l};\ldots;\zeta_d,u_{d1},\ldots,u_{dl})=\sat(F)$.
This $F$ is defined to be the {\em differential Chow form }of
$\V(\I)$ or $\I$. We list one of its properties which will be used
in this section.

\begin{theorem} \label{th-sf}\cite[Theorem 4.7]{li1}
Let $F(\bu_{0},\bu_{1},\ldots,\bu_{d})$ be the differential Chow
form of $V$ with $\ord(F)=h$ and $S_{F}=\frac{\partial F}{\partial
u_{00}^{(h)}}$ .
Suppose that $\bu_i$ are differentially specialized over $\qq$ to
sets $\bv_i\subset\mathcal{E}$ and $\overline{\P}_{i}$ are obtained
by substituting $\bu_i$ by $\bv_i$ in $\P_i\,(i=0,\ldots,d)$.
If\, $\overline{\P}_i=0\,(i=0,\ldots,d)$ meet $V$, then $\sat(F)$
vanishes at $(\bv_{0},\ldots,\bv_{d})$. Furthermore, if
$F(\bv_{0},\ldots,\bv_{d})=0$ and $S_{F}(\bv_{0},\ldots,\bv_{d})\neq
0$, then the $d+1$ differential hyperplanes $\overline{\P}_{i}=0$
$\,(i=0,\ldots,d)$ meet $V$.
\end{theorem}

With the above  preparations, we now proceed to show that the sparse
differential resultant is just the differential Chow form of
$X_\mathcal{A}$.

\begin{theorem}\label{th-toric-chow}
Let $\Res_{\mathcal{A}}$ be the sparse differential resultant of
$\P_0,\ldots,\P_n$. Then $\Res_{\mathcal{A}}$ is the differential
Chow form of $X_\mathcal{A}$ with respect to the generic hyperplanes
$\L_i\,(i=0,\ldots,n)$.
\end{theorem}
\proof By the proof of Theorem~\ref{th-XA}, $X_\mathcal{A}$ is an
irreducible projective differential variety of dimension $n$ with a
generic point
$(1,\frac{M_1(\eta)}{M_0(\eta)},\ldots,\frac{M_l(\eta)}{M_0(\eta)})$.
Let
$\zeta_i=-\sum_{k=1}^lu_{ik}\frac{M_k(\eta)}{M_0(\eta)}\,(i=0,\ldots,n)$.
Then
$\sat(\chow(X_\mathcal{A}))=\mathbb{I}((\zeta_0,u_{01},\ldots,u_{0l};\ldots;\zeta_n,u_{n1},\ldots,u_{nl}))$.
And by the definition of sparse differential resultant,
$\sat(\Res_{\mathcal{A}})=\mathbb{I}((\zeta_0,u_{01},\ldots,u_{0l};\ldots;\zeta_n,u_{n1},\ldots,u_{nl}))$.
By Lemma~\ref{le-char-codim1}, $\chow(X_\mathcal{A})$ and
$\Res_{\mathcal{A}}$ can only differ at most by a nonzero element in
$\qq$. Thus, $\Res_{\mathcal{A}}$ is just the differential Chow form
of $X_\mathcal{A}$. \qedd

Based on Theorem~\ref{th-sf}, we give another characterization of
the vanishing of sparse differential resultants below, where the
zeros are taken from $\ee$ instead of $\ee^{\wedge}$.
\begin{cor}
Let
$\bar{\L}_i=v_{i0}z_0+v_{i1}z_1+\cdots+v_{il}z_l=0\,(i=0,\ldots,n)$
be projective differential hyperplanes over $\ee$ with
$\bv_i=(v_{i0},\ldots,v_{il})$. Denote $\ord(\Res_{\mathcal{A}})=h$
and $S_\SR=\frac{\partial\Res_{\mathcal{A}}}{\partial
u_{00}^{(h)}}$. If $X_\mathcal{A}$ meets
$\bar{\L}_i=0\,(i=0,\ldots,n)$, then
$\Res_{\mathcal{A}}(\bv_0,\ldots,\bv_n)=0$. And if
$\Res_{\mathcal{A}}(\bv_0,\ldots,\bv_n)=0$ and
$S_\SR(\bv_0,\ldots,\bv_n)\ne0$, then $X_\mathcal{A}$ meets
$\bar{\L}_i=0\,(i=0,\ldots,n)$.
\end{cor}
\proof It follows directly from Theorems~\ref{th-toric-chow} and
\ref{th-sf}. \qedd

\begin{example}
Continue from Example~\ref{ex-z1z2-z2'}.
Following the proof of Theorem \ref{th-XA}, consider
$\J=[y_1z_1-y_1'z_0,y_1z_2-y_1^2z_0]:\mathbbm{m}$. It is easy to
show that $X_\mathcal{A}$ is the general component of
$z_1z_2-(z_0z_2'-z_0'z_2)$, that is,
$X_\mathcal{A}=\V(\sat(z_1z_2-(z_0z_2'-z_0'z_2)))$. And
$Res_\mathcal{A}$ is equal to the differential Chow form of
$X_\mathcal{A}$.
\end{example}

By Theorems \ref{th-XA} and \ref{th-toric-chow}, property 4) of
Theorem \ref{th-i1} is proved.

\subsection{Poisson-type product formulas}
In this section, we prove formulas for sparse differential
resultants, which are similar to the Poisson-type product formulas
for multivariate resultants \cite{Pedersen}.

Denote $\ord(\SR,\bu_i)$ by $h_i$ ($i=0,\ldots,n$), and suppose
$h_0\geq0$.
%We follow the steps in \cite{gao} to derive the formula.
%
Let $\tilde{\bu}=\cup_{i=0}^n \bu_i\setminus \{u_{00}\}$ and
$\qq_{0}=\qq\langle\tilde{\bu}\rangle(u_{00}^{(0)},\ldots,u_{00}^{(h_0-1)})$.
Consider $\SR$ as an irreducible algebraic polynomial
$r(u_{00}^{(h_0)})$ in $\qq_{0}[u_{00}^{(h_0)}]$.
In a suitable algebraic extension field of $\qq_{0}$,
$r(u_{00}^{(h_0)})=0$ has
$t_0=\deg(r,u_{00}^{(h_0)})=\deg(\SR,u_{00}^{(h_0)})$ roots
$\gamma_{1},\ldots,\gamma_{t_0}$. Thus
\begin{equation}\label{eq-fac00}
\SR(\bu_{0},\bu_{1},\ldots,\bu_{n})=A\prod^{t_0}_{\tau=1}(u_{00}^{(h_0)}-\gamma_{\tau})
\end{equation}
where
$A\in\qq\langle\bu_1,\ldots,\bu_n\rangle[\bu_0^{[h_0]}\backslash
u_{00}^{(h_0)}]$.
For each $\tau$ such that $1\le \tau\le t_0$, let
\begin{equation}\label{eq-ftau}
 \qq_\tau=\qq_0(\gamma_{\tau})=\qq\langle\tilde{\bu}\rangle(u_{00}^{(0)},\ldots,u_{00}^{(h_0-1)},\gamma_{\tau})
\end{equation}
be an algebraic extension field of $\qq_0$ defined by
$r(u_{00}^{(h)})=0$.
We will define a derivation operator $\delta_{\tau}$ on $\qq_\tau$
so that $\qq_\tau$ becomes a $\delta_\tau$-field.
This can be done in a very natural way. For $e\in
\Q\langle\tilde{\bu}\rangle$, define $\delta_{\tau} e = \delta
e=e'$. Define $\delta_{\tau}^{i} u_{00} = u_{00}^{(i)}$ for
$i=0,\ldots,h_0-1$ and $$\delta_{\tau}^{h_0} u_{00} =
\gamma_{\tau}.$$
Since $\SR$, regarded as an algebraic polynomial $r$ in
$u_{00}^{(h_0)}$, is a minimal polynomial of $\gamma_{\tau}$,
$\sep_{\SR}=\frac{\partial \SR}{\partial u_{00}^{(h_0)}}$ does not
vanish at $u_{00}^{(h_0)}=\gamma_{\tau}$. Now, we define the
derivatives of $\delta_{\tau}^{i} u_{00}$ for $i> h_0$ by induction.
Firstly, since $r(\gamma_{\tau})=0$, $\delta_\tau
(r(\gamma_{\tau}))=\sep_{\SR}\big|_{u_{00}^{(h_0)}=\gamma_\tau}\delta_\tau
(\gamma_{\tau})+T\big|_{u_{00}^{(h_0)}=\gamma_\tau}=0$, where
$T=\SR'-\sep_\SR u_{00}^{(h_0+1)}$. We define $\delta_{\tau}^{h_0+1}
u_{00}$ to be $\delta_\tau
(\gamma_{\tau})=-\frac{T}{\sep_{\SR}}\Big|_{u_{00}^{(h_0)}=\gamma_{\tau}}$.
Supposing the derivatives of $\delta_{\tau}^{h_0+j} u_{00}$ with
order less than $j<i$ have been defined, we now  define
$\delta_{\tau}^{h_0+i} u_{00}$. Since
$\SR^{(i)}=\sep_{\SR}u_{00}^{(h_0+i)}+T_{i}$ is linear in
$u_{00}^{(h_0+i)}$, we define $\delta_{\tau}^{h_0+i} u_{00}$ to be
$-\frac{T_{i}}{\sep_{\SR}}\Big|_{u_{00}^{(h_0+j)}=\delta_\tau^{h_0+j}u_{00},
j <i}$.

In this way, $(\qq_\tau,\delta_\tau)$ is a differential field which
can be considered as a finitely generated differential extension of
$\qq\langle\tilde{\bu}\rangle$. Recall that
$\qq\langle\tilde{\bu}\rangle$ is a finitely  generated
differential extension field of $\qq$ contained in $\ee$. By the
definition of universal differential extension field, there exists a
differential extension field $\mathcal {G}\subset\ee$ of
$\qq\langle\tilde{\bu}\rangle$ and a differential isomorphism
$\varphi_\tau$ over $\qq\langle\tilde{\bu}\rangle$ from
$(\qq_\tau,\delta_\tau)$ to $(\mathcal {G},\delta)$. Summing up the
above results, we have

\begin{lemma}\label{lm-ftau}
$(\qq_\tau, \delta_\tau)$ defined above is a finitely  generated
differential extension field of  $\qq\langle\tilde{\bu}\rangle$,
which is differentially $\qq\langle\tilde{\bu}\rangle$-isomorphic to
a subfield of $\ee$.
\end{lemma}

Let $p$ be a differential polynomial in
$\mathcal{F}\{\bu_0,\bu_1,\ldots,\bu_n\} =
\ff\{\tilde{\bu},u_{00}\}$.
For convenience,  by the symbol
$p\Big|_{u_{00}^{(h_0)}=\gamma_{\tau}}$, we mean substituting
$u_{00}^{(h_0+i)}$ by $\delta_\tau^{i}\gamma_{\tau}\,(i\geq0)$ in
$p$.
Similarly, by saying $p$ vanishes at $u_{00}^{(h_0)}=\gamma_{\tau}$,
we mean $p\Big|_{u_{00}^{(h_0)}=\gamma_{\tau}}=0$. It is easy to
prove the following lemma.

\begin{lemma}\label{lm-ftaup}
Let $p$ be a  differential polynomial in
$\ff\{\tilde{\bu},u_{00}\}$. Then $p\in\sat(\SR)$ if and only if $p$
vanishes at $u_{00}^{(h_0)}=\gamma_{\tau}$.
\end{lemma}

When a  differential polynomial
$p\in\qq\langle\tilde{\bu}\rangle\{\Y\}$ vanishes at a point
$\eta\in\qq_\tau^n$, it is easy to see that $p$ vanishes at
$\varphi_\tau(\eta)\in\ee^n$. For convenience, by saying $\eta$ is
in a differential variety $V$ over $\qq\langle\tilde{\bu}\rangle$,
we mean $\varphi_\tau(\eta)\in V$.

%\begin{remark}\label{rm-ftau}
%In order to make $\qq_\tau$ a differential field, we need to
%introduce a differential operator $\delta_\tau$ which is related to
%$\gamma_\tau$ and there does not exist a single differential
%operator to make all $\qq_\tau(\tau=1,\ldots,g)$ differential
%fields.
%%
%This natural phenomenon related with nonlinear differential
%equations seems not used before.
%%
%For instance, let $p=y'^2-4y$. Then $\CI=\sat(p)$ is a prime ideal
%in $\Q(t)\{y\}$ and let $\overline{\ff}$ be the differential
%rational function field of $\CI$.
%%
%By factoring $p = (y' -2\sqrt{y})(y' +2\sqrt{y})$, we can define two
%more differential fields: $\ff_1=\Q(t)(\sqrt{y})$ with a
%differential operator $\delta_1 y = 2\sqrt{y}$ and
%$\ff_2=\Q(t)(\sqrt{y})$ with a differential operator $\delta_2 y =
%-2\sqrt{y}$. Note that $\ff_1$ and $\ff_2$ are not compatible,
%although each of them is isomorphic to a subfield of $\ee$. Finally,
%both $\ff_1$ and $\ff_2$ are isomorphic to  $\overline{\ff}$.
%\end{remark}

With these preparations, we now give the following theorem.
\begin{theorem}\label{th-fac1}
Let  $\SR(\bu_{0},\bu_{1},\ldots,\bu_{n})$
 be the sparse differential resultant
of $\P_0,\ldots,\P_n$ with $\ord(\SR,\bu_{0})=h_0\geq0$. Let
$\deg(\SR,u_{00}^{(h_0)})=t_0$. Then there exist $\xi_{\tau k}$ in
an extension field $(\qq_\tau,\delta_\tau)$ of
$(\qq\langle\tilde{\bu}\rangle,\delta)$ for $\tau=1,\ldots,t_0$ and
$k=1,\ldots,l_0$ such that
\begin{equation} \label{eq-decom1} \SR=A\prod_{\tau=1}^{t_0} (u_{00}+\sum\limits_{k=1}^{l_0} u_{0
k}\xi_{\tau k})^{(h_0)},
\end{equation} where $A$ is a polynomial in
$\qq\langle\bu_1,\ldots,\bu_n\rangle[\bu_0^{[h_0]}\backslash
u_{00}^{(h_0)}]$.
 Note that equation \bref{eq-decom1} is formal and should be
understood in the following precise meaning:
$(u_{00}+\sum_{\rho=1}^n u_{0\rho}\xi_{\tau \rho})^{(h_0)}
\stackrel{\triangle}{=}
\delta^{h_0}u_{00}+\delta_\tau^{h_0}(\sum_{\rho=1}^n
u_{0\rho}\xi_{\tau \rho})$.
\end{theorem}
\proof We will follow the notations introduced in the proof of Lemma
\ref{lm-ftau}. Since $\SR$ is irreducible, we have $\SR_{\tau
0}=\frac{\partial \SR}{\partial u_{00}^{(h_0)}}\Big|
_{u_{00}^{(h_0)}=\gamma_{\tau}} \neq 0$.
Let $\xi_{\tau\rho}=\SR_{\tau
\rho}\big/\SR_{\tau0}\,(\rho=1,\ldots,l_0)$, where $\SR_{\tau
\rho}=\frac{\partial \SR}{\partial u_{0 \rho}^{(h_0)}}\Big|
_{u_{00}^{(h_0)}=\gamma_{\tau}}$. Note that $\SR_{\tau \rho}$ and
$\xi_{\tau\rho}$ are in $\ff_\tau$.
We will prove \[\gamma_{\tau}=-\delta_\tau^{h_0}(u_{01}\xi_{\tau
1}+u_{02}\xi_{\tau2}+\cdots+u_{0l_0}\xi_{\tau l_0}).\]
Differentiating the equality
$\SR(\bu;\zeta_{0},\zeta_{1},\ldots,\zeta_{n})=0$ w.r.t.
$u_{0\rho}^{(h_0)}$, we have
\begin{equation}\label{eq-partial0}\overline{\frac{\partial \SR}{\partial
u_{0\rho}^{(h_0)}}}+
 \overline{\frac{\partial \SR}{\partial
 u_{00}^{(h_0)}}}(-\frac{M_{0\rho}(\eta)}{M_{00}(\eta)})=0,\end{equation}
%\frac{\partial f}{\partial \zeta_{0}^{(h)}}(-\xi_{\rho})=0,\] where
where $\overline{\frac{\partial \SR}{\partial u_{0\rho}^{(h_0)}}}$
are obtained by substituting $\zeta_{i}$ to $u_{i0}\,(i=0, 1,
\ldots, n)$ in $\frac{\partial \SR}{\partial u_{ 0\rho}^{(h_0)}}$.
Multiplying $u_{0\rho}$ to the above equation and for $\rho$ from 1
to $l_0$, adding them together, we have
\[\sum_{\rho=1}^{l_0} u_{0\rho}\overline{\frac{\partial \SR}{\partial u_{0\rho}^{(h_0)}}}+
 \overline{\frac{\partial \SR}{\partial u_{00}^{(h_0)}}}\bigg(-\sum_{\rho=1}^{l_0} u_{0\rho}\frac{M_{0\rho}(\eta)}{M_{00}(\eta)}\bigg)
 =\sum_{\rho=1}^{l_0} u_{0\rho}\overline{\frac{\partial \SR}{\partial u_{0\rho}^{(h_0)}}}
  +\zeta_0 \overline{\frac{\partial \SR}{\partial u_{00}^{(h_0)}}}=0.\]
Thus, $q=\sum_{\rho=1}^{l_0} u_{0\rho}\frac{\partial \SR}{\partial
u_{0\rho}^{(h_0)}}+u_{00}\frac{\partial \SR}{\partial
u_{00}^{(h_0)}} \in \,\sat(\SR)$. Since $q$ is of order not greater
than $\SR$, it must be divisible by $\SR$. Since $q$ and $\SR$ have
the same degree, there exists an $a\in\qq$ such that $q=a\SR$.
Setting $u_{00}^{(h_0)}=\gamma_{\tau}$ in both sides of $q=a\SR$, we
have $\sum_{\rho=1}^{l_0} u_{0\rho}\SR_{\tau \rho}+u_{00}\SR_{\tau
0}=0$. Hence, as an algebraic equation, we have
\begin{equation}\label{eq-ff2}
%^aP_0(\xi_{\tau 1},\ldots,\xi_{\tau n}) =
u_{00}+\sum_{\rho=1}^{l_0} u_{0\rho}\xi_{\tau \rho}=0
\end{equation}
under the constraint $u_{00}^{(h_0)}=\gamma_\tau$. Equivalently, the
above equation is valid in $(\ff_\tau,\delta_\tau)$.
As a consequence,
$\gamma_{\tau}=-\delta_\tau^{h_0}(\sum_{\rho=1}^{l_0}
u_{0\rho}\xi_{\tau \rho})$. Substituting them into
equation~\bref{eq-fac00}, the theorem is proved.\qedd

Note that the quantities $\xi_{\tau \rho}$ are not expressions in
terms of $y_i$. In the following theorem, we will show that if
$\{\mathcal {A}_i\,(i=0,\ldots,n)\}$ satisfies certain conditions,
Theorem \ref{th-fac1} can be strengthened to make $\xi_{\tau \rho}$
as productions of certain values of $y_i$ and its derivatives.
Following the notations introduced before Lemma \ref{th-ifunique},
we have
\begin{theorem} \label{th-p1}
Assume that 1) any $n$  of the $\mathcal {A}_i\,(i=0,\ldots,n)$ form
a differentially independent set and 2) for each $j=1,\ldots,n$,
$\textbf{e}_j\in\Span_{\mathbb{Z}}\{\alpha_{ik}-\alpha_{i0}:k=1,\ldots,l_i;i=0,\ldots,n\}$.
Then there exist $\eta_{\tau k}$ \,$(\tau=1,\ldots,t_0;\,k=1,\ldots,n)$
such that
\begin{align} \label{eq-fac2}
\SR & = A\prod_{\tau=1}^{t_0} \bigg(u_{00}+\sum_{k=1}^{l_0}u_{0k}\frac{M_{0k}(\eta_\tau)}{M_{00}(\eta_\tau)}\bigg)^{(h_0)}\\
  & = A\prod_{\tau=1}^{t_0}\bigg[\frac{\P_0(\eta_\tau)}{M_{00}(\eta_\tau)}\bigg]^{(h_0)}, \quad \text{where~$\eta_\tau=(\eta_{\tau1},\ldots,\eta_{\tau n})$.} \nonumber
\end{align}
Moreover, $\eta_\tau\,(\tau=1,\ldots,t_0)$ lies on
$\P_1,\ldots,\P_n$.
\end{theorem}
\proof Follow the notations in this section and those introduced
before Lemma \ref{th-ifunique}.
By condition 1), each $h_i\geq0.$ Denote
$\P_i^N=M_i\P_i\,(i=0,\ldots,n)$ where $M_i$ are Laurent
differential polynomials.
 Then $(\eta;\zeta_0,u_{01},\ldots,u_{0l_0};\ldots;\zeta_n,u_{n1},\ldots,u_{nl_n})$
is a generic point of $[\P_0^N,\ldots,\P_n^N]:\mathbbm{m}$ where
$\eta=(\eta_1,\ldots,\eta_n)$ is a generic point of $[0]$ over
$\qq\langle \bu\rangle$ and
$\zeta_i=-\sum_{k=1}^{l_i}u_{ik}\frac{M_{ik}(\eta)}{M_{i0}(\eta)}$.
By the proof of Lemma~\ref{th-ifunique}, there exist  $S_j$ and
$T_j$ which are products of nonnegative powers of $\frac{\partial
\SR}{\partial u_{ik}^{(h_i)}}$ such that
$S_jy_j-T_j\in[\P_0^N,\ldots,\P_n^N]:\mathbbm{m}$. That is,
$\eta_j=\overline{T_j}/\overline{S_j}$ for $j=1,\ldots,n$, where
$\overline{T_j}, \overline{S_j}$ are obtained by substituting
$(u_{00},\ldots,u_{n0})=(\zeta_0,\ldots,\zeta_n)$ in $T_j, S_j$
respectively. Since $\SR$ is an irreducible polynomial, every
$\frac{\partial \SR}{\partial
 u_{ik}^{(h_i)}}$ does not vanishes at $u_{00}^{(h_0)}=\gamma_\tau$.
Let $\eta_{\tau
j}=\frac{T_j}{S_j}\big|_{u_{00}^{(h_0)}=\gamma_\tau}$ and
$\eta_\tau=(\eta_{\tau 1},\ldots,\eta_{\tau n})$. By
(\ref{eq-partial0}), $\frac{M_{0k}(\eta)}{M_{00}(\eta)}=
\prod\limits_{j=1}^n\prod\limits_{k=0}^{s_0}(\eta_j^{(k)})^{(\alpha_{0k}-\alpha_{00})_{jk}}
=\overline{\frac{\partial \SR}{\partial
u_{0k}^{(h_0)}}}\Big/\overline{\frac{\partial \SR}{\partial
 u_{00}^{(h_0)}}}$. So $\prod\limits_{j=1}^n\prod\limits_{k=0}^{s_0}\Big[\Big(\frac{\overline{T_j}}{\overline{S_j}}\Big)^{(k)}\Big]^{(\alpha_{0k}-\alpha_{00})_{jk}}
 =\overline{\frac{\partial \SR}{\partial
u_{0k}^{(h_0)}}}\Big/\overline{\frac{\partial \SR}{\partial
 u_{00}^{(h_0)}}}$. Let $\mathcal{S}$ be the differential polynomial set consisting of
$\frac{\partial \SR}{\partial u_{ik}^{(h_i)}}$ and
$(S_j)^{m+1}\big(\frac{T_j}{S_j}\big)^{(m)}$ for all
$i=0,\ldots,n;k=0,\ldots,l_i;j=1,\ldots,n$ and $m\in\mathbb{N}$.
By Corollary \ref{cor-r1}, there exists a finite set $\mathcal{S}_1$
of $\mathcal{S}$ and $a\in\mathbb{N}$ such that
$H=\big(\prod\limits_{S\in\mathcal{S}_1}S\big)^a
\Big(\prod\limits_{j=1}^n\prod\limits_{k=0}^{s_0}\big[\big(T_j/S_j\big)^{(k)}\big]^{(\alpha_{0k}-\alpha_{00})_{jk}}
 -\frac{\partial \SR}{\partial
u_{0k}^{(h_0)}}\Big/\frac{\partial \SR}{\partial
 u_{00}^{(h_0)}}\Big)\in \sat(\SR)$. By Lemma~\ref{lm-ftaup}, $H$ vanishes at $u_{00}^{(h_0)}=\gamma_\tau.$
And by the proof of Theorem~\ref{th-idealrelation},
 $\mathcal{S}\cap\sat(\SR)=\emptyset$. So $\xi_{\tau k}=\frac{M_{0k}(\eta_\tau)}{M_{00}(\eta_\tau)}$.
 By Theorem~\ref{th-fac1}, $\SR=A\prod_{\tau=1}^{t_0} (u_{00}+\sum\limits_{k=1}^{l_0} u_{0
k}\xi_{\tau k})^{(h_0)}$. Thus, (\ref{eq-fac2}) follows.

To prove the second part of this theorem, we need first to show that
$\delta_\tau^{k}\eta_{\tau j}\neq0$ for each $k\geq0.$ Suppose the
contrary, that is, there exists some $k$ such that
$\delta_\tau^{k}\eta_{\tau j}=0$. From $\eta_{\tau
j}=\frac{T_j}{S_j}\big|_{u_{00}^{(h_0)}=\gamma_\tau}$,
$\delta_\tau^{k}\eta_{\tau
j}=\big(\frac{T_j}{S_j}\big)^{(k)}\big|_{u_{00}^{(h_0)}=\gamma_\tau}=0$.
Thus, $S_j^{k+1}\big(\frac{T_j}{S_j}\big)^{(k)}\in\sat(\SR)$. It
follows that
$\eta_j^{(k)}=\big(\frac{\overline{T_j}}{\overline{S_j}}\big)^{(k)}=0$,
a contradiction to the fact that $\eta_j$ is a differential
indeterminate.

Follow the above procedure, we can show that
$\frac{M_{ik}(\eta_\tau)}{M_{i0}(\eta_\tau)}=\widehat{\frac{\partial
\SR}{\partial u_{ik}^{(h_i)}}}\big/\widehat{\frac{\partial
\SR}{\partial
 u_{i0}^{(h_i)}}}$ where $\widehat{\frac{\partial \SR}{\partial
u_{ik}^{(h_i)}}}=\frac{\partial \SR}{\partial
u_{ik}^{(h_i)}}\Big|_{u_{00}^{(h_0)}=\gamma_\tau}$. From
(\ref{eq-partial}), it is easy to show that
$\sum_{k=0}^{l_i}u_{ik}\frac{\partial \SR}{\partial
u_{ik}^{(h_i)}}=b\SR$ for some $b$ in $\Q$. So, for each $i\neq 0$,
$\sum_{k=0}^{l_i}u_{ik}\widehat{\frac{\partial \SR}{\partial
u_{ik}^{(h_i)}}}=0.$ It follows that for each $i\neq0$,
$\P_i(\eta_\tau)=\sum_{k=0}^{l_i}u_{ik}M_{ik}(\eta_\tau)
=\frac{M_{i0}(\eta_\tau)}{\widehat{\frac{\partial \SR}{\partial
 u_{i0}^{(h_i)}}}}\bigg(\sum_{k=0}^{l_i}u_{ik}\widehat{\frac{\partial \SR}{\partial
u_{ik}^{(h_i)}}}\bigg)=0$. So $\eta_\tau$ lies on $\P_1,\ldots,\P_n$.
\qedd

%-gao
% Here, we may further extend the results like the Chow form
% Possible related the degree deg(R,ui) with mixed volume
Under the conditions of Theorem \ref{th-p1},  we further have the
following theorem.
\begin{theorem} \label{th-p2}
The elements  $\eta_\tau\,(\tau=1,\ldots,t_0)$ defined in Theorem
\ref{th-p1} are generic points of the prime ideal
$[\P_1^N,\ldots,\P_n^N]:\mathbbm{m}\subset
\qq\langle\hat{\bu}\rangle\{\Y\}$, where
$\hat{\bu}=\cup_{i=1}^n\bu_i$.
\end{theorem}
\proof Follow the notations in this section.  Let
$\J_0=[\P_1^N,\ldots,\P_n^N]:\mathbbm{m}\subset\qq\{\Y,\hat{\bu}\}$
and
$\J=[\P_1^N,\ldots,\P_n^N]:\mathbbm{m}\subset\qq\langle\hat{\bu}\rangle\{\Y\}$.
Similar to the proof of Theorem~\ref{th-Mcodim1}, it is easy to show
that $\J_0$ is a prime differential ideal. And by condition 1),
 $\J_0\cap\qq\{\hat{\bu}\}=[0]$. Thus, $\J=[\J_0]$ is a prime
 differential ideal and $\J\cap\qq\{\Y,\hat{\bu}\}=\J_0$. Let $\xi=(\xi_1,\ldots,\xi_n)$ be a generic
 point of $\J$. Then $(\xi;\hat{\bu})$ is a generic point of
$\J_0$. Let $\beta=-\sum_{k=1}^{l_0}u_{0k}M_{0k}(\xi)/M_{00}(\xi)$.
Then $(\xi;\beta,u_{01},\ldots,u_{0l_0};\hat{\bu})$ is a generic
point of
$\I=[\P_0^N,\P_1^N,\ldots,\P_n^N]:\mathbbm{m}\subset\qq\{\Y;\bu_0,\hat{\bu}\}$.
Since $S_jy_j-T_j\in\I\,(j=1,\ldots,n)$,
$\xi_j=\frac{T_j}{S_j}(\beta,u_{01},\ldots,u_{0l_0};\hat{\bu})$.

By Theorem~\ref{th-p1}, $\eta_\tau$ is a common non-polynomial
solution of $\P_i^N=0\,(i=1,\ldots,n)$, thus also a differential
zero of $\J$. Recall $\eta_{\tau
j}=\frac{T_j}{S_j}\big|_{u_{00}^{(h_0)}=\gamma_\tau}$. If $f$ is any
differential polynomial in $\qq\langle\hat{\bu}\rangle\{\Y\}$ such
that $f(\eta_\tau)=0$, then
$f(\frac{T_1}{S_1},\ldots,\frac{T_n}{S_n})\big|_{u_{00}^{(h_0)}=\gamma_\tau}=0$.
There exist $a_j\in\mathbb{N}$ such that
$g=\prod_jS_j^{a_j}f(\frac{T_1}{S_1},\ldots,\frac{T_n}{S_n})\in\qq\{\Y;\bu_0,\hat{\bu}\}.$
Then $g|_{u_{00}^{(h_0)}=\gamma_\tau}=0$. By Lemma~\ref{lm-ftaup},
$g\in\sat(\SR)$ while $S_j\not\in\sat(\SR)$. As a consequence,
$g(\beta,u_{01},\ldots,u_{0l_0};\hat{\bu})=0$ and
$S_j(\beta,u_{01},\ldots,u_{0l_0};\hat{\bu})\neq0$. It follows that
$f(\xi_1,\ldots,\xi_n)=0$ and $f\in\J$. Thus, $\eta_\tau$ is a
generic point of $\J$. \qedd

With Theorems \ref{th-fac1}, \ref{th-p1}, and \ref{th-p2}, property
4) of Theorem \ref{th-i1} is proved.

\subsection{Structures of non-polynomial solutions}\label{sec-bp2}

In this section, we will analyze the structures of the
non-polynomial solutions. Firstly, we will give the following
theorem which shows the relation between the original differential
system and their sparse differential resultant.
%
%
%-gao
%Here we will need results of CM YUAN
%
%

Let  $\mathcal {A}_i\,(i=0,\ldots,n)$ be  a  Laurent differentially
essential system of monomial sets. Then by
Theorem~\ref{th-rankessential}, $\mathcal {A}_i\,(i=0,\ldots,n)$ can
be divided into two disjoint sets $\{\mathcal {A}_i:\,i\in\TT\}$ and
$\{\mathcal {A}_i:\,i\in\{0,1,\ldots,n\}\backslash\TT\}$, where
$\TT\subseteq\{0,1,\ldots,n\}$ is rank essential. In this section,
we will assume that $\{0,1,\ldots,n\}$ is rank essential, that is,
any $n$ of the $\mathcal {A}_i\,(i=0,\ldots,n)$ form a
differentially independent set, which is equivalent to the fact that
each $\bu_i$ occurs in $\SR$ effectively.

\begin{theorem}\label{th-idealrelation}
Let $\P_i=\sum_{k=0}^{l_i}u_{ik}M_{ik}\,(i=0,\ldots,n)$ be a  system
of differential polynomials such that any $n$ of the $\P_i$ form a
differentially independent set. Let $\SR(\bu_0,\ldots,\bu_n)$ be the
sparse differential resultant of $\P_i$ with $\ord(\SR,\bu_i)=h_i$.
Denote $Q_{ik}=\frac{\partial \SR}{\partial
u_{i0}^{(h_i)}}M_{ik}-\frac{\partial \SR}{\partial
u_{ik}^{(h_i)}}M_{i0}$ and $S$ to be the set consisting of
$\frac{\partial \SR}{\partial u_{i0}^{(h_i)}}\,(i=0,\ldots,n)$ and
$(y_i^{(k)})_{1\leq i\leq n;k\geq0}$.  Then
$$[\P_0,\ldots,\P_n]:\mathbbm{m}=[\SR,(Q_{ik})_{0\leq i\leq n;1\leq
k\leq l_i}]:S^\infty$$ in $\qq\{\Y,\bu_0,\ldots,\bu_n\}$.
\end{theorem}
\proof Let $\mathcal{I}=[\P_0,\ldots,\P_n]:\mathbbm{m}$. Following
the notations in the proof of Theorem~\ref{th-Mcodim1},
$\mathcal{I}$ is a prime differential ideal with a generic point
$(\eta;\zeta_0,u_{01},\ldots,u_{0l_0};\ldots;\zeta_n,u_{n1},\ldots,u_{nl_n})$
where
$\zeta_i=-\sum_{k=1}^{l_i}u_{ik}\frac{M_{ik}(\eta)}{M_{i0}(\eta)}$.
By the definition of sparse differential resultant,
$\mathcal{I}\cap\qq\{\bu_0,\ldots,\bu_n\}=\sat(\SR)$.
Differentiating the equality
$\SR(\bu;\zeta_{0},\zeta_{1},\ldots,\zeta_{n})=0$ w.r.t.
$u_{ik}^{(h_i)}$, we have \begin{equation}\label{eq-partial}
\overline{\frac{\partial \SR}{\partial u_{ik}^{(h_i)}}}+
 \overline{\frac{\partial \SR}{\partial
 u_{i0}^{(h_i)}}}(-\frac{M_{ik}(\eta)}{M_{i0}(\eta)})=0
 \end{equation}
%\frac{\partial f}{\partial \zeta_{0}^{(h)}}(-\xi_{\rho})=0,\] where
where $\overline{\frac{\partial \SR}{\partial u_{ik}^{(h_i)}}}$ are
obtained by substituting $\zeta_{i}$ to $u_{i0}\,(i=0, 1, \ldots,
n)$ in $\frac{\partial \SR}{\partial u_{ ik}^{(h_i)}}$. Let
$Q_{ik}=\frac{\partial \SR}{\partial
u_{i0}^{(h_i)}}M_{ik}-\frac{\partial \SR}{\partial
u_{ik}^{(h_i)}}M_{i0}$. Clearly, $Q_{ik}\in\mathcal{I}$.

Since any $n$ of the $\P_i$ form a differentially independent set,
$\ord(\SR,\bu_i)\geq0$.
Substituting $M_{ik}$ by $\big(Q_{ik}+M_{i0}\frac{\partial
\SR}{\partial u_{ik}^{(h_i)}}\big)/\frac{\partial \SR}{\partial
u_{i0}^{(h_i)}}$ in each $\P_i$, we have
$\P_i=\sum_{k=0}^{l_i}u_{ik}M_{ik}=u_{i0}M_{i0}+\sum_{k=1}^{l_i}u_{ik}\big(Q_{ik}+M_{i0}\frac{\partial
\SR}{\partial u_{ik}^{(h_i)}}\big)/\frac{\partial \SR}{\partial
u_{i0}^{(h_i)}}$. So $\frac{\partial \SR}{\partial
u_{i0}^{(h_i)}}\P_i=\sum_{k=1}^{l_i}
u_{ik}Q_{ik}+(\sum_{k=0}^{l_i}u_{ik}\frac{\partial \SR}{\partial
u_{ik}^{(h_i)}})M_{i0}$. Since $Q_{ik}\in\mathcal{I}$,
$\sum_{k=0}^{l_i}u_{ik}\frac{\partial \SR}{\partial
u_{ik}^{(h_i)}}\in\mathcal{I}$. Thus, there exists some $a\in\qq$
such that $\sum_{k=0}^{l_i}u_{ik}\frac{\partial \SR}{\partial
u_{ik}^{(h_i)}}=a\SR$. It follows that $\P_i\in[\SR,(Q_{ik})_{0\leq
i\leq n;1\leq k\leq l_i}]:S^\infty$. For any differential polynomial
$f\in\mathcal{I}$, there exists a differential monomial
$M\in\mathbbm{m}$ such that
$Mf\in[\P_0,\ldots,\P_n]\subset([\SR,(Q_{ik})_{0\leq i\leq n;1\leq
k\leq l_i}]:S^\infty)$. Thus, $f\in [\SR,(Q_{ik})_{0\leq i\leq
n;1\leq k\leq l_i}]:S^\infty$ and
$\mathcal{I}\subseteq[\SR,(Q_{ik})_{0\leq i\leq n;1\leq k\leq
l_i}]:S^\infty$ follows. Conversely, for any differential polynomial
$g\in[\SR,(Q_{ik})_{0\leq i\leq n;1\leq k\leq l_i}]:S^\infty$, there
exist some differential monomial $M$ and some $b\in\mathbb{N}$ such
that
 $M(\prod_{i}\frac{\partial \SR}{\partial
u_{i0}^{(h_i)}})^bg\in[\SR,Q_{ik}]\subset\mathcal{I}$. Since
$\mathcal{I}$ is a prime differential ideal, $g\in\I$. Hence,
$\I=[\SR,(Q_{ik})_{0\leq i\leq n;1\leq k\leq l_i}]:S^\infty.$ \qedd

%-gao
% Again, here can we introduce the concept of toric
%

Theorem \ref{th-idealrelation} shows that under the condition
$\SR=0$, the non-polynomial solutions of $\P_i=0 (i=0,\ldots,n)$ are
the solutions of some differential polynomials of two terms.

\begin{cor} \label{th-generalized-sf}
Let $\P_0,\ldots,\P_n$ be a Laurent differentially essential system
of the form \bref{eq-sparseLaurent} and $\SR(\bu_{0},$
$\ldots,\bu_{n})$ its sparse differential resultant. Suppose
$\ord(\SR,\bu_0)=h_0\geq0$ and denote $S_\SR=\frac{\partial
\SR}{\partial u_{00}^{(h_0)}}$. Suppose that when
$\bu_i\,(i=0,\ldots,n)$ are specialized to sets $\bv_i$ which are
elements in an extension field of $\mathcal{F}$, $\P_i$ are
specialized to $\overline{\P}_{i}\,(i=0,\ldots,n)$. Suppose
$S_\SR(\bv_0,\ldots,\bv_n)\neq 0$. If
$\overline{\P}_i=0(i=0,\ldots,n)$ have a common non-polynomial
differential solution $\xi$, then for each  $k$, we have
\begin{equation} \label{eq-root}
\frac{M_{0k}(\xi)}{M_{00}(\xi)}=\frac{\partial \SR}{\partial
u_{0k}^{(h_0)}}(\bv_0,\ldots,\bv_n)\big/S_\SR(\bv_0,\ldots,\bv_n).
\end{equation}
\end{cor}
\proof Denote $\P_i^N=M_i\P_i\,(i=0,\ldots,n)$ where  $M_i$ are
Laurent differential monomials.. By the proof of
Theorem~\ref{th-idealrelation}, for each $k=1,\ldots,l_0$, the
polynomial $S_\SR M_0M_{0k}- \frac{\partial \SR}{\partial u_{0k
}^{(h_0)}}M_0M_{00}\in[\P_0^N,\ldots,\P_n^N]:\mathbbm{m}$. Thus, if
$\xi$ is a common non-polynomial differential solution of
$\overline{\P_i}=0$, then $S_\SR(\bv_0,\ldots,\bv_n)\cdot
M_{0k}(\xi)-\frac{\partial \SR}{\partial u_{0k
}^{(h_0)}}(\bv_0,\ldots,\bv_n)M_{00}(\xi)=0$. Since
$S_\SR(\bv_0,\ldots,\bv_n)\neq 0$, \eqref{eq-root} follows. \qedd

\vskip 5pt We conclude this section by giving a sufficient condition
for a differentially essential system to have a unique
non-polynomial solution.

Follow the notations in section \ref{sec-defresultant}, that is,
 $\mathcal
{A}_i=\{M_{i0},M_{i1},\ldots,M_{il_i}\}$ are finite sets of Laurent
differential monomials where $M_{ik}= (\Y^{[s_i]})^{\alpha_{ik}}$,
and $\P_i=\sum\limits_{k=0}^{l_i}u_{ik} M_{ik}\,(i=0,\ldots,n)$.
$\alpha_{ik}\in\mathbb{Z}^{n(s_i+1)}$ is an exponent vector written
in terms of the degrees of $y_1,\ldots,y_n,y'_1,\ldots,y'_n,$
$\ldots,y_1^{(s_i)},\ldots,y_{n}^{(s_i)}$. Let $o=\max_i\{s_i\}$. Of
course, every vector in $\mathbb{Z}^{n(s_i+1)}$ can be embedded in
$\mathbb{Z}^{n(o+1)}$. Let $\textbf{e}_i$ be the exponent vector for
$y_i$ in $\mathbb{Z}^{n(o+1)}$ whose $i$-th coordinate is 1 and
other coordinates are equal to zero.

\begin{lemma}\label{th-ifunique}
Assume that 1) any $n$  of the $\mathcal {A}_i\,(i=0,\ldots,n)$ form
a differentially independent set and 2) for each $j=1,\ldots,n$,
$\textbf{e}_j\in\Span_{\mathbb{Z}}\{\alpha_{ik}-\alpha_{i0}:k=1,\ldots,l_i;i=0,\ldots,n\}$.
Denote $\ord(\SR,\bu_i)=h_i$ where
$\SR(\bu_0,\ldots,\bu_n)=\Res_{\mathcal {A}_0,\ldots,\mathcal
{A}_n}$. Let $\overline{\P}_{i}$ be a specialization of $\P_i$ with
coefficient vector $\bv_i\,(i=0,\ldots,n)$.  If
$\SR(\bv_0,\ldots,\bv_n)=0$ and $\frac{\partial \SR}{\partial
u_{ik}^{(h_i)}}(\bv_0,\ldots,\bv_n)\neq0$ for each $i$ and $k$, then
$\overline{\P}_{i}=0\,(i=0,\ldots,n)$ have at most one common
non-polynomial solution.
\end{lemma}
\proof By hypothesis 1), each $h_i\geq0$. Denote $\P_i^N=M_i\P_i$
where  $M_i$ are Laurent differential monomials.
 Similar to procedures
to derive equation~(\ref{eq-partial}), we have
$\frac{M_{ik}(\eta)}{M_{i0}(\eta)}=\overline{\frac{\partial
\SR}{\partial u_{ik}^{(h_i)}}}\big/\overline{\frac{\partial
\SR}{\partial u_{i0}^{(h_i)}}}$, where $\overline{\frac{\partial
\SR}{\partial u_{ik}^{(h_i)}}}$ are obtained by substituting
$\zeta_{i}$ to $u_{i0}\,(i=0, 1, \ldots, n)$ in $\frac{\partial
\SR}{\partial u_{ik}^{(h_i)}}$. By hypothesis 2), there exist
$t_{jik}\in\mathbb{Z}$ such that
$\sum_{i,k}t_{jik}(\alpha_{ik}-\alpha_{i0})=\textbf{e}_j$ for
$j=1,\ldots,n$. So
$\prod_{i,k}\Big(\frac{M_{ik}}{M_{i0}}\Big)^{t_{jik}}=y_j$. Thus,
$\prod_{i,k}\Big(\frac{M_{ik}(\eta)}{M_{i0}(\eta)}\Big)^{t_{jik}}=\eta_j=\prod_{i,k}\big(\overline{\frac{\partial
\SR}{\partial u_{ik}^{(h_i)}}}\big/\overline{\frac{\partial
\SR}{\partial u_{i0}^{(h_i)}}}\big)^{t_{jik}}$. It follows that
$S_jy_j-T_j\in[\P_0^N,\ldots,\P_n^N]:\mathbbm{m}$ where $S_j$ and
$T_j$ are products of nonnegative powers of $\frac{\partial
\SR}{\partial u_{ik}^{(h_i)}}$ obtained from the above identity.
 Since
$\frac{\partial \SR}{\partial
u_{ik}^{(h_i)}}(\bv_0,\ldots,\bv_n)\neq0$ for each $i$ and $k$,
$T_j(\bv_0,\ldots,\bv_n)\cdot S_j(\bv_0,\ldots,\bv_n)\neq0$.  Let
$\bar{y}_j=T_j(\bv_0,\ldots,\bv_n)/S_j(\bv_0,\ldots,\bv_n)$. If
$\xi=(\xi_1,\ldots,\xi_n)$ is any  common non-polynomial solution of
$\overline{\P}_{i}=0\,(i=0,\ldots,n)$, then
$\xi_j=T_j(\bv_0,\ldots,\bv_n)/S_j(\bv_0,\ldots,\bv_n)=\bar{y}_j$.
Thus $\xi=(\bar{y}_1,\ldots,\bar{y}_n)$ and
$\overline{\P}_{i}=0\,(i=0,\ldots,n)$ have at most one common
non-polynomial solution.
  \qedd

\begin{theorem} \label{th-unique}
Assume that 1) any $n$  of the $\mathcal {A}_i\,(i=0,\ldots,n)$ form
a differentially independent set and 2) for each $j=1,\ldots,n$,
$\textbf{e}_j\in\Span_{\mathbb{Z}}\{\alpha_{ik}-\alpha_{i0}:k=1,\ldots,l_i;i=0,\ldots,n\}$.
Denote $\ord(\SR,\bu_i)=h_i$ where
$\SR(\bu_0,\ldots,\bu_n)=\Res_{\mathcal {A}_0,\ldots,\mathcal
{A}_n}$. Let $\overline{\P}_{i}$ be a specialization of $\P_i$ with
coefficient vector $\bv_i\,(i=0,\ldots,n)$.  Then there exists a
differential polynomial set $\mathcal {S}\subset
\Q\{\bu_0,\ldots,\bu_n\}$ such that
$\V(\SR)\backslash\bigcup\limits_{S\in\mathcal{S}}\V(S)\neq
\emptyset$ and whenever
$(\bv_0,\ldots,\bv_n)\in\V(\SR)\backslash\bigcup\limits_{S\in\mathcal{S}}\V(S)$,
$\overline{\P}_{i}=0\,(i=0,\ldots,n)$ have a unique common
non-polynomial solution.
\end{theorem}
\proof Follow the notations in the proof of Lemma~\ref{th-ifunique}.
Denote $\P_i^N=M_i\P_i\,(i=0,\ldots,n)$ where $M_i$ are Laurent
differential monomials.
 Then by the proof of Lemma~\ref{th-ifunique},
there exist  $S_j$ and $T_j$ which are products of nonnegative
powers of $\frac{\partial \SR}{\partial u_{ik}^{(h_i)}}$ such that
$S_jy_j-T_j\in[\P_0^N,\ldots,\P_n^N]:\mathbbm{m}$. Thus,
$\SR,S_1y_1-T_1,\ldots,S_ny_n-T_n$ is a characteristic set of
$[\P_0^N,\ldots,\P_n^N]:\mathbbm{m}$ w.r.t. any elimination ranking
$u_{ik}\prec y_1 \prec \cdots \prec y_n$.

Let $\mathcal{S}$ be the differential polynomial set consisting of
$\frac{\partial \SR}{\partial u_{ik}^{(h_i)}}$ and
$(S_j)^{m+1}\big(\frac{T_j}{S_j}\big)^{(m)}$ for all
$i=0,\ldots,n;k=0,\ldots,l_i;j=1,\ldots,n$ and $m\in\mathbb{N}$.
Firstly, we show that
$\V(\SR)/\bigcup\limits_{S\in\mathcal{S}}\V(S)\neq\emptyset$.
Suppose the contrary, viz.
$\V(\SR)\subset\bigcup\limits_{S\in\mathcal{S}}\V(S)$. In
particular, there exists one $S\in\mathcal{S}$ such that $S$
vanishes at the generic point $\zeta$ of $\sat(\SR)$. It is obvious
that $\frac{\partial \SR}{\partial u_{ik}^{(h_i)}}$ does not vanish
at $\zeta$. If $(S_j)^{m+1}\big(\frac{T_j}{S_j}\big)^{(m)}$ vanishes
at $\zeta$ for some $m$,
$(S_j)^{m+1}\big(\frac{T_j}{S_j}\big)^{(m)}\in\sat(\SR)$. Since
$S_j^{m+1}y_j^{(m)}-(S_j)^{m+1}\big(\frac{T_j}{S_j}\big)^{(m)}\in[\P_0^N,\ldots,\P_n^N]:\mathbbm{m}$,
it follows that $S_j^{m+1}\in[\P_0^N,\ldots,\P_n^N]:\mathbbm{m}$, a
contradiction.

 Suppose
 $(\bv_0,\ldots,\bv_n)\in\V(\SR)/\bigcup\limits_{S\in\mathcal{S}}\V(S)$.
 Since $\frac{\partial \SR}{\partial
u_{ik}^{(h_i)}}(\bv_0,\ldots,\bv_n)\neq0$ for each $i$ and $k$,
$T_j(\bv_0,\ldots,\bv_n)\cdot S_j(\bv_0,\ldots,\bv_n)\neq0$. Let
$\bar{y}_j=\frac{T_j}{S_j}(\bv_0,\ldots,\bv_n)$ and denote
$\bar{y}=(\bar{y}_1,\ldots,\bar{y}_n)$. And for each
$m\in\mathbb{N},$
$\bar{y}_j^{(m)}=(\frac{T_j}{S_j})^{(m)}(\bv_0,\ldots,\bv_n)\neq0$.
Thus, $\bar{y}\in(\ee^\wedge )^n.$ Since
$\SR,S_1y_1-T_1,\ldots,S_ny_n-T_n$ is a characteristic set of
$[\P_0^N,\ldots,\P_n^N]:\mathbbm{m}$, $H\cdot M_i\P_i\equiv 0,
\mod\,[\SR,S_1y_1-T_1,\ldots,S_ny_n-T_n]$ where $H$ is a product of
powers of $\frac{\partial \SR}{\partial u_{ik}^{(h_i)}}$. Hence,
$M_i(\bar{y})\cdot\overline{\P}_{i}(\bar{y})=0$, which follows that
$\overline{\P}_{i}(\bar{y})=0$. Thus, $\bar{y}$ is a non-polynomial
common solution of $\overline{\P}_{i}$. On the other hand, for every
non-polynomial common solution of $\overline{\P}_{i}$,
$S_j(\bv_0,\ldots,\bv_n)y_j-T_j(\bv_0,\ldots,\bv_n)$ vanishes at it.
Thus, it must be equal to the point $\bar{y}$. As a consequence, we
have proved that $\overline{\P}_{i}=0$ have a unique common
non-polynomial solution.
  \qedd

Theorem \ref{th-unique} can be rephrased as the following geometric
form.
\begin{cor}
Let $\mathcal {Z}_1(\mathcal{A}_0,\ldots,\mathcal{A}_n)$ be a subset
of $\ee^{l_0+1}\times\cdots\times\ee^{l_n+1}$ consisting of points
$(\bv_0,\ldots,\bv_n)$ for which the corresponding Laurent
differential polynomials $F_i=0\,(i=0,\ldots,n)$ have a unique
non-polynomial common solution and
$\overline{\mathcal{Z}_1(\mathcal{A}_0,\ldots,\mathcal{A}_n)}$ the
Kolchin closure of
$\mathcal{Z}_1(\mathcal{A}_0,\ldots,\mathcal{A}_n)$.
Then under the condition of Theorem \ref{th-unique}, we have
$\overline{\mathcal{Z}_1(\mathcal{A}_0,\ldots,\mathcal{A}_n)}=\V\big(\sat(\Res_{\mathcal{A}_0,\ldots,\mathcal{A}_n})\big)$.
\end{cor}

%\begin{remark}
%By the above theorem,
%$\overline{\V(\SR)/\bigcup\limits_{S\in\mathcal{S}}\V(S)}=\V(\sat(\SR))$
%follows. Indeed, by Theorem~\ref{th-generalized-sf} and
%Theorem~\ref{th-unique}, $\emptyset\neq
%\V(\SR)/\bigcup\limits_{S\in\mathcal{S}}\V(S)\subseteq\V(\sat(\SR))$, so
%$\overline{\V(\SR)/\bigcup\limits_{S\in\mathcal{S}}\V(S)}\subseteq\V(\sat(\SR))$.
%For a generic point $\zeta$ of $\V(\sat(\SR))$, by the proof of
%Theorem~\ref{th-unique}, we obtain that
%$\zeta\in\V(\SR)/\bigcup\limits_{S\in\mathcal{S}}\V(S)$. Thus,
%$\V(\sat(\SR))\subseteq\overline{\V(\SR)/\bigcup\limits_{S\in\mathcal{S}}\V(S)}$
%and the result follows.
%\end{remark}

\begin{example}
Continue from Example~\ref{ex-2}. In this example, the sparse
differential resultant $\SR$ of $\P_0,\P_1,\P_2$ is free from the
coefficients of $\P_2.$ The system can be solved as follows: $y_1$
can be solved from $\P_0=\P_1=0$ and $\P_2=u_{10}+u_{11}y'_2$ is of
order one in $y_2$ which will lead to an infinite number of
solutions. Thus the system can not have a unique solution
This shows the importance of the first condition in
Theorem~\ref{th-unique}.
\end{example}

\begin{example}
Continue from Example~\ref{ex-3}. In this example, the
characteristic set of $[\P_0,\P_1,\P_2]$ w.r.t. the elimination
ranking $u_{ik}\prec y_2\prec y_1$ is
$\SR,u_{11}u_{00}y'_2-u_{01}u_{10}y_2,u_{01}y_2y_1+u_{00}$. Here
$\mathcal{A}_0,\mathcal{A}_1,\mathcal{A}_2$ do not satisfy condition
2) and the system $\{\P_0,\P_1,\P_2\}$ does not have a unique
solution under the condition $\SR=0$.
\end{example}

\subsection{Sparse differential resultant for differential
polynomials with non-vanishing degree zero terms}

As pointed out in the preceding sections, for Laurent differential
polynomials, non-polynomial zeros are considered. But, for certain
differential polynomials, this condition seems to be too demanding.
In this section, we restrict to consider the sparse differential
resultant for differential polynomials with non-vanishing degree
zero terms.
To be more precise, consider $n+1$ differential polynomials  of the
form
 \begin{equation}\label{eq-qgpol}
 \P_i=u_{i0}+\sum\limits_{k=1}^{l_i}u_{ik} M_{ik}\,(i=0,\ldots,n)\end{equation}
where $M_{ik} = (\Y^{[s_i]})^{\alpha_{ik}}$  is a monomial in
$\{y_1,\ldots,y_n,\ldots,y_1^{(s_i)},$ $\ldots,y_n^{(s_i)}\}$  whose
exponent vector $\alpha_{ik}\in\mathbb{Z}_{\geq0}^{n(s_i+1)}$ with
$|\alpha_{ik}|\geq1$, and all the $u_{ik}$ are differentially
independent over $\Q$. Denote
$\mathcal{B}_i=\{1,M_{i1},\ldots,M_{il_i}\}\,(i=0,\ldots,n)$. The
set of exponent vectors $\SP_i =\{\textbf{0}, \alpha_{ik}:\,
k=1,\ldots,l_i\}$ is called the {\em support} of $\P_i$, where
$\textbf{0}$ is the exponent vector for the constant term. Denote
$\bu_i=(u_{i0},\ldots,u_{il_i})\,(i=0,\ldots,n)$ and
$\bu=\cup_{i}\bu_i\backslash\{u_{i0}\}$.

\begin{definition}
Let $\P_i(i=0,\ldots,n)$ be a differential polynomial system of the
form (\ref{eq-qgpol}).  $\{\P_0,\ldots,\P_n\}$ is called a {\em
differentially essential system} if they form a Laurent
differentially essential system when considered as Laurent
differential polynomials. In this case, we also call
$\mathcal{B}_0,\ldots,\mathcal{B}_n$ a differentially essential
system.
%if there exist $(i_k,j_k)$ with $i_{k_1}\neq
%i_{k_2}(k_1\neq k_2)$ and $j_k\in\{1,\ldots,l_{i_k}\}$ such that
%$M_{i_k,j_k}$ are differentially independent over $\qq$.
\end{definition}

%\subsection{Properties of sparse differential resultant}
All results for sparse differential resultants proved in the
previous sections can be naturally rephrased in this case by just
setting $M_{i0}$ in \bref{eq-sparseLaurent} to $1$.
The main difference is that we do not need to consider
non-polynomial solutions and hence results in section \ref{sec-bp2}
could be modified.

First, we show that Theorem \ref{th-Mcodim1} can be modified as
follows.

\begin{theorem}\label{th-0codim1}
Let $\P_0,\ldots,\P_n$ be differential polynomials as defined in
(\ref{eq-qgpol}). Then $[\P_0,\ldots,$ $\P_n]$ is a prime
differential ideal in $\Q\{\Y,\bu_0,$ $\ldots,\bu_n\}$. And
$([\P_0,\P_1,\ldots,\P_n])\cap\Q\{\bu_0,\ldots,\bu_n\}$ is of
codimension 1 if and only if $\{\P_i, i=0,\ldots,n\}$ is  a
differentially essential system.
\end{theorem}
\proof Let $\eta=(\eta_1,\ldots,\eta_n)$ be a generic point of $[0]$
over $\qq\langle\bu\rangle$. Denote
$\zeta_i=-\sum_{k=1}^{l_i}u_{ik}M_{ik}(\eta)$. It is easy to show
that $(\eta;\zeta)$ is a generic point of
$[\P_0,\P_1,\ldots,\P_n]\subset\Q\{\Y,\bu_0,$ $\ldots,\bu_n\}$ where
$\zeta=(\zeta_0,u_{01},\ldots,$ $u_{0l_0};\ldots;$
$\zeta_n,u_{n1},\ldots,u_{nl_n})$, and it follows that
$[\P_0,\P_1,\ldots,\P_n]$ is a prime differential ideal. So
$[\P_0,\P_1,\ldots,\P_n]:\mathbbm{m}=[\P_0,\P_1,\ldots,\P_n]$. By
Theorem~\ref{th-Mcodim1}, the second part follows.  \qedd

Then, for a differentially essential system $\{\P_0,\ldots,\P_n\}$
of form \bref{eq-qgpol}, its sparse differential resultant $\SR$ can
be defined as
 \begin{equation}\label{eq-sres}
 [\P_0,\ldots,\P_n]\cap\Q\{\bu_{0},\ldots,\bu_{n}\} =
  \sat(\SR).\end{equation}
This equation is different from equation~(\ref{eq-lsres}) in that
the differential ideal  $[\P_0,\ldots,\P_n]$ here
  is a differential ideal in $\Q\{\Y,\bu_0,$
$\ldots,\bu_n\}$ while the other  one is generated in the Laurent
differential polynomial ring.

%
%-gao
% Repeat The main results in Section 4.2
% Here we donot need to consider non-polynomial solutions
Now we first introduce some notations similar to Section
\ref{sec-bp2}. Let $\mathcal{B}_0,\ldots,\mathcal{B}_n$ be a
differentially essential system of monomial sets. For every specific
differential polynomial set $(F_0,\ldots,F_n)$ with
$F_i=v_{i0}+\sum_{k=1}^{l_i}v_{ik}M_{ik}\in\ee\{\Y\}$, we also
represent it by
$(\bv_0,\ldots,\bv_n)\in\ee^{l_0+1}\times\cdots\times\ee^{l_n+1}$
 where $\bv_i=(v_{i0},v_{i1},\ldots,v_{il_i})$. Let
% $\mathcal
%{Z}_0(\mathcal{A}_0,\ldots,\mathcal{A}_n)$ be a subset of
%$\ee^{l_0+1}\times\cdots\times\ee^{l_n+1}$ consisting of points
%$(\bv_0,\ldots,\bv_n)$ for which $F_i=0\,(i=0,\ldots,n)$ have common
%solutions. That is,
\begin{eqnarray}\label{eq-zA1} \quad&\quad&\mathcal
{Z}_0(\mathcal{B}_0,\ldots,\mathcal{B}_n)=\{(\bv_0,\ldots,\bv_n)\in\ee^{l_0+1}\times\cdots\times\ee^{l_n+1}:
F_0=\cdots=F_n=0 \,\,\text{have} \nonumber\\&\quad&\qquad\qquad
\qquad\qquad\text{ a common solution in}\, \ee^n\}
\end{eqnarray} Let $\overline{\mathcal {Z}(\mathcal{B}_0,\ldots,\mathcal{B}_n)}$ be the
Kolchin differential closure of $\mathcal
{Z}_0(\mathcal{B}_0,\ldots,\mathcal{B}_n)$ in
$\ee^{l_0+1}\times\cdots\times\ee^{l_n+1}$.
Note that zeros from $\ee$ are considered, instead of $\ee^{\wedge}$
as in \bref{eq-zA}.

 The following result shows that the vanishing of sparse
differential resultant gives a necessary condition for the existence
of solutions, and as well as gives a sufficient condition in some
sense.

\begin{theorem}\label{th-0kolclosure}
Suppose $\mathcal{B}_i\,(i=0,\ldots,n)$ form a differentially
essential system. Then we have $\mathcal
{Z}(\mathcal{B}_0,\ldots,\mathcal{B}_n)=\V\big(\sat(\Res_{\mathcal{B}_0,\ldots,\mathcal{B}_n})\big)$.
\end{theorem}
\proof  Since
$\sat(\Res_{\mathcal{B}_0,\ldots,\mathcal{B}_n})\subset\,[\P_0,\ldots,\P_n]\subset\qq\{\Y;\bu_0,\ldots,\bu_n\}$,
it follows directly that $\mathcal
{Z}_0(\mathcal{B}_0,\ldots,\mathcal{B}_n)\subseteq\V\big(\sat(\Res_{\mathcal{B}_0,\ldots,\mathcal{B}_n})\big)$.
Consequently, $\mathcal
{Z}(\mathcal{B}_0,\ldots,\mathcal{B}_n)\subseteq\V\big(\sat(\Res_{\mathcal{B}_0,\ldots,\mathcal{B}_n})\big)$.

For the other direction, follow the notations in the proof of
Theorem~\ref{th-0codim1}. By Theorem~\ref{th-0codim1},
$[\P_0,\ldots,\P_n]$ is a prime differential ideal with a generic
point $(\eta,\zeta)$ where $\eta=(\eta_1,\ldots,\eta_n)$ is a
generic point of $[0]$ over $\qq\langle\bu\rangle$ and
$\zeta=(\zeta_0,u_{01},\ldots,$ $u_{0l_0};\ldots;$
$\zeta_n,u_{n1},\ldots,u_{nl_n})$.  Let $(F_0,\ldots,F_n)$ be a set
of  differential polynomials represented by $\zeta$. Clearly, $\eta$
is a  solution of $F_i=0$. Thus, $\zeta\in\mathcal
{Z}_0(\mathcal{B}_0,\ldots,\mathcal{B}_n)\subset\mathcal
{Z}(\mathcal{B}_0,\ldots,\mathcal{B}_n)$. Since
 $\zeta$ is a generic point of
$\sat(\Res_{\mathcal{B}_0,\ldots,\mathcal{B}_n})$, it follows that
$\V\big(\sat(\Res_{\mathcal{B}_0,\ldots,\mathcal{B}_n})\big)\subseteq\mathcal
{Z}(\mathcal{B}_0,\ldots,\mathcal{B}_n)$. As a consequence,
$\V\big(\sat(\Res_{\mathcal{B}_0,\ldots,\mathcal{B}_n})\big)=\mathcal
{Z}(\mathcal{B}_0,\ldots,\mathcal{B}_n)$. \qedd

%The above theorem shows that the sparse differential resultant gives
%a sufficient and necessary condition for a differentially essential
%over-determined system to have non-polynomial solutions over an open
%set of $\prod_{i=0}^n \mathcal{L}(\mathcal{B}_0)$ in the sense of
%Kolchin topology.
%

In the following, we will analyze the properties of the solutions as
we did in section \ref{sec-bp2}. The following lemma  shows the
relation between the original differential system and their sparse
differential resultant, which is a direct consequence of
Theorem~\ref{th-idealrelation}.

\begin{lemma}\label{le-0idealrelation}
Let $\P_i=u_{i0}+\sum_{k=1}^{l_i}u_{ik}M_{ik}\,(i=0,\ldots,n)$ be a
system of differential polynomials
  satisfying that any $n$ of the $\P_i$
form a differentially independent set. Let $\SR(\bu_0,\ldots,\bu_n)$
be the sparse differential resultant of $\P_i$ with
$\ord(\SR,\bu_i)=h_i$. Denote $Q_{ik}=\frac{\partial \SR}{\partial
u_{i0}^{(h_i)}}M_{ik}-\frac{\partial \SR}{\partial u_{ik}^{(h_i)}}$
and $S$ to be the set consisting of $\frac{\partial \SR}{\partial
u_{i0}^{(h_i)}}\,(i=0,\ldots,n)$.  Then
$$[\P_0,\ldots,\P_n]=[\SR,(Q_{ik})_{0\leq i\leq n;1\leq
k\leq l_i}]:S^\infty$$ in $\qq\{\Y,\bu_0,\ldots,\bu_n\}$.
\end{lemma}
\proof It is a direct consequence of Theorem~\ref{th-idealrelation}
by setting $M_{i0}=1$ and from the fact that
$[\P_0,\ldots,\P_n]:\mathbbm{m}=[\P_0,\ldots,\P_n]$ as differential
ideals in $\qq\{\Y,\bu_0,\ldots,\bu_n\}$.  \qedd

%-gao
% Again, here can we introduce the concept of toric
%

Lemma \ref{le-0idealrelation} shows that under the condition
$\SR=0$, the  solutions of $\P_i=0\, (i=0,\ldots,n)$ generally are
the solutions of some differential polynomials of two terms.

\begin{cor} \label{cor-0generalized-sf}
Let $\P_0,\ldots,\P_n$ be a differentially essential system of the
form \bref{eq-qgpol} and $\SR(\bu_{0},$ $\ldots,\bu_{n})$ their
sparse differential resultant. Suppose $\ord(\SR,\bu_0)=h_0\geq0$
and denote $S_\SR=\frac{\partial \SR}{\partial u_{00}^{(h_0)}}$.
Suppose that when $\bu_i\,(i=0,\ldots,n)$ are specialized to sets
$\bv_i$ over $\Q$ which are elements in an extension field of
$\mathcal{F}$, $\P_i$ are specialized to
$\overline{\P}_{i}\,(i=0,\ldots,n)$. If
$S_\SR(\bv_0,\ldots,\bv_n)\neq 0$, in the case that
$\overline{\P}_i=0(i=0,\ldots,n)$ have a common
 differential solution $\xi$, then for each  $k$, we
have
\begin{equation} \label{eq-0root}
M_{0k}(\xi)=\frac{\partial \SR}{\partial
u_{0k}^{(h_0)}}(\bv_0,\ldots,\bv_n)\big/S_\SR(\bv_0,\ldots,\bv_n).
\end{equation}
\end{cor}
\proof By Lemma~\ref{le-0idealrelation}, for each $k=1,\ldots,l_0$,
the polynomial $S_\SR M_{0k}- \frac{\partial \SR}{\partial u_{0k
}^{(h_0)}}\in[\P_0,\ldots,\P_n]$. Thus, if $\xi$ is a common
differential solution of $\overline{\P_i}=0$, then
$S_\SR(\bv_0,\ldots,\bv_n)\cdot M_{0k}(\xi)-\frac{\partial
\SR}{\partial u_{0k }^{(h_0)}}(\bv_0,\ldots,\bv_n)=0$. Since
$S_\SR(\bv_0,\ldots,\bv_n)\neq 0$, \eqref{eq-0root} follows. \qedd

\vskip 5pt In the rest of this section, we will gives a sufficient
condition for a differentially essential system to have a unique
solution.

\begin{theorem}\label{th-0unique}
Assume that 1) any $n$  of the $\mathcal {B}_i\,(i=0,\ldots,n)$ form
a differentially independent set and 2) for each $j=1,\ldots,n$,
$\textbf{e}_j\in\Span_{\mathbb{Z}}\{\alpha_{ik}:k=1,\ldots,l_i;i=0,\ldots,n\}$.
Denote $\ord(\SR,\bu_i)=h_i$ where
$\SR(\bu_0,\ldots,\bu_n)=\Res_{\mathcal {A}_0,\ldots,\mathcal
{A}_n}$. Let $\overline{\P}_{i}$ be a specialization of $\P_i$ over
$\Q$ with coefficient vector $\bv_i\,(i=0,\ldots,n)$.  If
$\SR(\bv_0,\ldots,\bv_n)=0$ and $\frac{\partial \SR}{\partial
u_{ik}^{(h_i)}}(\bv_0,\ldots,\bv_n)\neq0$ for each $i$ and $k$, then
$\overline{\P}_{i}=0\,(i=0,\ldots,n)$ have a unique common solution.
\end{theorem}
\proof By hypothesis 1), each $h_i\geq0$. By
Lemma~\ref{le-0idealrelation}, $Q_{ik}=\frac{\partial \SR}{\partial
u_{i0}^{(h_i)}}M_{ik}-\frac{\partial \SR}{\partial
u_{ik}^{(h_i)}}\in\,[\P_0,\ldots,\P_n]$
$\subset\qq\{\Y;\bu_0,\ldots,\bu_n\}$.
 Since $(\eta;\zeta)$ is a generic point of $[\P_0,\ldots,\P_n]$,
${M_{ik}(\eta)}=\overline{\frac{\partial \SR}{\partial
u_{ik}^{(h_i)}}}\big/\overline{\frac{\partial \SR}{\partial
u_{i0}^{(h_i)}}}$, where $\overline{\frac{\partial \SR}{\partial
u_{ik}^{(h_i)}}}$ are obtained by substituting $\zeta_{i}$ to
$u_{i0}\,(i=0, 1, \ldots, n)$ in $\frac{\partial \SR}{\partial
u_{ik}^{(h_i)}}$. By hypothesis 2), there exist
$t_{jik}\in\mathbb{Z}$ such that
$\sum_{i,k}t_{jik}\alpha_{ik}=\textbf{e}_j$ for $j=1,\ldots,n$. So
$\prod_{i,k}\big(M_{ik}\big)^{t_{jik}}=y_j$. Thus,
$\prod_{i,k}\big(M_{ik}(\eta)\big)^{t_{jik}}=\eta_j=\prod_{i,k}\big(\overline{\frac{\partial
R}{\partial u_{ik}^{(h_i)}}}\big/\overline{\frac{\partial
R}{\partial u_{i0}^{(h_i)}}}\big)^{t_{jik}}$. It follows that
$S_jy_j-T_j\in[\P_0,\ldots,\P_n]$ where $S_j$ and $T_j$ are products
of nonnegative powers of $\frac{\partial \SR}{\partial
u_{ik}^{(h_i)}}$ obtained from the above identity. Thus,
$R,S_1y_1-T_1,\ldots,S_ny_n-T_n$ is a characteristic set of
$[\P_0,\ldots,\P_n]$ w.r.t. any elimination ranking $u_{ik}\prec y_1
\prec \cdots \prec y_n$. For each $\P_i$, there exists a product
$A_i$ of nonnegative powers of $\frac{\partial \SR}{\partial
u_{ik}^{(h_i)}}$ such that
$A_i\P_i\in\,[\SR,S_1y_1-T_1,\ldots,S_ny_n-T_n]$. Now specialize
$\bu_i$ to $\bv_i$ over $\Q$\,($i=0,\ldots,n$).
 Since
$\frac{\partial \SR}{\partial
u_{ik}^{(h_i)}}(\bv_0,\ldots,\bv_n)\neq0$ for each $i$ and $k$,
$T_j(\bv_0,\ldots,\bv_n)\cdot S_j(\bv_0,\ldots,\bv_n)\neq0$.  Let
$\bar{y}_j=T_j(\bv_0,\ldots,\bv_n)/S_j(\bv_0,\ldots,\bv_n)$ and
$\bar{y}=(\bar{y}_1\ldots,\bar{y}_n)$. Clearly,
$\overline{\P}_{i}(\bar{y})=0$. Thus, $\bar{y}$ is a common solution
of $\overline{\P}_{i}\,(i=0,\ldots,n)$.

 On the other hand, for any solution $\xi$
of $\overline{\P}_{i}$,
$S_j(\bv_0,\ldots,\bv_n)y_j-T_j(\bv_0,\ldots,\bv_n)\,(j=1,\ldots,n)$
vanishes at it.
 So $\xi=\bar{y}$. As a consequence, we have proved
that  in this case, $\overline{\P}_{i}=0\,(i=0,\ldots,n)$ have a
unique solution.
  \qedd

\section{A single exponential algorithm to compute the sparse differential resultant}
In this section, we give an algorithm to compute the sparse
differential resultant for a  Laurent differentially essential
 system with single exponential complexity.
The idea is to estimate the degree bounds for the resultant and then
to use linear algebra to find the coefficients of the resultant.

\subsection{Degree of algebraic elimination ideal}
\label{sec-adeg}
In this section, we will prove several properties about the degrees
of elimination ideals  in the algebraic case, which will be used to
estimate the degree bound for sparse differential resultants.

\vskip5pt Let $P$ be a polynomial in $K[\X]$ where $K$ is an
algebraic field and $\X=\{x_1,\ldots,x_n\}$ a set of algebraic
indeterminates. We use $\deg(P)$ to denote the total degree of $P$.
Let $\I$ be a prime algebraic ideal in $K[\X]$ with dimension $d$.
We use $\deg(\I)$ to denote the {\em degree} of $\I$, which is
defined to be the number of solutions of the zero dimensional prime
ideal $(\I,\L_1,\ldots,\L_d)$ in $K(U)[\X]$, where $\L_i =
u_{i0}+\sum_{j=1}^n u_{ij} x_j \,(i=1,\ldots,d)$ are $d$ generic
hyperplanes \cite{hodge} and $U =
\{u_{ij}\,(i=1,\ldots,d,j=0,\ldots,n)\}$. That is,
 \begin{equation}\label{eq-deg1}
  \deg(\I)=|\V(\I,\L_1,\ldots,\L_d)|.\end{equation}
Clearly, $\deg(\I)=\deg(\I,\L_1,\ldots,\L_i)$ for $i=1,\ldots,d$.
$\deg(\I)$ is also equal to the maximal number of intersection
points of $\V(\I)$ with $d$ hyperplanes under the condition that the
number of these points is finite \cite{Gabriela}. That is,
 \begin{eqnarray}\label{eq-deg2}
  \deg(\I)=\max\{|\V(\I)\cap H_1\cap\cdots\cap H_d|: \,H_i \,\hbox{are}\, \hbox{affine hyperplanes} \nonumber \\
   \hbox{ such that }\, |\V(\I)\cap
H_1\cap\cdots\cap H_d|<\infty\}\end{eqnarray}
 We investigate the relation
between the degree of an ideal and that of its elimination ideal by
proving Theorem \ref{th-elimination}.

\begin{lemma} \label{le-deg-0dimen}
Let $\I$ be a prime ideal of dimension zero in $K[\X]$ and
$\I_k=\I\cap K[x_1,\ldots,x_k]$ the elimination ideal of $\I$ with
respect to  $x_1,\ldots,x_k$. Then $\deg(\I_k)\le \deg(\I)$.
\end{lemma}
\proof Since both $\I$ and $\I_k$ are prime ideals of dimension
zero, $\deg(\I_k)=|\V(\I_k)| $ and $\deg(\I)=|\V(\I)|$. To show
$\deg(\I_k)\le \deg(\I)$, it suffices to prove that every point of
$\V(\I_k)$ can be extended to a  point of $\V(\I)$. Let
$(\xi_1,\ldots,\xi_k)\in\V(\I_k)$. For any point
$(\eta_1,\ldots,\eta_n)\in \V(\I)$, $(\eta_1,\ldots,\eta_k)$ is a
zero point of $\I_k$. So we have $K(\xi_1,\ldots,\xi_k)\cong
K(\eta_1,\ldots,\eta_k)$. By \cite[Proposition 9, Chapter 1, \S
3]{weil}, there exist $\xi_{k+1},\ldots,\xi_n$ such that
$K(\xi_1,\ldots,\xi_n)\cong K(\eta_1,\ldots,\eta_n)$. Thus,
$(\xi_1,\ldots,\xi_n)$ is a zero of $\I$, which completes the proof.
\qedd

\begin{theorem}\label{th-elimination} Let $\I$ be a prime ideal in $K[\X]$
and $\I_k=\I\cap K[x_1,\ldots,x_k]$ for any $1\leq k\leq n$. Then
$\deg(\I_k)\leq \deg(\I)$.
\end{theorem}
\proof Suppose $\dim(\I)=d$ and $\dim(\I_k)=d_1$. Two cases are
considered: \vskip5pt Case (a): $d_1=d$. Let
$\P_i=u_{i0}+u_{i1}x_1+\cdots+u_{ik}x_k\,(i=1,\ldots,d)$. Denote
$\bu=\{u_{ij}:\,i=1,\ldots,d;j=0,\ldots,k\}$. Then by \cite[Theorem
1, p. 54]{hodge}, $\J=(\I_k,\P_1,\ldots,\P_d)$ is a prime ideal of
dimension zero in $K(\bu)[x_1,\ldots,x_k]$ and has the same degree
as $\I_k$. We claim that \vskip2.5pt i) $(\I,\P_1,\ldots,\P_d)\cap
K(\bu)[x_1,\ldots,x_k]=\J.$

\vskip2.5pt ii) $(\I,\P_1,\ldots,\P_d)$ is a 0-dimensional prime
ideal over $K(\bu)$.

To prove i), it suffices to show that  whenever $f$ is in the left
ideal, $f$ belongs to $\J$. Without loss of generality, suppose
$f\in K[\bu][x_1,\ldots,x_k]$. Then there exist
 $h_l,q_i\in K[\bu
][\X]$ and $g_l\in\I$ such that $f=\sum_{l}h_lg_l+\sum_{i=1}^d q_i
\P_i$. Substituting $u_{i0}=-u_{i1}x_1-\cdots-u_{ik}x_k$ into the
above equality, we get $\bar{f}=\sum_{l}\bar{h_l}g_l\in\I$. Thus,
$\bar{f}\in\I_k$. But $f\equiv\bar{f}\, \mod (\P_1,\ldots,\P_d)$, so
$f\in \J$, and i) follows.

To prove ii), let $(\xi_1,\ldots,\xi_n)$ be a generic point of $\I$.
Denote $U_0=\{u_{10},\ldots,u_{d0}\}$. Then
$\J_0=(\I,\P_1,\ldots,\P_d)\subseteq K(\bu \backslash
U_0)[\X,U_{0}]$ is a prime ideal of dimension $d$ with a generic
point $(\xi_1,\ldots,\xi_n,\zeta_1,$ $\ldots,\zeta_d)$, where
$\zeta_i=-\sum_{j=1}^k u_{ij}\xi_j\,(i=1,\ldots,d)$. Since $d_1=d$,
there exist $d$ elements in $\{\xi_1,\ldots,$$\xi_k\}$ algebraically
independent over $K$.  So by \cite[p.168-169]{hodge1},
$\zeta_1,\ldots,\zeta_d$ are algebraically independent over $K(\bu
\backslash U_0)$. Thus, $\J_0\cap K(\bu \backslash U_0)[U_{0}]=(0)$ and
ii) follows.

By Lemma~\ref{le-deg-0dimen},
$\deg(\J)\le\deg(\I,\P_1,$ $\ldots,\P_d)$. So by \bref{eq-deg2},
$\deg(\I)\geq |\V(\I,\P_1,\ldots,\P_d)|\geq \deg(\J)=\deg(\I_k)$.

\vskip5pt Case (b): $d_1<d$. Let
$\L_i=u_{i0}+u_{i1}x_1+\cdots+u_{in}x_n\,(i=1,\ldots,d-d_1)$. By
\cite[Theorem 1, p. 54]{hodge}, $\J=(\I,\L_1,\ldots,$
$\L_{d-d_1})\subseteq K(\bu)[\X]$ is a prime ideal of dimension
$d_1$ and $\deg(\J)=\deg(\I)$, where $\bu
=\{u_{ij}:i=1,\ldots,d-d_1;j=0,\ldots,n\}$. Let $\J_k=\J\cap
K(\bu)[x_1,\ldots,x_k].$ We claim that $\J_k=(\I_k)$ in
$K(\bu)[x_1,\ldots,x_k]$. Of course, $\J_k\supseteq (\I_k)$. Since
both $\J_k$ and $(\I_k)$  are prime ideals and $\dim((\I_k))=d_1$,
it suffices to prove that $\dim(\J_k)=d_1$.

Suppose $(\xi_1,\ldots,\xi_n)$ is a generic point of $\I$, then
$(\xi_1,\ldots,\xi_k)$ is that of $\I_k$. Let
$\J_0=(\I,\L_1,\ldots,\L_{d-d_1})\subseteq K(\bu\backslash
U_0)[\X,U_0]$,  then $(\xi_1,\ldots,$ $\xi_{n},-\sum_{j=1}^n
u_{1j}\xi_j,$ $\ldots,-\sum_{j=1}^n u_{d-d_1,j}\xi_j)$ is a generic
point of it, where $U_0=\{u_{10},\ldots,u_{d-d_1,0}\}$.
%Now by  proving   $\J_k\cap K(\bu)[x_1,\ldots,x_{d_1}]=(0)$, we show
%$\dim(\J_k)=d_1$. Suppose the contrary, i.e.
%$\J_k\cap K(\bu)[x_1,\ldots,x_{d_1}]\neq(0)$.  Then $\J_0\cap K(\bu
%_1)[x_1,\ldots,x_{d_1},U_{0}]\neq(0)$. So $\xi_1,\ldots,\xi_{d_1}, -\sum_{j=1}^n
%u_{1j}\xi_j,\ldots,$ $-\sum_{j=1}^n u_{d-d_1,j}\xi_j$ are algebraically
%dependent over $K(\bu\backslash \U_0)$. When $u_{ij}$ specializes to
%$-\delta_{k+i,j}$ for $i=1,\ldots,d-d_1$ and $j=1,\ldots,n$, by
%\cite[Lemma 2.12]{Gao},
%$\xi_1,\ldots,\xi_{d_1},\xi_{k+1},\ldots,\xi_{k+(d-d_1)}$ are
%algebraically dependent over $K$, a contradiction obtained.
Since $\dim(\I_k)=d_1$, without loss of generality, suppose
$\xi_1,\ldots,\xi_{d_1}$ is a transcendence basis of
$K(\xi_1,\ldots,\xi_k)/K$ and $\xi_1,\ldots,\xi_{d_1},$
$\xi_{k+1},\ldots,\xi_{k+(d-d_1)}$ is that of
$K(\xi_1,\ldots,\xi_n)/K$. Then by \cite[p.168-169]{hodge1}, it is
easy to show that $\J_0\cap K(\bu\backslash
U_0)[x_1,\ldots,x_{d_1},$ $U_{0}]=(0)$, and $\J_k\cap
K(\bu)[x_1,\ldots,x_{d_1}]=0$ follows. So $\dim(\J_k)=d_1$ and
$\J_k=(\I_k)$.

Since $\dim(\J_k)=\dim(\J)$, by case (a), we have
$\deg(\J_k)\leq\deg(\J)=\deg(\I)$. Due to the fact that
$\deg(\J_k)=\deg((\I_k))=\deg(\I_k)$, $\deg(\I_k)\leq\deg(\I)$
follows. \qedd

In this article, we will use the following two results.

\begin{lemma} \cite[Corollary 2.28]{vogel} \label{le-vogel}
Let $V_1,\ldots,V_r\subset\textbf{P}^n\,(r\geq 2)$ be pure
dimensional projective varieties in $\textbf{P}^n$. Then
$$\prod_{i=1}^r\deg(V_i)\geq \sum_{C}\deg(C)$$
where $C$ runs through all irreducible components of
$V_1\cap\cdots\cap V_r$.
\end{lemma}

\begin{lemma}\cite[Proposition 1, p.151]{lazard} \label{le-deg-pol}
Let $F_1,\ldots,F_m \in K[\X]$ be  polynomials generating an ideal
$\I$ of dimension $r$. Suppose $\deg(F_1)\geq\cdots\geq\deg(F_m)$
and let $D:=\prod_{i=1}^{n-r}\deg(F_i)$. Then $\deg(\I)\leq D$.
\end{lemma}

%\subsubsection{Degree of algebraic generalized Chow form}

\subsection{Degree bound for sparse differential resultant}

In this section,  we give an upper bound for the degree of
the sparse differential resultant, which will be crucial to our
algorithm to compute the sparse resultant.

%$\ord(\P_i^N,y_j)=e_{ij}$,
\begin{theorem} \label{th-spardeg}
Let $\P_0,\ldots,\P_n$ be a Laurent differentially essential system
of form
 \bref{eq-sparseLaurent} with  $\Eord(\P_i)=e_i$ and $\deg(\P_i^N,\Y)$
$=m_i$.  Suppose $\P_{i}^N=\sum_{k=0}^{l_i}u_{ik}N_{ik}$ and $J_i$
is the modified Jacobi number of
$\{\P_0^N,\ldots,\P^N_{n}\}\backslash\{\P^N_{i}\}$. Denote
$m=\max_i\{m_i\}$.
 Let $\SR(\bu_0,\ldots,\bu_n)$ be the sparse differential
resultant of $\P_i\,(i=0,\ldots,n)$.  Suppose $\ord(\SR,\bu_i)=h_i$
for each $i$. Then the following assertions hold:
\begin{description}
\item[1)]
$\deg(\SR)\leq \prod_{i=0}^n
    (m_i+1)^{h_i+1}\leq (m+1)^{\sum_{i=0}^n(J_i+1)}$, where $m=\max_i\{m_i\}$.
%There exists one differential monomial $M$ such that when multiplied
%by $M$, $\SR$ can be written as a linear combination of $\P_i^N$ and
%their derivatives up to order $h_i$.
%   Precisely,
%  $$M\cdot R=\sum_{i=0}^n\sum_{k=0}^{h_i}G_{ik}\big(\P_i^N\big)^{(k)}$$
%   for some $G_{ik}\in\Q[\bu_0^{[h_0]},\ldots,\bu_n^{[h_n]},\Y^{[h]}]$ and $\ord(M)\leq h$ where $h=\max_{i}\{h_i+s_i\}$.
\item[2)]
$\SR$ has a representation
\begin{equation}\label{eq-deg5}
\prod_{i=0}^n N_{i0}^{(h_i+1)\deg(\SR)}\cdot
\SR=\sum_{i=0}^n\sum_{j=0}^{h_i}G_{ij}\big(\P_{i}^N\big)^{(j)}
\end{equation}
where
 $G_{ij}\in \Q[\bu_0^{[h_0]},\ldots,\bu_n^{[h_n]},\Y^{[h]}]$
and $h=\max\{h_i+e_i\}$ such that $\deg(G_{ij}(\P_{i}^N)^{(j)})\leq
[m+1+\sum_{i=0}^n(h_i+1)\deg(N_{i0})]\deg(\SR)$.
\end{description}
\end{theorem}
\proof  1) By the definition of sparse differential resultant,
    $[\P_0^{\text{N}},\ldots,\P_n^{\text{N}}]:\mathbbm{m}\cap\,\mathbb{Q}\{\bu_0,\ldots,\bu_n\}=\sat(\SR)$.
     Let $\eta=(\eta_1,\ldots,\eta_n)$ be a generic point of $[0]$.
     Denote
     $\zeta_i=-\sum_{k=1}^{l_i}u_{ik}\frac{M_{ik}(\eta)}{M_{i0}(\eta)}\,(i=0,\ldots,n)$.
     Then
     $(\eta;\zeta_0,u_{01},\ldots,u_{0l_0};\ldots;\zeta_n,u_{n1},\ldots,u_{nl_n})$
     is a generic point of
     $[\P_0^{\text{N}},\ldots,\P_n^{\text{N}}]:\mathbbm{m}$.
     Clearly, $\P_0^{\text{N}},\ldots,\P_n^{\text{N}}$ is a
     characteristic set of
     $[\P_0^{\text{N}},\ldots,\P_n^{\text{N}}]:\mathbbm{m}$ w.r.t.
     the elimination ranking $u_{n0}\succ \cdots\succ u_{10}\succ u_{00}\succ \bu\succ
     \Y$. Taking the differential remainder of $\SR$ w.r.t. this characteristic set,  we have
     $$\prod N_{i0}^{a_i}\SR=\sum_{i=0}^n\sum_{k=0}^{h_i}G_{ik}\big(\P_{i}^{\text{N}}\big)^{(k)}$$ for
$a_i\in\mathbb{N}$. Denote $h=\max_i\{h_i+e_i\}$ and by
$\mathbbm{m}^{[h]}$ we mean the set of     all monomials in
$\Y^{[h]}$.  Let $\mathcal
{J}=\big((\P_{0}^{\text{N}})^{[h_0]},\ldots,(\P_{n}^{\text{N}})^{[h_n]}\big):\mathbbm{m}^{[h]}$
be an ideal  in
$\mathcal{R}=\qq[\Y^{[h]},\bu_0^{[h_0]},\ldots,\bu_n^{[h_n]}]$. Then
$\SR\in\mathcal{J}$.     Furthermore, it is easy to show that
     $\mathcal {J}$ is a   prime ideal in $\mathcal{R}$ with a generic point
     $(\eta^{[h]};\widetilde{\bu},\zeta_0^{[h_0]},\ldots,\zeta_n^{[h_n]})$
     and $\mathcal
     {J}\cap\Q[\bu_0^{[h_0]},\ldots,\bu_n^{[h_n]}]=(\SR)$, where $\widetilde{\bu}=\cup_i\bu_i^{[h_i]}\backslash
     \{u_{i0}^{[h_i]}\}$.
    Let $H_{ik}$ be the homogenous polynomial corresponding to
    $\big(\P_{i}^{\text{N}}\big)^{(k)}$ with $x_0$ the variable of
    homogeneity. Then $\mathcal {J}^0=((H_{ik})_{1\leq i\leq n;0\leq k\leq h_i}):\widetilde{\mathbbm{m}}$
    is a prime ideal in $\Q[x_0,\Y^{[h]},\bu_0^{[h_0]},\ldots,\bu_n^{[h_n]}]$ with a generic point
    $(v,v\eta^{[h]};v\widetilde{\bu},v\zeta_0^{[h_0]},\ldots,v\zeta_n^{[h_n]})$
    where $\widetilde{\mathbbm{m}}$ is the whole set of
      monomials in $\Y^{[h]}$ and $x_0$. Then $\deg(\mathcal {J}^0)
      =\deg(\mathcal{J})$.

Since $\V((H_{ik})_{1\leq i\leq n;0\leq k\leq h_i})=\V(\mathcal
{J}^0)\cup\V(H_{ik},x_0)\bigcup
    \cup_{j,l}\V(H_{ik},y_{j}^{(l)})$,  $\V(\mathcal {J}^0)$ is an irreducible component of
$\V((H_{ik})_{1\leq i\leq n;0\leq k\leq h_i})$. By
Lemma~\ref{le-vogel}, $\deg(\mathcal {J}^0)\leq
\prod_{i=0}^n\prod_{k=0}^{h_i}(m_i+1)=\prod_{i=0}^n(m_i+1)^{h_i+1}$.
Thus, $\deg(\mathcal {J})\leq\prod_{i=0}^n(m_i+1)^{h_i+1}$.    Since
$\mathcal {J}\cap\Q[\bu_0^{[h_0]},\ldots,\bu_n^{[h_n]}]=(\SR)$, by
Theorem~\ref{th-elimination},    $\deg(\SR)\leq
    \deg(\mathcal {J})\leq\prod_{i=0}^n(m_i+1)^{h_i+1}\leq (m+1)^{\sum_{i=0}^n(J_i+1)}$
    follows. The last inequality holds because  $h_i\leq J_i$ by Theorem~\ref{th-jacobi-order3}.

2)
 To obtain the degree bounds for this  representation for $\SR$, we
 first substitute $u_{i0}$ by $\big(\P_{i}^N-\sum_{k=1}^{l_i}u_{ik}N_{ik}\big)/N_{i0}$
 into
 $\SR$ and then expand it. To be more precise, we take
 one  monomial $M(\bu;u_{00},\ldots,$ $u_{n0})$ in $\SR(\bu_0,\ldots,\bu_n)$ for an example.
Denote
 $M=\bu^{\gamma}\prod_{i=0}^n\prod_{k=0}^{h_i}(u_{i0}^{(k)})^{d_{ik}}$
 with $|\gamma|+\sum_{i=0}^n\sum_{k=0}^{h_i}d_{ik}=
 \deg(\SR)$, where $\bu^{\gamma}$ represents a monomial in $\bu$ and their derivatives with exponent vector $\gamma$.
 Substitute $u_{i0}$ by $\big(\P_{i}^N-\sum_{k=1}^{l_i}u_{ik}N_{ik}\big)/N_{i0}$ into $M$,
 we have
 {\small$$M(\bu;u_{00},\ldots,u_{n0})=
 \bu^{\gamma}\prod_{i=0}^n\prod_{k=0}^{h_i}\bigg(\Big(\big(\P_{i}^N-\sum_{k=1}^{l_i}u_{ik}N_{ik}\big)/N_{i0}\Big)^{(k)}\bigg)^{d_{ik}}.$$}
When expanded, the denominator is of the form $\prod_{i=0}^n
N_{i0}^{\sum_{k}(k+1)d_{ik}}$ and  every term of the numerator has
total degree
$|\gamma|+\sum_{i=0}^n\sum_{k=0}^{h_i}[(k+1)\deg(N_{i0})+(m_i+1-\deg(N_{i0}))]d_{ik}$
in $\bu_0^{[h_0]},\ldots,\bu_n^{[h_n]}$ and
 $\Y^{[h]}$ with $h=\max\{h_i+e_i\}$. For every monomial $M$ in $\SR$,
 $\sum_{k=0}^{h_i}(k+1)d_{ik}\leq (h_i+1)\deg(\SR)$. Thus,
 $\prod_{i=0}^n
N_{i0}^{(h_i+1)\deg(\SR)}\cdot
\SR=\sum_{i=0}^n\sum_{j=0}^{h_i}G_{ij}\big(\P_{i}^N\big)^{(j)}+T$
where $G_{ij},T\in\Q[\bu_0^{[h_0]},\ldots,\bu_n^{[h_n]},\Y^{[h]}]$
and $T$ is free from $u_{i0}$ for $i=0,\ldots,n$. It is easy to see
that $T\in[\P_{0}^N,\ldots,\P_{n}^N]:\mathbbm{m}$ and $T=0$ follows.
By the above substitution for every monomial in $\SR$, we can see
that
 \begin{eqnarray*}
&& \deg(G_{ij}\big(\P_{i}^N\big)^{(j)})\\
 &\leq& \max_{(\gamma,d_{ik})}\{
|\gamma|+\sum_{i=0}^n\sum_{k=0}^{h_i}[(k+1)\deg(N_{i0})+(m_i+1-\deg(N_{i0}))]d_{ik}\\
&&\quad
+\sum_{i=0}^n(h_i+1)\deg(\SR)\deg(N_{i0})-\sum_{i=0}^n(\sum_{k}(k+1)d_{ik})\deg(N_{i0})\}\\
&\leq&\deg(\SR)+\sum_{i,k}(m_i-\deg(N_{i0}))d_{ik}+\deg(\SR)\sum_{i=0}^n(h_i+1)\deg(N_{i0})\\
&\leq&(m+1)\deg(\SR)+\deg(\SR)\sum_{i=0}^n(h_i+1)\deg(N_{i0})\\
&=&[m+1+\sum_{i=0}^n(h_i+1)\deg(N_{i0})]\deg(\SR).
 \end{eqnarray*}
 \qedd

% -gao
% why removed the following contents?

\begin{example} \label{ex-13}
Continue from Example~\ref{ex-3}. In this example, $J_0=2$,
$J_1=J_2=1$ and $m_0=m_1=m_2=2$. The expression of $\SR$ shows that
$h_0=\ord(\SR,\bu_0)=1< J_0$, $h_i=\ord(\SR,\bu_i)=0<J_i\,(i=1,2)$
and $\deg(\SR)=5<<3^4=\prod_{i=0}^2(m_i+1)^{h_i+1}.$
\end{example}

\vt For a differentially essential system of form
 \bref{eq-qgpol}, the second part of
Theorem \ref{th-spardeg} can be improved as follows.

\begin{theorem} \label{th-pspardeg}
Let $\P_0,\ldots,\P_n$ be a differentially essential system of form
 \bref{eq-qgpol} with $m=\max_i\{\deg(\P_i^N,\Y)\}$ and $J_i$ the modified Jacobi number of $\{\P_0^N,\ldots,\P^N_{n}\}\backslash\{\P^N_{i}\}$.
 Let $\SR(\bu_0,\ldots,$ $\bu_n)$ be the sparse differential
resultant of $\P_i\,(i=0,\ldots,n)$.  Suppose $\ord(\SR,\bu_i)=h_i$
for each $i$ and $h=\max\{h_i+s_i\}$. Then  $\SR$ has a
representation
$$\SR(\bu_0,\ldots,\bu_n)=\sum_{i=0}^n\sum_{j=0}^{h_i}G_{ij}\P_{i}^{(j)}$$
where $G_{ij}\in \Q[\bu_0^{[h_0]},\ldots,\bu_n^{[h_n]},\Y^{[h]}]$
 such that $\deg(G_{ij}\P_{i}^{(j)})\leq
(m+1)\deg(\SR)\leq(m+1)^{\sum\limits_{i=0}^n(J_i+1)+1}$.
\end{theorem}
\proof Regarding $\P_i$ as Laurent differential polynomials,
$\P_i^N=\P_i$ and $N_{i0}=1$. By setting $N_{i0}=1$ in
Theorem~\ref{th-spardeg} these three assertions  directly follow.
\qedd

The following result gives an effective differential Nullstellensatz
under certain conditions.

\begin{cor}
Let $f_0,\ldots,f_n\in\F\{y_1,\ldots,y_n\}$ have no common solutions
with $\deg(f_i)\leq m$. Let
$\Jac(\{f_0,\ldots,f_{n}\}\backslash\{f_{i}\})=J_i$. If the sparse
differential resultant of $f_0,\ldots,f_n$ is nonzero, then there
exist $H_{ij}\in\F\{y_1,\ldots,y_n\}$ s.t.
$\sum_{i=0}^n\sum_{j=0}^{J_i}H_{ij}f_{i}^{(j)}$ $=1$ and
$\deg(H_{ij}f_{i}^{(j)})\leq (m+1)^{\sum_{i=0}^n(J_i+1)+1}$.
\end{cor}
\proof The hypothesis implies that $\P(f_i)$ form a differentially
essential system. Clearly, $\SR(\bu_0,\ldots,\bu_n)$ has the
property stated in Theorem~\ref{th-pspardeg}, where $\bu_i$ are
coefficients of $\P(f_i)$. The result follows directly from
Theorem~\ref{th-pspardeg} by specializing $\bu_i$ to the
coefficients of $f_i$. \qedd

With Theorem \ref{th-spardeg}, properties 6) and 7) of Theorem
\ref{th-i1} are proved.

\subsection{A single exponential algorithm to compute sparse differential resultant}

If a polynomial $R$ is the linear combination of some known
polynomials $F_i(i=1,\ldots,s)$, that is $R=\sum_{i=1}^s H_i F_i$,
 and we know the upper bounds of the degrees of $R$ and $H_iF_i$, then a general
idea to estimate the computational complexity of $R$ is  to use
linear algebra to find the coefficients of $R$.

For sparse differential resultant, we already gave its degree bound and the degrees of the
expressions in the linear combination in
Theorem \ref{th-spardeg}.

Now, we give the algorithm {\bf SDResultant} to compute sparse
differential resultants based on the linear algebra techniques.  The
algorithm works adaptively by searching for $\SR$ with an order
vector $(h_0,\ldots,h_n)\in\mathbb{N}^{n+1}$ with $h_i \leq J_i$ by
Theorem \ref{th-spardeg}. Denote $o=\sum_{i=0}^n h_i$. We start with
$o=0$. And for this $o$, choose one vector $(h_0,\ldots,h_n)$ at a
time. For this $(h_0,\ldots,h_n)$, we search for $\SR$ from degree
$d=1$. If we cannot find an $\SR$ with such a degree, then we repeat
the procedure with degree $d+1$ until
$d>\prod_{i=0}^n(m_i+1)^{h_i+1}$. In that case, we choose another
$(h_0,\ldots,h_n)$ with $\sum_{i=0}^nh_i=o$. But if for all
$(h_0,\ldots,h_n)$ with $h_i\leq J_i$ and $\sum_{i=0}^nh_i=o$, $\SR$
cannot be found, then we repeat the procedure with $o+1.$ In this
way, we will find an $\SR$ with the smallest order satisfying
equation \bref{eq-deg5}, which is the sparse resultant.

\begin{algorithm}\label{alg-dresl}
  \caption{\bf --- SDResultant($\P_0,\ldots,\P_n$)} \smallskip
  \Inp{A generic Laurent differentially essential system $\P_0,\ldots,\P_n$.}\\
  \Outp{The sparse differential resultant $\SR(\bu_0,\ldots,\bu_n)$ of $\P_0,\ldots,\P_n$.}\medskip

  \noindent
  1. For $i=0,\ldots,n$, set $P_i^N=\sum_{k=0}^{l_i}u_{ik}N_{ik}$ with $\deg(N_{i0})\leq\deg(N_{ik})$.\\
      \SPC Set $e_{ij}=\ord(\P_i^N,y_j)$,  $m_i=\deg(\P_i^N)$, $m_{i0}=\deg(N_{i0})$, $\bu_i=\coeff(\P_i)$ and $|\bu_i|=l_i+1$.\\
      \SPC Set $A=(e_{ij})$ and compute $J_i=\Jac(A_{\hat{i}})$.\\
  2. Set $\SR=0$, $o=0$,  $m=\max_i\{m_i\}$.\\
  3. While $\SR=0$ do\\
   \SPC 3.1. For each vector $(h_0,\ldots,h_n)\in \mathbb{N}^{n+1}$ with
   $\sum_{i=0}^n h_i$$=o$ and $h_i\leq J_i$  do \\
   \SPC\SPC 3.1.1. $U=\cup_{i=0}^n\bu_i^{[h_i]}$,
   $h=\max_i\{h_i+e_i\}$, $d=1$.  \\
   \SPC\SPC 3.1.2. While $\SR=0$ and $d\leq
   \prod_{i=0}^n(m_i+1)^{h_i+1}$ do \\
   \SPC\SPC\SPC 3.1.2.1. Set $\SR_0$ to be a homogenous GPol of degree $d$ in $U$.\\
   \SPC\SPC\SPC 3.1.2.2. Set $\bc_{0}=\coeff(\SR_0,U)$.\\
   \SPC\SPC\SPC 3.1.2.3. Set $H_{ij}(i=0,\ldots,n;j=0,\ldots,h_i)$ to be GPols of degree \\
  \SPC\SPC\SPC\SPC\SPC\quad$[m+1+\sum_{i=0}^n(h_i+1)m_{i0}]d-m_i-1$
   in $\Y^{[h]},U$.\\
   \SPC\SPC\SPC 3.1.2.4. Set $\bc_{ij}=\coeff(H_{ij}, \Y^{[h]}\cup U)$.
   \\
   \SPC\SPC\SPC 3.1.2.5. Set $\mathcal {P}$ to be the set of coefficients of $\prod_{i=0}^n
N_{i0}^{(h_i+1)d}\SR_0(\bu_0,\ldots,\bu_n)-$
\\
  \SPC\SPC\SPC \SPC \SPC\quad $\sum_{i=0}^n\sum_{j=0}^{h_i}H_{ij}(\P_{i}^N)^{(j)}$  as a  polynomial in
  $\Y^{[h]},U$.\\
   \SPC\SPC\SPC 3.1.2.6. Solve the linear equation $\mathcal {P}=0$ in
   variables $\bc_{0}$ and
  $\bc_{ij}.$\\
   \SPC\SPC\SPC 3.1.2.7. If $\bc_0$ has a nonzero solution, then
   substitute it into  $\SR_0$ to get $\SR$ and go to  \\
  \SPC\SPC\SPC \SPC \SPC\quad Step 4, else $\SR=0$.\\
  \SPC\SPC\SPC 3.1.2.8. d:=d+1.\\
  \SPC 3.2. o:=o+1.\\
  4. Return $\SR$.\medskip

  \noindent /*/\;  GPol stands for generic algebraic polynomial.\smallskip

  \noindent /*/\; $\coeff(P,V)$ returns the set of coefficients of $P$ as an ordinary  polynomial in
  variables $V$.
\smallskip
\end{algorithm}

\begin{theorem}\label{th-cdres}
Let $\P_0,\ldots,\P_n$ be a Laurent differentially essential system
of form \bref{eq-sparseLaurent}. Denote
$\P=\{\P_0^N,\ldots,\P_n^N\}$, $J_i=\Jac(\P_{\hat{i}})$,
$J=\sum_{i=0}^nJ_i$ and $m=\max_{i=0}^n \deg(\P_i,\Y)$.
Algorithm {\bf SDResultant} computes sparse differential resultant
$\SR$ of $\P_0,\ldots,\P_n$ with the following complexities: \vf

 1) In terms of the degree bound $D$ of $\SR$, the algorithm needs at most
$O\big(\frac{(mD(J+n+2))^{O(l(J+1))}}{n^{n}}\big)$ $\Q$-arithmetic
operations, where $l=\sum_{i=0}^n(l_i+1)$
 is the
size of all $\P_i$.

 \vf
 2) The algorithm needs at most $O\big(\frac{(J+n+2)^{O(l(J+1))}m^{O(l(J+1)(J+n+1))}}{n^{n}}\big)$
$\Q$-arithmetic operations.
\end{theorem}
\proof The algorithm finds a differential polynomial $P$ in
$\Q\{\bu_0,\ldots,\bu_n\}$ satisfying equation \bref{eq-deg5}, which
has the smallest order and the smallest degree in those with the
same order. Existence for such a differential polynomial is
guaranteed by Theorem \ref{th-spardeg}. Such a $P$ must be in
$\sat(\SR)$ by equation \bref{eq-lsres}. Since each differential
polynomial in $\sat(\SR)$ not equal to $\SR$ either has  greater
order than $\SR$ or has the same order but greater degree than
$\SR$, $P$ must be $\SR$.

We will estimate the complexity of the algorithm below. Denote $D$
to be the degree bound of $\SR.$ By Theorem \ref{th-spardeg}, $D\leq
(m+1)^{\sum_{i=0}^n(J_i+1)}=(m+1)^{J+n+1}$, where
$J=\sum_{i=0}^nJ_i$.
In each loop of Step 3, the complexity of the algorithm is clearly
dominated by Step 3.1.2, where we need to solve a system of linear
equations $\mathcal {P}=0$ over $\Q$ in $\bc_{0}$ and $\bc_{ij}$.
It is easy to show that $|\bc_{0}|={d+L-1\choose L-1}$ and
$|\bc_{ij}|={d_1-m_i-1+L+n(h+1)\choose L+n(h+1)}$, where
$L=\sum_{i=0}^n (h_i+1)(l_i+1)$ and
$d_1=[m+1+\sum_{i=0}^n(h_i+1)m_{i0}]d$. Then $\mathcal {P}=0$ is a
linear equation system with $N={d+L-1\choose
L-1}+\sum_{i=0}^n(h_i+1){d_1-m_i-1+L+n(h+1)\choose L+n(h+1)}$
variables and $M={d_1+L+n(h+1)\choose L+n(h+1)}$ equations. To solve
it, we need at most $(\max\{M,N\})^{\omega}$ arithmetic operations
over $\Q$, where $\omega$ is the matrix multiplication exponent and
the currently best known $\omega$ is 2.376.

The iteration in Step 3.1.2 may go through $1$ to
$\prod_{i=0}^n(m_i+1)^{h_i+1}\leq(m+1)^{\sum_{i=0}^n(J_i+1)}$, and
the iteration in Step 3.1 at most will repeat $\prod_{i=0}^n(J_i+1)$
times. And by Theorem \ref{th-spardeg}, Step 3 may loop from $o=0$
to $\sum_{i=0}^n(J_i+1)$. The whole algorithm needs at most
%\begin{eqnarray}\quad &&
{\small \begin{eqnarray}& &\sum_{o=0}^{\sum_{i=0}^n J_i}\sum_{
h_i\leq
J_i\atop\sum_{i}h_i=o}\sum_{d=1}^{\prod_{i=0}^n(m_i+1)^{h_i+1}}\big(\max\{M,
N\}\big)^{2.376}\nonumber\\
 & \leq& (J+1)
\Big(\prod_{i=0}^n(J_i+1)\Big)\cdot D \bigg[(J+n+2)\binom{[m+1+\sum_{i=0}^n(J_i+1)m_{i0}]D+L+n(h+1)}{L+n(h+1)}\bigg]^{2.376}\nonumber\\
 & \leq&
(J+n+2)^{3.376}\big(\frac{\sum_{i=0}^n(J_i+1)}{n+1}\big)^{n+1}\cdot D\cdot\big[\big(m(J+n+2)D\big)^{2.376(L+n(h+1))}\nonumber\\
 & \leq& (J+n+2)^{3.376}\frac{(J+n+1)^{n+1}}{n^{n}}\cdot D\cdot\big(m(J+n+2)D\big)^{2.376((l+n)(J+1)+n)}\nonumber
% & \leq& O\big(\frac{(mD(J+n+2))^{O(l(J+1))}}{n^{n}}\big)\nonumber
\end{eqnarray} }
%\nonumber\end{eqnarray}

\noindent arithmetic operations over $\Q$. The above inequalities
follow from the fact that $h\le J$, $L=\sum_{i=0}^n
(h_i+1)(l_i+1)\leq lJ+l$, $L+n(h+1)\leq(l+n)J+l+n=(l+n)(J+1)$ and
the assumption $[m+1+\sum_{i=0}^n(J_i+1)m]D\geq L+n(h+1)$.

Since $l\ge 2(n+1)$, the complexity bound is
$O\big(\frac{(mD(J+n+2))^{O(l(J+1))}}{n^{n}}\big)$.
%
%When $J\geq n$, the complexity is $O(\frac{J^{O(lJ)}m^{O(lJ^2)}}{n^{n}})$.
%
Our complexity assumes an $O(1)$-complexity cost for all field
operations over $\mathbb{Q}$. Thus, the complexity follows.
Now 1) is proved. To prove 2), we just need to replace $D$ by the
degree bound for $\SR$ in Theorem \ref{th-spardeg} in the complexity
bound in 1).\qedd

%& \leq& O\big(\frac{(mD(J+n+2))^{O(lJ+n)}}{n^{n}}\big)\leq
% O\big(\frac{(J+n+2)^{O(lJ+n)}m^{O(lJ^2+n^2)}}{n^{n}}\big)\nonumber

\begin{remark}
As we indicated at the end of Section 3.3, if we  first use
Algorithm~\ref{algo-red} to compute the rank-essential set $\TT$,
then the algorithm can be improved by only considering the Laurent
differential polynomials $\P_i\,(i\in\TT)$ in the linear combination
of the sparse resultant.
\end{remark}

\begin{remark}
Algorithm {\bf SDResultant} can be improved by using a better search
strategy. If $d$ is not big enough, instead of checking $d+1$, we
can check $2d$. Repeating this procedure, we may find a $k$ such
that $2^k\leq\deg(\SR)\leq2^{k+1}$. We then bisecting the interval
$[2^k,2^{k+1}]$ again to find the proper degree for $\SR$. This will
lead to a better complexity, which is still single exponential.
\end{remark}

\begin{remark}
If the given system is algebraic, that is $J=0$, then the complexity
bound given in 1) of Theorem \ref{th-cdres} is essentially the same
as that given in \cite{sturmfels}[p. 288] since $D\gg m$ and $D\gg
n$.
\end{remark}

\vskip 10pt For differential polynomials with non-vanishing degree
terms given in \bref{eq-qgpol}, a better degree bound is given in
Theorem~\ref{th-pspardeg}. Based on this bound, we can simplify the
Algorithm {\bf SDResultant} to compute the sparse differential
resultant by removing the computation for $P_i^N$ and $N_{i0}$ in
the first step where $N_{i0}$ is exactly equal to 1.
\begin{theorem}\label{th-0cdres}
Algorithm {\bf SDResultant} computes sparse differential resultants
for a differentially essential system of form \bref{eq-qgpol} with
at most
$O\big(\frac{(J+n+1)^{O(n)}m^{O(l(J+1)(J+n+1))}}{n^{n}}\big)$
%$O(n^{3.376}(s+1)^{O(n)}(m+1)^{O(nls^2)})$.
 $\Q$-arithmetic
operations.
\end{theorem}
\proof Follow the proof process of Theorem~\ref{th-cdres}, it can be
shown that the complexity is one mentioned in the theorem.
 \qedd

With Theorem \ref{th-cdres}, Theorem \ref{th-i2} is proved.

\subsection{Degree bound for differential resultant in terms of mixed volumes}

The degree bound given in Theorem \ref{th-spardeg} is essentially a
B\'{e}zout type bound.
In this section,  a BKK style degree bound for differential
resultant will be given, which is the sum of the mixed volumes of
certain polytopes generated by the supports of certain differential
polynomials and their derivatives.

We first recall results about the degree of algebraic sparse
resultant given by Sturmfels  (\cite{sturmfels2}).
%
%\vskip10pt
%{\bf A.  Degree of algebraic sparse resultant}
%
Let $\kk[\X]=\kk[x_1,\ldots,x_n]$ be the polynomial ring defined
over a field $\kk.$ For any vector
$\alpha=(a_1,\ldots,a_n)\in\mathbb{Z}^n$, denote the Laurent
monomial $x_1^{a_1}x_2^{a_2}\cdots x_n^{a_n}$ by $\X^{\alpha}$. Let
$\mathcal{B}_0,\ldots,\mathcal{B}_n\subset\mathbb{Z}^n$ be subsets
which jointly span the affine lattice $\mathbb{Z}^n$. Suppose ${\bf
0}=(0,\ldots,0)\in\mathcal{B}_i$ for each $i$ and
$|\mathcal{B}_i|=l_i+1\geq 2$. Let
\begin{equation}\label{eq-algsparse}
\BF_i(x_1,\ldots,x_n)=c_{i0}+\sum_{\alpha\in\mathcal{B}_i\backslash
\{{\bf 0}\}}c_{i,\alpha}\X^{\alpha}\,(i=0,1,\ldots,n)\end{equation}
be generic sparse Laurent polynomials defined w.r.t
$\mathcal{B}_i\,(i=0,1,\ldots,n)$. $\mathcal{B}_i$ or
$\{\X^{\alpha}:\alpha\in\mathcal{B}_i\}$ are called the support of
$\BF_i$. Denote $\bc_i=(c_{i\alpha})_{\alpha\in\mathcal{B}_i}$ and
$\bc=\cup_i(\bc_i\backslash \{c_{i0}\})$. Let $\CQ_i$ be the convex
hull of $\mathcal{B}_i$ in $\mathbb{R}^n$, which is the smallest
convex set containing $\mathcal{B}_i$. $\CQ_i$ is also called the
{\em Newton polytope} of $\BF_i$, denoted by $\NP(\BF_i)$. In
\cite{sturmfels2}, Sturmfels gave the definition of algebraic
essential set and proved that a necessary and sufficient condition
for the existence of sparse resultant is that there exists a unique
subset $\{\mathcal{B}_i\}_{i\in\text{ I}}$ which is essential. Now,
we restate the definition of essential sets in our words for the
sake of later use.

\begin{definition}
Follow the notations introduced above.
\begin{itemize}
\item A collection of  $\{\mathcal{B}_i\}_{i\in\text{
J}}$, or $\{\BF_i\}_{i\in\text{J}}$ of the form
(\ref{eq-algsparse}),
 is said
to be algebraically independent if
$\trdeg\,\mathbb{Q}(\bc)(\BF_i-c_{i0}:\,i\in\text{
J})/\mathbb{Q}(\bc)=|\text{J}|$. Otherwise, they are said to be
algebraically dependent.
\item A collection of $\{\mathcal{B}_i\}_{i\in\text{ I}}$ is said to be
essential if $\{\mathcal{B}_i\}_{i\in\text{I}}$ is algebraically
dependent and for each proper subset $\text{J}$ of \text{I},
$\{\mathcal{B}_i\}_{i\in\text{ J}}$ are algebraically independent.
\end{itemize}

\end{definition}

In the case that $\{\mathcal{B}_0,\ldots,\mathcal{B}_n\}$ is
essential, the degree of the sparse resultant can be described by
mixed volumes.

\begin{theorem}[\cite{sturmfels2}]\label{le-alg-sparse-deg}
Suppose that $\{\mathcal{B}_0,\ldots,\mathcal{B}_n\}$ is essential.
For each $i\in\{0,1,\ldots,n\}$, the degree of the sparse resultant
in $\bc_i$ is a positive integer, equal to the {\em mixed volume}
\begin{eqnarray} \quad&\quad&\mathcal
{M}(\CQ_0,\ldots,\CQ_{i-1},\CQ_{i+1},\ldots,\CQ_n)
=\sum_{\text{J}\subset
\{0,\ldots,i-1,i+1,\ldots,n\}}(-1)^{n-|\text{J}|}\vol(\sum_{j\in\text{J}}\CQ_j)
\nonumber
\end{eqnarray} where $\vol(\CQ)$ means the $n$-dimensional volume of
$\CQ\subset\mathbb{R}^n$ and $\CQ_1+\CQ_2$ means the Minkowski sum
of $\CQ_1$ and $\CQ_2$.
\end{theorem}

The mixed volume of the Newton polytopes of a polynomial system is
important in that it relates to the number of solutions of these
polynomial equations contained in $(\mathbb{C}^\ast)^n $, which is
the famous BKK bound. The following theorem explains it.
\begin{theorem} [Bernstein's Theorem](\cite{bkk})\label{th-BKK}
Given polynomials $f_1,\ldots,f_n$ over $\mathbb{C}$ with finitely
many common zeroes in $(\mathbb{C})^n$, let $\CQ_i$ be the Newton
polytope of $f_i$ in $\mathbb{R}^n$. Then the number of common
zeroes of the $f_i$ in $(\mathbb{C}^\ast)^n$ is bounded by the mixed
volume $\mathcal {M}(\CQ_1,\ldots,\CQ_n)$. Moreover, for generic
choices of the coefficients in the $f_i$, the number of common
solutions in $(\mathbb{C}^\ast)^n$ is exactly $\mathcal
{M}(\CQ_1,\ldots,\CQ_n)$.
\end{theorem}

It is well known that for a given polynomial system over
$\mathbb{C}$, the B\'{e}zout bound gives a bound for the number of
isolated solutions in $(\mathbb{C})^n $. Comparing the BKK bound
with B\'{e}zout bound, we have the following lemma.

\begin{lemma}\label{le-BKK-BEZOUT}
 Follow the  notations in Theorem~\ref{th-BKK}.
Then $\mathcal {M}(\CQ_1,\ldots,\CQ_n)\leq\prod_{i=1}^n\deg(f_i)$.
\end{lemma}
\proof Suppose $f_i\,(i=1,\ldots,m)$ are a system with generic
coefficients. Then by Theorem~\ref{th-BKK}, the number of common
zeroes of the $f_i$ in $(\mathbb{C}^\ast)^n$ is equal to the mixed
volume $\mathcal {M}(\CQ_1,\ldots,\CQ_n)$. And by
Lemma~\ref{le-deg-pol}, the number of common zeroes of the $f_i$ in
$(\mathbb{C})^n$ is bounded by $\prod_{i=1}^n\deg(f_i)$. Thus,
$\mathcal {M}(\CQ_1,\ldots,\CQ_n)\leq\prod_{i=1}^n\deg(f_i)$
follows. \qedd

\vskip10pt
% {\bf B. Degree bound for the differential resultant in terms of mixed volume}

\vskip5pt In the rest of this section, the degree of sparse
resultant will be used to give a degree bound for differential
resultant in terms of mixed volumes.
A system of $n+1$ generic differential polynomials with degrees
$m_0,\ldots,m_n$ and orders $s_0,\ldots,s_n$ respectively of the
form
\begin{equation}\label{eq-generic}
 \P_i=u_{i0}+\sum_{
 \begin{array}{c} \alpha \in \mathbb{Z}^{n(s_i+1)}_{\geq 0} \\ 1\leq |\alpha|\leq
 m_i
 \end{array}}u_{i \alpha}(\Y^{[s_i]})^{\alpha}\, (i=0,\ldots,n),
\end{equation}
of course forms a differentially essential system and their sparse
differential resultant is exactly equal to their differential
resultant defined in \cite{gao}. So Theorem \ref{th-pspardeg} also
gives a degree bound for differential resultant. But when we use
Theorem \ref{th-pspardeg} to estimate the degree of $\SR$, not only
Be\'{z}out bound is used, but also the degrees of $\P_i$ in both
$\Y$ and $\bu_i$ are considered.

The following theorem  gives a better upper bound for  degrees of
differential resultants, the proof of which is not valid for sparse
differential resultants. Precisely, in the following result, when
estimate the degree of $\SR$, the BKK bound is used rather than the
Be\'{z}out bound as did in Theorem \ref{th-pspardeg}.

\begin{theorem}\label{th-resultant-deg}
Let $\P_i\,(i=0,\ldots,n)$ be generic differential polynomials in
$\Y=\{y_1,\ldots,y_n\}$ with order $s_i$, degree $m_i$, and
coefficients $\bu_i$ respectively.  Let $\SR(\bu_0,\ldots,\bu_n)$ be
the differential resultant of $\P_0,\ldots,\P_n$. Denote
$s=\sum_{i=0}^n s_i$. Then for each $i\in\{0,1,\ldots,n\}$,
\begin{equation}\deg(\SR,\bu_i)\leq\sum_{k=0}^{s-s_i}\mathcal
{M}\big((\CQ_{jl})_{j\neq i,0\leq l\leq
s-s_j},\CQ_{i0},\ldots,\CQ_{i,k-1},\CQ_{i,k+1},\ldots,\CQ_{i,s-s_i}\big)
\end{equation}
where $\CQ_{jl}$ is the Newton polytope of $\P_j^{(l)}$ as a
polynomial in $y^{[s]}_1,\ldots,y^{[s]}_n$.
\end{theorem}
\proof By \cite[Theorem 6.8]{gao},
$\ord(\SR,\bu_i)=s-s_i\,(i=0,\ldots,n)$ and $(\SR)=(\P_0^{[s-s_0]},$
$\ldots,$ $\P_n^{[s-s_n]})\cap \Q[\bu_0^{[s-s_0]},$
$\ldots,\bu_n^{[s-s_n]}]$. Regard
$\P_i^{(k)}~(i=0,\ldots,n,k=0,\ldots,s-s_i)$ as polynomials in the
$n(s+1)$ variables
$\Y^{[s]}=\{y_1,\ldots,y_n,y'_1,\ldots,y'_n,\ldots,y^{(s)}_1,$
$\ldots,y^{(s)}_n\}$, and we denote its support by
$\mathcal{B}_{ik}$. Let $\BF_{ik}$ be the generic sparse polynomial
with support $\mathcal{B}_{ik}$. Denote $\bv_{ik}$ to be the set of
coefficients of $\BF_{ik}$ and in particular, suppose $v_{ik0}$ is
the coefficient of the monomial $1$ in $\BF_{ik}$. Now we claim that

\vskip5pt C1) $\overline{\mathcal{B}}=\{\mathcal{B}_{ik}:0\leq i\leq
n; 0\leq k \leq
 s-s_i\}$ is an essential set.

\vskip5pt

  C2) $\overline{\mathcal{B}}=\{\mathcal{B}_{ik}:0\leq i\leq n; 0\leq k \leq
 s-s_i\}$ jointly span the affine lattice $\mathbb{Z}^{n(s+1)}$.

\vskip5pt

Note that $|\overline{\mathcal{B}}|=n(s+1)+1$. To prove C1), it
suffices to show that any $n(s+1)$ of distinct $\BF_{ik}$ are
algebraically independent.
 Without loss of generality, we prove that for a fixed $k\in\{0,\ldots,s-s_0\}$,
 $$S_k=\{(\BF_{jl})_{1\leq j \leq n;0\leq l\leq s-s_j},\BF_{00},\ldots,
   \BF_{0,k-1},\BF_{0,k+1},\ldots,\BF_{0,s-s_0}\}$$
is an algebraically independent set. Remember that
$\{y_1,\ldots,y_n,y'_1,\ldots,y'_n,\ldots,y^{(s_i+l)}_1,\ldots,$
$y^{(s_i+l)}_n\}$ is a subset of the support of $\BF_{il}$. Now we
choose a monomial from each $\BF_{jl}$ and denote it by
$m(\BF_{jl})$. For each $j\in\{1,\ldots,n\}$ and
$l\in\{0,\ldots,s-s_j\}$, let $m(\BF_{jl})=y_j^{(s_j+l)}$ which
belongs to the support of $\BF_{jl}$. For the fixed $k$, there
exists a $\tau\in\{0,1,\ldots,n-1\}$ such that either
$\sum_{i=1}^{\tau}s_{i}\leq k\leq \sum_{i=1}^{\tau+1}s_{i}-1$ for
some $\tau\in\{0,1,\ldots,n-2\}$ or $\sum_{i=1}^{\tau}s_{i}\leq
k\leq \sum_{i=1}^{\tau+1}s_{i}$ for $\tau=n-1$. Here when $\tau=0$,
it means $0\leq k\leq s_1-1$. Then for $l\neq k$, let
\begin{eqnarray*} m(\BF_{0l})=\left\{\begin{array}{lll}
y_1^{(l)}&\quad&0\leq
l\leq s_1-1 \\ y_2^{(l-s_1)}&\quad&s_1\leq l\leq s_1+s_2-1 \\
\quad\vdots &\quad&\quad\vdots\\
y_{\tau+1}^{(l-\sum_{i=1}^{\tau}s_i)}&\quad&\sum_{i=1}^{\tau}s_{i}\leq
l\leq k-1\\
y_{\tau+1}^{(l-\sum_{i=1}^{\tau}s_i-1)}&\quad&k+1\leq l \leq
\sum_{i=1}^{\tau+1}s_{i}\\y_{\tau+2}^{(l-\sum_{i=1}^{\tau+1}s_i-1)}&\quad&\sum_{i=1}^{\tau+1}s_{i}+1\leq
l\leq \sum_{i=1}^{\tau+2}s_{i}\
\\\quad\vdots&\quad&\quad\vdots\\
y_{n}^{(l-\sum_{i=1}^{n-1}s_i-1)}&\quad&\sum_{i=1}^{n-1}s_{i}+1\leq l\leq \sum_{i=1}^{n}s_{i}=s-s_0 \\
\end{array}\right.
\end{eqnarray*}

So $m(S_k)$ is equal to $\{y_j^{[s]}:1\leq j\leq n\}$, which are
algebraically independent over $\mathbb{Q}$. Thus, the $n(s+1)$
members of $S_k$ are algebraically independent over $\Q$. For if
not, $\BF_{jl}-v_{jl0}$ are algebraically dependent over $\Q(\bv)$
where
$\bv=\cup_{i=0}^{n}\sum_{k=0}^{s-s_i}\bv_{ik}\backslash\{v_{ik0}\}$.
Now specialize the coefficient of $m(\BF_{jl})$ in $\BF_{jl}$ to 1,
and all the other coefficients of $\BF_{jl}-v_{jl0}$ to 0, by the
algebraic version of Lemma~\ref{lm-special},
$\{m(\BF_{jl}):\,\BF_{jl}\in S_k\}$ are algebraically dependent,
which is a contradiction. Thus, claim C1) is proved. Claim C2)
follows from the fact that $\{1,y_j^{[s]}:1\leq j\leq n\}$ is
contained in the support of $\BF_{0,s-s_0}$.

From the claims C1) and C2), the sparse resultant of
$(\BF_{ik})_{0\leq i\leq n; 0\leq k\leq s-s_i}$ exists and we denote
it by $G$. Then $(G)=\big((\BF_{ik})_{0\leq i\leq n; 0\leq k\leq
s-s_i}\big)\bigcap$ $\mathbb{Q}[(\bv_{ik})_{0\leq i\leq n; 0\leq
k\leq s-s_i}]$, and by Theorem~\ref{le-alg-sparse-deg},
$\deg(G,\bv_{ik})=\mathcal {M}\big((\CQ_{jl})_{j\neq i,0\leq l\leq
s-s_j},\CQ_{i0},\ldots,$
$\CQ_{i,k-1},\CQ_{i,k+1},\ldots,\CQ_{i,s-s_i}\big)$.

Now suppose $\xi$ is a generic point of the zero ideal  $(0)$ in
$\mathbb{Q}(\bv)[\Y^{[s]}]$. Let $\zeta_{ik}=-\BF_{ik}(\xi)+v_{ik0}$
and
$\overline{\zeta}_{ik}=-\P_i^{(k)}(\xi)+u_{i0}^{(k)}$\,($i=0,\ldots,n;k=0,\ldots,s-s_i$).
Clearly, $\zeta_i$ and $\overline{\zeta}_i$ are free of $v_{ik0}$
and $u_{i0}^{(k)}$ respectively. It is easy to see that
$(\xi;\bv,\zeta_{00},\ldots,\zeta_{0,s-s_0},\ldots,\zeta_{n0},\ldots,\zeta_{n,s-s_n})$
is a generic point of the algebraic prime ideal
$\big((\BF_{ik})_{0\leq i\leq n; 0\leq k\leq
s-s_i}\big)\subset\Q[\Y^{[s]},(\bv_{ik})_{0\leq i\leq n; 0\leq k\leq
s-s_i}]$, while $(\xi;\cup_{i=0}^n(\bu_i\backslash
\{u_{i0}\})^{[s-s_i]},$
$\overline{\zeta}_{00},\ldots,\overline{\zeta}_{0,s-s_0},$ $\ldots,
\overline{\zeta}_{n0},$ $\ldots,\overline{\zeta}_{n,s-s_n})$ is a
generic point of the algebraic prime ideal
$\big((\P_{i}^{(k)})_{0\leq i\leq n; 0\leq k\leq
s-s_i}\big)\subset\Q[\Y^{[s]},$
$\bu_0^{[s-s_0]},\ldots,\bu_n^{[s-s_n]}]$. If we regard $G$ as a
polynomial in $v_{ik0}$ over $\mathbb{Q}(\bv)$, then $G$ is the
vanishing polynomial of
$(\zeta_{00},\ldots,\zeta_{0,s-s_0},\ldots,\zeta_{n0},$
$\ldots,\zeta_{n,s-s_n})$ over $\Q(\bv)$.
Now specialize the coefficients $\bv_{ik}$ of $\BF_{ik}$ to the
corresponding coefficients of $\P_i^{(k)}$. Then $\zeta_i$ are
specialized to $\overline{\zeta}_i$. In particular, $v_{ik0}$ are
specialized to $u_{i0}^{(k)}$ which are algebraically independent
over the field $\Q(\xi,\cup_{i=0}^n\bu_i^{[s-s_i]}\backslash
u_{i0}^{[s-s_i]})$. We claim that there exists a nonzero polynomial
$H(\cup_{i=0}^n\bu_i^{[s-s_i]}\backslash u_{i0}^{[s-s_i]};$
$u_{00},\ldots,u_{00}^{(s-s_0)},\ldots,$
$u_{n0},\ldots,u_{n0}^{(s-s_n)})\in \Q[\bu_0^{[s-s_0]},\ldots,$
$\bu_n^{[s-s_n]}]$ such that

\vskip5pt
C3) $H(\cup_{i=0}^n\bu_i^{[s-s_i]}\backslash
u_{i0}^{[s-s_i]};\overline{\zeta}_{00},\ldots,\overline{\zeta}_{0,s-s_0},\ldots,\overline{\zeta}_{n0},\ldots,\overline{\zeta}_{n,s-s_n})=0$
and

\vskip5pt C4) $\deg(H,\bu_i^{[s-s_i]})\leq
\deg(G,\cup_{k=0}^{s-s_i}\bv_{ik})$. \vskip5pt We obtain $H$ by
specializing  $\bv$ one by one in $G$. For each $v\in\bv$, denote
$u$ to be its corresponding coefficient in $\P_i^{(k)}$. Now we
first specialize $v $ to $u $ and suppose $\zeta_{ik}$ is
specialized to $\tilde{\zeta}_{ik}$ correspondingly. Clearly,
$G(\bv\backslash\{v\},u;\tilde{\zeta}_{00},\ldots,\tilde{\zeta}_{0,s-s_0},\tilde{\zeta}_{n0},\ldots,\tilde{\zeta}_{n,s-s_n})=0.$
If
$\bar{G}=G(\bv\backslash\{v\},u;v_{000},v_{010},\ldots,v_{0,s-s_0,0},\ldots,v_{n00},v_{n10}\ldots,v_{n,s-s_n,0})\neq0$,
denote $\bar{G}$ by $H_1$. Otherwise, there exists some
$a\in\mathbb{N}$ such that $G=(v-u)^aG_1$ with $G_1|_{v=u}\neq0$.
But
$G(\bv\backslash\{v\},u;\tilde{\zeta}_{00},\ldots,\tilde{\zeta}_{0,s-s_0},\tilde{\zeta}_{n0},$
$\ldots,\tilde{\zeta}_{n,s-s_n})=0=
(v-u)^aG_1(\bv\backslash\{v\},u;\tilde{\zeta}_{00},\ldots,\tilde{\zeta}_{0,s-s_0},\tilde{\zeta}_{n0},$
$\ldots,\tilde{\zeta}_{n,s-s_n})$, so
$G_1(\bv\backslash\{v\},u;\tilde{\zeta}_{00},\ldots,\tilde{\zeta}_{0,s-s_0},\tilde{\zeta}_{n0},\ldots,\tilde{\zeta}_{n,s-s_n})=0$.
Denote $G_1|_{v=u}$ by $H_1$. Clearly,
$\deg(H_1,\bu_i^{[s-s_i]}\bigcup\cup_{k}\bv_{ik})\leq\deg(G,\cup_{k}\bv_{ik})$
for each $i$. Continuing this process for $|\bv|$ times till each
$v\in\bv$ is specialized to its corresponding element $u$, we will
obtain a nonzero polynomial $H_{|\bv|}(\cup_{i=0}^n(\bu_i\backslash
\{u_{i0}\})^{[s-s_i]};v_{000},v_{010},\ldots,v_{0,s-s_0,0},\ldots,v_{n00},v_{n10},$
$\ldots,v_{n,s-s_n,0})$ satisfying
$H_{|\bv|}(\cup_{i=0}^n(\bu_i\backslash
\{u_{i0}\})^{[s-s_i]};\overline{\zeta}_{00},\ldots,\overline{\zeta}_{0,s-s_0},\overline{\zeta}_{n0},\ldots,\overline{\zeta}_{n,s-s_n})=0$
and moreover, for each $i$,
$\deg(H_{|\bv|},\bu_i^{[s-s_i]}\bigcup\cup_{k}\{v_{ik0}\})\leq\deg(G,\cup_{k}\bv_{ik})$.
Since $u_{i0}^{(k)}$ are algebraically independent over the field
$\Q(\xi,\cup_{i=0}^n(\bu_i\backslash \{u_{i0}\})^{[s-s_i]})$,
$H=H_{|\bv|}(\cup_{i=0}^n(\bu_i\backslash
\{u_{i0}\})^{[s-s_i]};u_{00},\ldots,u_{00}^{(s-s_0)},\ldots,$
$u_{n0},$ $\ldots,u_{n0}^{(s-s_n)})\in \Q[\bu_0^{[s-s_0]},\ldots,$
$\bu_n^{[s-s_n]})$ is a polynomial satisfying C3) and C4).

From C3), $H\in (\P_0^{[s-s_0]},\ldots,$ $\P_n^{[s-s_n]}).$ Since
$(\P_0^{[s-s_0]},\ldots,\P_n^{[s-s_n]})\cap\Q[\bu_0^{[s-s_0]},\ldots,\bu_n^{[s-s_n]}]=(\SR)$
and $\SR$ is irreducible, $\SR$ divides $H$. It follows that
$\deg(\SR,\bu_i^{[s-s_i]})\leq \deg(H,\bu_i^{[s-s_i]})$ $\leq
\deg(G,\cup_{k}\bv_{ik})=\sum\limits_{k=0}^{s-s_i}\deg(G,\bv_{ik})=\sum\limits_{k=0}^{s-s_i}\mathcal
{M}\big((\CQ_{jl})_{j\neq i,0\leq l\leq
s-s_j},\CQ_{i0},\ldots,\CQ_{i,k-1},\CQ_{i,k+1},\ldots,$
$\CQ_{i,s-s_i}\big)$.
 \qedd

%In the above theorem, we give a degree bound for differential
%resultant using the BKK bound.
%
As a corollary, we give another degree bound for differential
resultant by using B\'{e}zout bound, which is better than the bound
given in Theorem \ref{th-pspardeg} in that only the degrees of
$\P_i$ in $\Y$ are considered in the bound.

\begin{cor}\label{cor-dbb}
Let $\P_i\,(i=0,\ldots,n)$ be generic differential polynomials in
$\Y=\{y_1,\ldots,y_n\}$ with order $s_i$, degree $m_i$ and
coefficients $\bu_i$ respectively.  Let $\SR(\bu_0,\ldots,\bu_n)$ be
the differential resultant of $\P_0,\ldots,\P_n$. Denote
$s=\sum_{i=0}^n s_i$. Then for each $i\in\{0,1,\ldots,n\}$,
$\deg(\SR,\bu_i)\le \frac{s-s_i+1}{m_i}\prod_{j=0}^n m_j^{s-s_j+1}$.
\end{cor}
\proof Follow the notations in the proof of
Theorem~\ref{th-resultant-deg}. Since $\{\mathcal{B}_{ik}:0\leq
i\leq n; 0\leq k \leq
 s-s_i\}$ is an essential set, for any fixed $k\in\{0,\ldots,s-s_i\}$,  the polynomials in
 $S_k$ together generate an ideal of dimension zero in $\Y^{[s]}$. By
 lemma~\ref{le-BKK-BEZOUT}, $\mathcal
{M}\big((\CQ_{jl})_{j\neq i,0\leq l\leq
s-s_j},\CQ_{i0},\ldots,\CQ_{i,k-1},\CQ_{i,k+1},$
$\ldots,\CQ_{i,s-s_i}\big)\leq\frac{1}{m_i}\prod_{j=0}^nm_j^{s-s_j+1}$.
 Hence, by Theorem~\ref{th-resultant-deg}, \begin{eqnarray}\deg(\SR,\bu_i)&\leq&\sum_{k=0}^{s-s_i}\mathcal
{M}\big((\CQ_{jl})_{j\neq i,0\leq l\leq
s-s_j},\CQ_{i0},\ldots,\CQ_{i,k-1},\CQ_{i,k+1},\ldots,\CQ_{i,s-s_i}\big)\nonumber \\
&\leq&\sum_{k=0}^{s-s_i}\frac{1}{m_i}\prod_{j=0}^nm_j^{s-s_j+1}
%\nonumber\\&=&
=\frac{s-s_i+1}{m_i}\prod_{j=0}^nm_j^{s-s_j+1}.\nonumber
 \end{eqnarray}\qedd

\begin{example}
Consider two generic differential polynomials of order  one and
degree two in one indeterminate $y$:
\begin{eqnarray*}
 \P_0 &=& u_{00}+u_{01}y+u_{02}y'+u_{03}y^2+u_{04}yy'+u_{05}(y')^2,\\
 \P_1 &=& u_{10}+u_{11}y+u_{12}y'+u_{13}y^2+u_{14}yy'+u_{15}(y')^2.
\end{eqnarray*}
Then the degree bound given by Theorem \ref{th-spardeg} is
$\deg(\SR)\le (2+1)^4=81$.
The degree bound given by Corollary \ref{cor-dbb} is
$\deg(\SR,\bu_0)\le 2^4=16$ and hence $\deg(\SR)\le 32$.
The degree bound $\deg(\SR,\bu_0)$ given by Theorem
\ref{th-resultant-deg} is  $\mathcal{M}(\CQ_{10},\CQ_{11},\CQ_{00})
+
 \mathcal{M}(\CQ_{10},\CQ_{11},\CQ_{01})=4+6=10$  and
 consequently $\deg(\SR)\leq20$,
 where
 $\CQ_{01}=\CQ_{10}=\conv\{(0,0,0),(2,0,0),(0,2,0)\}$,
 $\CQ_{01}=\CQ_{11}=\conv\{(0,0,0),(2,0,0),(0,2,0),$ $(0,0,1),(1,0,1),$
  $(0,1,1)\}$,
%
%$\CQ_{00}=\conv\{(0,0,0),(2,0,0),(0,2,0)\}$ and
%$\CQ_{01}=\conv\{(0,0,0),(2,0,0),(0,2,0),$ $(0,0,1),$
%$(1,0,1),(0,1,1)\}$,
%
and $\conv(\cdot)$ means taking the convex hull in
$\mathbb{R}^3$.\end{example}

\vf We will end this section by giving Algorithm {\bf DResultant} to
compute differential resultant based on the degree bound given in
Theorem~\ref{th-resultant-deg}.

\begin{algorithm}\label{alg-dres2}
  \caption{\bf --- DResultant($\P_0,\ldots,\P_n$)} \smallskip
  \Inp{A generic differential polynomial system $\P_0,\ldots,\P_n$.}\\
  \Outp{The differential resultant $\SR(\bu_0,\ldots,\bu_n)$ of $\P_0,\ldots,\P_n$.}\medskip

  \noindent
  1. For $i=0,\ldots,n$, set $s_i=\ord(\P_i)$, $m_i=\deg(\P_i,\Y)$ and $\bu_i=\coeff(\P_i)$.\\
  %2. For each $\mathcal{S}_{ik}=\{(\NP(\P_j^{(l)}))_{0\leq j\neq i \leq n;0\leq l\leq s-s_j},(\NP(\P_i^{(l)}))_{0\leq l\neq k\leq s-s_i}\}$,
%   compute $\mathcal{M}(\mathcal{S}_{ik})$.\\
%\SPC
%Set $D_i=\sum_{k=0}^{s-s_i}\mathcal{M}(\mathcal{S}_{ik}).\,\,(i=0,\ldots,n)$\\
  2. Set $\SR=0$,  $s=\sum_{i=0}^ns_i$, $m=\max_i\{m_i\}$, $d=n+1$, $U=\cup_{i=0}^n\bu_i^{[s-s_i]}$.\\
  3. While $\SR=0$ do\\
\SPC 3.1. Set $\SR_0$ to be a homogenous GPol   of degree $d$ in $U$.\\
\SPC 3.2. Set $\bc_{0}=\coeff(\SR_0,U)$.\\
   \SPC 3.3. Set $G_{ik}(i=0,\ldots,n;k=0,\ldots,s-s_i)$ to be GPols of degree $(m+1)d-m_i-1$ \\
  \SPC\SPC\,\,\, in $\Y^{[s]},U$.\\
\SPC 3.4. Set $\bc_{ik}=\coeff(G_{ik}, \Y^{[s]}\cup U)$.
   \\
\SPC 3.5. Set $\mathcal {P}$ to be the set of coefficients of
$\SR_0(\bu_0,\ldots,\bu_n)-\sum_{i=0}^n\sum_{k=0}^{s-s_i}G_{ik}\P_{i}^{(k)}$
\\   \SPC \SPC\,\, as a  polynomial in
  $\Y^{[s]},U$.\\
  \SPC 3.6. Solve the linear equation $\mathcal {P}=0$ in
   variables $\bc_{0}$ and
  $\bc_{ik}.$\\
\SPC 3.7. If $\bc_0$ has a nonzero solution, then
   substitute it into  $\SR_0$ to get $\SR$ and go to  \\
\SPC \SPC\,\,\, Step 4, else $\SR=0$.\\
    \SPC 3.8. d:=d+1.\\
  4. Return $\SR$.\medskip

  \noindent /*/\; GPol stands for generic algebraic polynomial.\smallskip

  \noindent /*/\; $\coeff(P,U)$ returns the set of coefficients of $P$ as an ordinary  polynomial in
  variables $U$.

\smallskip
\end{algorithm}

\begin{theorem}\label{th-cdres2}
Let $\P_0,\ldots,\P_n$ be a generic differential polynomial system
of the form~(\ref{eq-generic}). Denote $s=\sum_{i=0}^n
\ord(\P_i,\Y)$ and $m=\max_{i=0}^n \deg(\P_i,\Y)$.
Algorithm {\bf DResultant} computes the  differential resultant
$\SR$ of $\P_0,\ldots,\P_n$ with the following complexities: \vf

 1) In terms of $\deg(\SR)$, the algorithm needs at most
$O\big((ns+n)^{2.376}[m\deg(\SR)]^{O(ln(s+1))}\big)$ $\Q$-arithmetic
operations where $l=\max_{i=0}^n{m_i+n(s_i+1)\choose n(s_i+1)}$
 is the
size of system $\P_i$.

 \vf

 2) The algorithm needs at most $O\big((ns+n)^{2.376}[mD]^{O(ln(s+1))}\big)$ $\Q$-arithmetic operations, where $D$ is the
degree bound of $\SR$ given by Theorem~\ref{th-resultant-deg}.
\end{theorem}
\proof The algorithm terminates  by Theorem~\ref{th-resultant-deg},
and returns a differential polynomial $P$ in
$(\P_0^{[s-s_0]},\ldots,\P_n^{[s-s_n]})\cap\Q[\bu_0^{[s-s_0]},\ldots,$
$\bu_n^{[s-s_n]}]$ with the smallest degree, which is exactly the
differential resultant.

We will estimate the complexity of the algorithm below. Denote
$l_i=|\bu_i|={m_i+n(s_i+1)\choose n(s_i+1)}\,$ $(i=0,\ldots,n),$ and
$l=\max_{i=0}^nl_i$. So $|U|=\sum_{i=0}^nl_i(s-s_i+1)$.
In each loop of Step 3, the complexity of the algorithm is clearly
dominated by Step 3.5., where we need to solve a system of linear
equations $\mathcal {P}=0$ over $\Q$ in $\bc_{0}$ and $\bc_{ik}$.
It is easy to show that $|\bc_{0}|={d+|U|-1\choose |U|-1}$ and
$|\bc_{ik}|={(m+1)d-m_i-1+|U|+n(s+1)\choose |U|+n(s+1)}$. Then
$\mathcal {P}=0$ is a linear equation system with
$N=|\bc_{0}|+\sum_{i=0}^n\sum_{k=0}^{s-s_i}|\bc_{ik}|$ variables and
$M={(m+1)d+|U|+n(s+1)\choose |U|+n(s+1)}$ equations. To solve it, we
need at most $(\max\{M,N\})^{\omega}$ arithmetic operations over
$\Q$, where $\omega$ is the matrix multiplication exponent and the
currently best known $\omega$ is 2.376.

Step 3 may loop from $d=n+1$ to $\deg(\SR)\leq D$, where $D$ is the
degree bound of $\deg(\SR)$ given by Theorem~\ref{th-resultant-deg}.
The whole algorithm needs at most {
\begin{eqnarray}\sum_{d=n+1}^{\deg(\SR)}\big(\max\{M, N\}\big)^{2.376}
%\leq \deg(\SR)\{(ns+n+1)[(m+1)\deg(\SR)]^{l(ns+n+1)+ns+n}\}^{2.376}
&\leq &O\big((n s+n)^{2.376}[m\deg(\SR)]^{O(ln(s+1))}\big)\nonumber
\\ &\leq& O\big((ns+n)^{2.376}[mD]^{O(ln(s+1))}\big)\nonumber\end{eqnarray}
}
arithmetic operations over $\Q$. In the above inequalities, we
assume that $(m+1)\deg(\SR)\geq l(ns+n+1)+n(s+1)$, which is
generally true.
Otherwise, $m\deg(\SR)$ need to be  replaced by $lns$ to give a
single exponential complexity bound.
Our complexity also assumes an $O(1)$-complexity cost for all field
operations over $\mathbb{Q}$. Thus, the complexity follows. \qedd

\begin{remark}
One might suggest to use an approach similar to Algorithm {\bf
DResultant} to compute the sparse differential resultant and to
obtain a complexity bound similar to that given in Theorem
\ref{th-cdres2}. In this way, a differential polynomial
$P\in\sat(\SR)$ will be obtained, which cannot be proved to be $\SR$
due to the reason that $P$ might have a higher order and a lower
degree than that of $\SR$.
\end{remark}

With Theorem \ref{le-alg-sparse-deg}, Theorem \ref{th-i3} is proved.

\section{Conclusion}
\label{sec-conc}

In this paper,  we first introduce the concepts of Laurent
differential polynomials and Laurent differentially essential
systems, and give a criterion for Laurent differentially essential
systems in terms of their supports. Then the sparse differential
resultant for Laurent differentially essential system is defined and
its basic properties are proved, such as the differential
homogeneity, necessary and sufficient conditions for the existence
of solutions, differential toric variety, and the Poisson-type
product formulas. Furthermore, order and degree bounds for the
sparse differential resultant are given. Based on these bounds, an
algorithm to compute the sparse differential resultant is proposed,
which is single exponential in terms of the order, the number of
variables, and the size of the Laurent differentially essential
system.

In the rest of this section, we propose several questions for
further study.

It is useful to represent the sparse differential resultant as the
quotient of two determinants, as done in \cite{dandrea1,emiris1} in
the algebraic case. In the differential case, we do not have such
formulas, even in the simplest case of the resultant for two generic
differential polynomials in one variable.
The treatment in \cite{dres1} is not complete. For instance, let
$f,g$ be two generic differential polynomials in one variable $y$
with order one and degree two. Then, the differential resultant for
$f,g$ defined in \cite{dres1} is zero, because all elements in the
first column of the matrix $M(\delta,n,m)$ in \cite[p.543]{dres1}
are zero.
Although using the idea of Dixon resultants, the algorithm in
\cite{yang-dixon} does not give a matrix representation for the
differential resultant.

From \bref{eq-deg5}, a natural idea to find a matrix representation
is trying to define the sparse differential resultant as the
algebraic sparse resultant of
$\P=\{\P_{i}^{(k)}\,(i=0,\ldots,n,k=0,\ldots,h_i)\}$ considered as
Laurent polynomials in $y^{(j)}_{l}$, which will lead to a matrix
representation for the sparse differential  resultant.
As far as we know, this is actually very difficult even in the case
of the resultant for two generic differential polynomials in one
variable.
%
%The major difficulty is to show that the algebraic sparse resultant
%of $\P$ is nonzero, which in turn is based on the analysis of the
%support of $\P$.

The degree of the algebraic sparse resultant is equal to the  mixed
volume of certain polytopes generated by the supports of the
polynomials \cite{Pedersen} or \cite[p.255]{gelfand}. A similar
degree bound is given in Theorem \ref{th-i3} for the differential
resultant. We conjecture that the bound given in Theorem \ref{th-i3}
is also a degree bound for the sparse differential resultant.

There exist very efficient algorithms to compute algebraic sparse
resultants\,\cite{emiris0,emiris1,emiris2005}, which are based on
matrix representations for the resultant. How to apply the
principles behind these algorithms to compute sparse differential
resultants is an important problem.
A reasonable goal is to find an algorithm whose complexity depends
on $\deg(\SR)$, but not on the worst case bound of $\deg(\SR)$ which
is the case in Algorithm {\bf SDResultant}.

In the algebraic case, it is shown that the sparse polynomials
$\P_i\,(i=0,\ldots,n)$ can be re-parameterized to a new system
$\SS_i\,(i=0,\ldots,n)$ with the help of the Newton polygon
associated with $\P_i$ such that vanishing of the sparse resultant
gives a sufficient and necessary condition for
$\SS_i\,(i=0,\ldots,n)$ to have solutions in $\CC^N$, where $\CC$ is
the field of complex numbers \cite[page 312]{cox}. It is interesting
to extend this result to the differential case. To do that we  need
a deeper study of differential toric variety introduced in Section
\ref{sec-toric}.

As a less important problem, we guess that assuming the first
condition in Theorem \ref{th-unique}, the second condition
$\textbf{e}_j\in\Span_{\mathbb{Z}}\{\alpha_{ik}-\alpha_{i0}:k=1,\ldots,l_i;i=0,\ldots,n\}$
is also a necessary condition for the system $\P_i$ to have a unique
solution under the condition of $\SR=0$ in the generic case.

% toric
% bkk via Connection between degree
%

\end{document}